%% file: paper.tex
\documentclass[letterpaper,twocolumn,10pt]{article}
\usepackage{usenix-2020-09}

\usepackage{amsmath}

\usepackage{filecontents}

\usepackage[title]{appendix}
\usepackage{amsfonts}
\usepackage{amssymb}
\usepackage{algorithmic}
\usepackage{graphicx}
\usepackage{textcomp}
\usepackage{xcolor}
\usepackage[most]{tcolorbox}

\usepackage{mathrsfs}
\usepackage{amsthm}
\usepackage{balance}
\usepackage{multirow}

\usepackage{epsfig,endnotes}
\usepackage{subfig}

\usepackage{grffile}
\usepackage{caption}

\usepackage{color, url}  
\usepackage[ruled,linesnumbered]{algorithm2e}

\usepackage[numbers,sort&compress,comma]{natbib}
\usepackage{balance}
\usepackage{bm}
\usepackage{rotating}

\usepackage{enumitem}
\usepackage{makecell}
\usepackage{pdfx}
\usepackage{bm}
\usepackage{booktabs}
\usepackage{float}
\usepackage{colortbl}

\newtheorem{corollary}{Corollary}

\newtheorem{proposition}{Proposition}
\newtheorem{remark}{Remark}

\newcommand{\argmax}{\operatornamewithlimits{argmax}}

\newcommand{\ours}{\texttt{ACE}}

\newcommand{\myparatight}[1]{\smallskip\noindent{\bf {#1}.}~}

\allowdisplaybreaks

\begin{document}

\date{}

\title{\Large \bf \ours: A Model Poisoning Attack on Contribution Evaluation Methods\\
in Federated Learning}

\author{
{\rm Zhangchen Xu}\\
University of Washington\\
\texttt{zxu9@uw.edu}
\and
{\rm Fengqing Jiang}\\
University of Washington\\
\texttt{fqjiang@uw.edu}
\and
{\rm Luyao Niu}\\
University of Washington\\
\texttt{luyaoniu@uw.edu}
\and
{\rm Jinyuan Jia}\\
Pennsylvania State University\\
\texttt{jinyuan@psu.edu}
\and
{\rm Bo Li}\\
University of Chicago\\
\texttt{bol@uchicago.edu}
\and
{\rm Radha Poovendran}\\
University of Washington\\
\texttt{rp3@uw.edu}
} 


\maketitle

\input{abstract}

\input{intro}

\input{background}

\input{problem}

\input{method}

\input{theory}

\input{exp}

\input{defense}

\input{discussion}

\input{conclusion}

\input{ack}


\bibliographystyle{plain}
\bibliography{ref}
\clearpage

\input{appx}

\end{document}

%% file: abstract.tex
\begin{abstract}
In \emph{Federated Learning (FL)}, a set of clients collaboratively train a machine learning model (called \emph{global model}) without sharing their local training data.
The local training data of clients is typically non-i.i.d. and heterogeneous, resulting in varying contributions from individual clients to the final performance of the global model. 
In response, many contribution evaluation methods were proposed, where the server could evaluate the contribution made by each client and incentivize the high-contributing clients to sustain their long-term participation in FL.
Existing studies mainly focus on developing new metrics or algorithms to better measure the contribution of each client. 
However, the security of contribution evaluation methods of FL operating in adversarial environments is largely unexplored. 
In this paper, we propose the \emph{first} model poisoning attack on contribution evaluation methods in FL, termed \ours.
Specifically, we show that any malicious client utilizing \ours~could manipulate the parameters of its local model such that it is evaluated to have a high contribution by the server, even when its local training data is indeed of low quality.
We perform both \emph{theoretical} analysis and \emph{empirical} evaluations of \ours. Theoretically, we show our design of \ours~can effectively boost the malicious client's perceived contribution when the server employs the widely-used cosine distance metric to measure contribution.
Empirically, our results show \ours~effectively and efficiently deceive five state-of-the-art contribution evaluation methods. In addition, \ours~preserves the accuracy of the final global models on testing inputs.
We also explore six countermeasures to defend \ours. Our results show they are inadequate to thwart \ours, highlighting the urgent need for new defenses to safeguard the contribution evaluation methods in FL.
\end{abstract}

%% file: intro.tex
\section{Introduction}


Federated learning (FL) \cite{mcmahan2017communication} enables a set of clients to collaboratively train a machine learning model, denoted as the \emph{global model}, using their local training data in an iterative manner. 
At each communication round, a cloud server first broadcasts the current global model to the clients.
Each client then adopts the global model as its local model, locally minimizes an empirical loss function (e.g., cross-entropy function) over its local training data to compute a \emph{local model update}, and finally sends the local model update to the server. 
The server aggregates the local model updates from the clients according to an aggregation rule (e.g., FedAvg~\cite{mcmahan2017communication}) to update the current global model.
FL has been widely deployed in real-world cross-silo settings where data is spread across multiple isolated organizations \cite{long2020federated,zheng2022applications,sheller2020federated}.

In practice, the local training data possessed by the clients in FL is non-i.i.d. and heterogeneous \cite{li2020federated,li2021fedsae,chai2019towards,tan2022fedproto}, and thus inherently of varying qualities.
Therefore, it is crucial to understand and evaluate the contribution of each client toward the performance (e.g., accuracy on testing inputs) of the global model.
Accurate contribution evaluation facilitates the designs of incentive mechanisms to encourage the clients, especially those owning high-quality data, to participate into FL \cite{xu2021gradient,gao2021fifl,deng2021fair,hu2022incentive}, which could further enhance the performance of the global model.
To this end, contribution evaluation methods in FL has been extensively studied \cite{shi2023towards,soltani2023survey}.
The existing studies~\cite{xu2021gradient,gao2021fifl,zhang2021refiner,deng2021fair,hu2022incentive, fan2022improving, wang2020principled,liu2022gtg, wang2019measure} primarily focus on the development of novel metrics or algorithms to measure the contributions of clients in FL.
At present, however, the security of contribution evaluation methods in FL remains largely unexplored.


\myparatight{Our Contribution}
In this paper, we propose the \emph{first} model poisoning attack on contribution evaluation methods in FL.
We term this attack as \ours.
We consider that an attacker owns a subset of clients in FL, denoted as malicious clients.
These malicious clients are evaluated as low-contributing participants by the server.
Specifically, the malicious clients manipulate the parameters of their local models, with the objective of elevating their contributions evaluated by the server. 
Accomplishing this attack goal can result in monetary advantages for the attacker when contribution-based incentive mechanisms are employed in FL.
For example, the FL server may distribute a certain amount of budget to the clients in proportion to their individual contributions in order to encourage clients with high contributions to remain in the FL \cite{nguyen2022trade}.
The attack studied in this paper can boost the contributions of malicious clients, and thus increase their shares of the budget.
This lowers the shares available to the other clients, thereby jeopardizing their interests.

We highlight that the attack goal of \ours~is significantly different from existing attacks in FL, including untargeted poisoning attacks \cite{shejwalkar2021manipulating,fang2020local,karimireddy2020byzantine, xie2020fall,cao2022mpaf,zhou2021deep} and backdoor attacks \cite{xie2019dba,bagdasaryan2020backdoor,sun2019can,wang2020attack, zhang2022neurotoxin, baruch2019little}. In particular, existing untargeted poisoning attacks and backdoor attacks aim to make the global model exhibit low performance for indiscriminate testing inputs or predict an attacker-chosen target class for any inputs embedded with a backdoor trigger \cite{jere2020taxonomy}. 
By contrast, \ours~aims to deceive the contribution evaluation method employed by the server to increase the contributions of malicious clients (with low-quality local training data). 
Our empirical studies demonstrate that \ours~retains the performance of the global models in different settings.
We defer the detailed discussion on the difference to Section~\ref{sec-discussion}.

\input{float_element/fig_overview}

A major challenge in developing \ours~is how the malicious clients should strategically manipulate the parameters of malicious clients' local models to increase the perceived contribution by the server.
Our insight to address this challenge is that the FL procedure provides the malicious clients information on the global model as shown in Figure \ref{fig: attack overview}, allowing the malicious client to predict how the global model evolves over communication rounds.
Therefore, the malicious clients can craft local model updates to better align with the prediction of the global model, making them more likely to be perceived by the server as having higher contributions.
\ours~uses the Cauchy mean value theorem \cite{lang2012real} to predict the global model at each communication round.
We show that predicting the global model significantly reduces the computation complexity for the malicious clients compared to iteratively learning local model updates from the local training datasets, allowing \ours~to boost the perceived contributions at negligible cost.


We \emph{theoretically} analyze the effectiveness of \ours~when the server measures contributions using cosine distance.
We prove that \ours~allows the malicious clients to always increase the perceived contribution by appropriately scaling up the predicted global model.
We further \emph{empirically} evaluate the effectiveness and efficiency of \ours~using five state-of-the-art contribution evaluation methods in FL \cite{lyu2020collaborative, xu2020reputation, xu2021gradient, wang2020principled, zhang2021refiner}.
We compare \ours~with four baselines using three models across three datasets including MNIST \cite{MNIST}, CIFAR-10 \cite{cifar10}, and Tiny-ImageNet \cite{tinyimagenet}.
We show that \ours~consistently outperforms all baselines when the local training data of clients is non-i.i.d., yielding the highest perceived contribution by the FL server. 
This demonstrates the severity of \ours~on contribution evaluation methods in FL.
We evaluate \ours~against countermeasures including extended Multi-Krum \cite{blanchard2017machine} and Trimmed-Mean \cite{yin2018byzantine}, which have been widely used in FL.
Our empirical evaluations demonstrate that these countermeasures are not effective against \ours.
These results underscore the need for the development of new defenses to thwart \ours.

To summarize, this paper makes the following major contributions:
\begin{itemize}
    \item We propose the first model poisoning attack on contribution evaluation methods in FL, termed as \ours.
    \item We present theoretical analysis and perform extensive empirical evaluations of \ours~to demonstrate its effectiveness and efficiency.
    \item We investigate the countermeasures to mitigate \ours. We show that \ours~can remain stealthy against the existing mitigation strategies, highlighting the needs for new defense mechanisms.
\end{itemize}


%% file: float_element/fig_overview.tex
\begin{figure}[t]
  \centering 
  \includegraphics[width=0.45\textwidth]{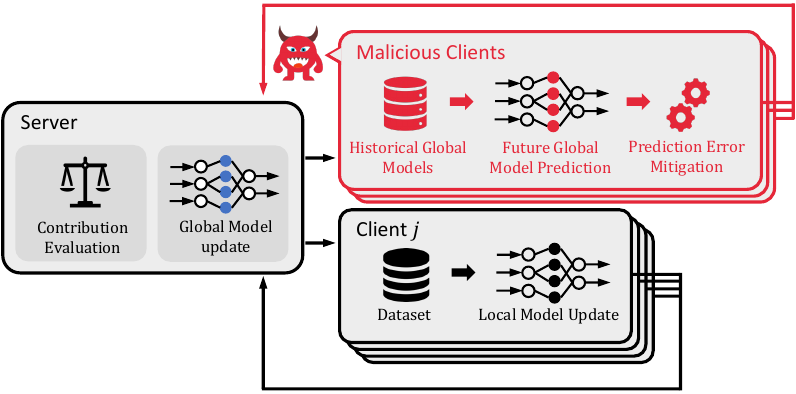}
  \caption{An illustration of \ours~consisting of two components: future global model prediction and prediction error mitigation.}                          
  \label{fig: attack overview} 
\end{figure}

%% file: background.tex
\section{Background and Related Work}\label{sec:background}
In this section, we present background on federated learning and contribution evaluation methods in FL.

\subsection{Federated Learning}
\label{Section: Federated Learning}

We consider an FL setting where $N$ clients collaboratively train a machine learning model, called \emph{global model}. Let $\mathcal{D}_i$ represent the local training dataset of the $i$-th client and $|\mathcal{D}_i|$ denote its size, where $i=1,2,\cdots, N$. We denote the set of clients as $\Gamma$, and thus the joint training dataset can be represented as $\mathcal{D}=\cup_{i\in\Gamma}\mathcal{D}_i$. 
To learn the global model, the set of clients collaboratively minimizes a loss function over their local training datasets as follows:
\begin{equation*}
    \min_{\mathbf{w}}\sum_{i\in\Gamma}L(\mathcal{D}_i;\mathbf{w}),
\end{equation*}
where $\mathbf{w}$ represents the parameters of the global model and $L(\mathcal{D}_i;\mathbf{w})$ is the empirical loss (e.g., cross-entropy loss) evaluated using the global model with parameters $\mathbf{w}$  on the local training dataset $\mathcal{D}_i$.
The clients iteratively solve the optimization problem through multiple communication rounds with an FL server.  
Specifically, there are three steps at each communication round $t$, which are detailed below.

\myparatight{Step I} 
The server broadcasts the current global model, denoted as $\mathbf{w}^t$, to the clients.

\myparatight{Step II} For each client $i$, it first uses the global model $\mathbf{w}^t$ to initialize its local model, and then uses the local training dataset $\mathcal{D}_i$ to update its local model by minimizing the empirical loss function $L$, i.e., $\mathbf{w}_i^{t+1} = \mathbf{w}^t - \eta_i \nabla L(\mathcal{D}_i;\mathbf{w}^t)$, where $\eta_i$ is the learning rate and $\nabla L(\mathcal{D}_i;\mathbf{w}^t)=\frac{\partial L(\mathcal{D}_i;\mathbf{w}^t)}{\partial \mathbf{w}^t}$. Finally, it sends the \emph{local model update} $\mathbf{g}^t_i = \mathbf{w}^{t} - \mathbf{w}_i^{t+1}$ back to the server. Note that it is equivalent for the client to send the local model update or local model due to the above relationship.

\myparatight{Step III} The server aggregates the local model updates from the clients, and updates the global model of the $(t+1)$-th communication round as $\mathbf{w}^{t+1} \gets \mathbf{w}^t - \mathbf{g}^t$, where $\mathbf{g}^t = \mathcal{A}(\mathbf{g}_1^t, \mathbf{g}_2^t, \cdots, \mathbf{g}_{N}^t)$ is the global model update at communication round $t$ and $\mathcal{A}$ is an aggregation rule.
A typical example of $\mathcal{A}$ is FedAvg \cite{mcmahan2017communication}, defined as $\mathcal{A}(\mathbf{g}_1^t, \mathbf{g}_2^t, \cdots, \mathbf{g}_{N}^t) = \sum_{i\in\Gamma} \frac{\left|\mathcal{D}_i\right|}{\left|\mathcal{D}\right|}\mathbf{g}_i^t$.

\subsection{Contribution Evaluation Methods in FL}
\label{Section: Contribution Evaluation in FL}

Contribution evaluation in FL aims to quantify the \emph{contribution} made by each client to the performance (e.g., accuracy on testing inputs) of the global model.
Following previous studies \cite{liu2022gtg,shi2023towards}, we divide the existing contribution evaluation methods in FL into three categories: \emph{self-reporting based contribution evaluation}~\cite{zeng2020fmore,yu2020sustainable}, \emph{individual performance based contribution evaluation}~\cite{lyu2020collaborative,xu2020reputation, xu2021gradient, shi2022fedfaim, jiang2023fair, zhang2021incentive, zhang2021quality, gao2021fifl, chen2020dealing}, and \emph{game theory based contribution evaluation}~\cite{fan2022improving,wang2019measure,zhang2021refiner,liu2022gtg,jia2019towards, ghorbani2019data,wang2020principled,zheng2023secure,wang2022efficient}.


\myparatight{Self-Reporting based Contribution Evaluation} The methods in this category utilize the information reported by each client to measure its contribution. For instance, Zeng et al.~\cite{zeng2020fmore} propose to use self-reported local data size and communication bandwidth to evaluate the contribution. Yu et al.~\cite{yu2020sustainable} proposed to use the data quantity and quality (e.g., measured by the marginal revenue generated by the global model) reported by each client to measure its contribution.
Since those methods heavily rely on self-reported information, they are susceptible to malicious clients who can untruthfully report their information to the server. 


\myparatight{Individual Performance based Contribution Evaluation} 
This category of methods proposes some performance metrics defined on the local model updates of clients to measure their contributions.
Some studies~\cite{lyu2020collaborative,chen2020dealing} assume the server has a clean validation dataset in order to calculate the proposed performance metrics.
For instance, Lyu et al.~\cite{lyu2020collaborative} evaluate the contribution of a client at each communication round using the validation accuracy of its local model on the validation dataset of the server.
Chen et al.~\cite{chen2020dealing} propose to quantify the contribution of a client at each individual communication round by computing the mutual cross-entropy loss between the local model of the client and the global model on the validation dataset. 
\textcolor{black}{However, these methods assume the server has a clean, small validation dataset, which has been shown to be impractical \cite{rieger2022deepsight}.}
When the server does not have any validation data, existing studies propose to utilize the distance (e.g., cosine distance~\cite{xu2020reputation, xu2021gradient, shi2022fedfaim, jiang2023fair, zhang2021incentive} and Euclidean distance~\cite{zhang2021quality, gao2021fifl}) between the global model and the local model of a client to measure its contribution at each communication round. 
The key assumption is that a client whose local model is more similar to the aggregated global model makes more contribution to it.
Therefore, the contribution of a client is smaller if the distance becomes larger. 

\myparatight{Game Theory based Contribution Evaluation} 
This category of methods formulates the FL as a cooperative game, where all clients collaboratively learn a global model using their local training data.
Then the aggregated contribution of the clients is represented by the utility of the game (e.g., accuracy  \cite{liu2022gtg} or empirical loss \cite{fan2022improving,wang2019measure,zhang2021refiner} of the global model on testing inputs), and the contribution of each individual client is modeled by its payoff received in this game.

Wang et al. \cite{wang2019measure} propose to quantify the contribution (averaged over all communication rounds) of each client in FL using its marginal loss, which is defined as the utility difference when the client joins the game as opposed to not joining the game. 
Similarly, Zhang et al. \cite{zhang2021refiner} measure the contributions of clients at each communication round using their marginal performance loss evaluated on a held-out validation dataset.
However, these methods are sensitive to the order in which the clients join the game, and may not yield consistent results for contribution evaluation when the order changes.

To eliminate the effect from the order of joining the FL, Shapley value (SV) is adopted to measure the contribution of each client under the cooperative game framework \cite{jia2019towards, ghorbani2019data}. 
Wang et al. \cite{wang2020principled} propose the federated SV (FedSV) to quantify the contribution of each client at each communication round, which retains the key features of the traditional SV without introducing additional communication costs.
However, computing SVs is generally computationally expensive. Consequently, numerous efforts have been made to reduce the computation complexity of SV \cite{liu2022gtg, fan2022improving, fan2022fair,zheng2023secure,wang2022efficient}.

%% file: problem.tex
\section{Problem Formulation}\label{sec:problem}

In this section, we characterize the threat model by presenting the capabilities, background knowledge, and goals of an attacker.
We then formally formulate the model poisoning attack on contribution evaluation methods in FL.
We finally describe the design goals of the attack.

\subsection{Threat Model}

\myparatight{Attacker's capabilities and background knowledge} 
We consider an attacker owns a set of clients, referred to as \emph{malicious clients}.
As a result, the attacker could (1) access the local training datasets of malicious clients and the global model sent by the server, (2) control the training processes of local models of malicious clients, and (3) manipulate the parameters of the malicious clients' local models before sending the local model updates to the server. 
However, the attacker lacks the necessary background knowledge (e.g., local training datasets) of all other clients and the capabilities to manipulate parameters in other clients' local models.
Moreover, we assume the attacker knows the contribution evaluation method employed by the server.
This assumption is realistic in practice since the server normally shares such information with all clients for transparency and trustworthiness purposes during the initiation of the FL system \cite{fang2020local, mcmahan2017communication}.
However, we consider that the attacker does not know the hyperparameters of the contribution evaluation method and thus does not know the contributions of malicious clients computed by the server in each communication round. 
For example, the server may use a validation dataset to evaluate the local model of a client, and use the validation accuracy as the contribution of the client, which is not broadcast to the clients \cite{lyu2020collaborative, wang2020principled}. 
We also assume that the attacker possesses some storage capacity to retain the global models broadcast by the server in the previous $m$ (e.g., $m=3$) communication rounds.



\myparatight{Attacker's goal} Suppose the server employs a contribution evaluation method (as reviewed in Section \ref{sec:background}) to quantify the contribution of each client.
We consider that the goal of the attacker is to elevate the malicious clients' contributions computed by the server, compared to truthfully sending the local model updates learned using their local training data to the server.
By accomplishing the attack goal, the attacker can get extra rewards from the existing incentive mechanisms deployed in FL \cite{lyu2020collaborative, hu2022incentive, deng2021fair, zhang2021incentive}, even though it holds low-quality local training data. 
\textcolor{black}{For instance, in a cross-silo FL system where multiple banks collaborate to jointly learn a global model for commercial purposes \cite{long2020federated}, if the profit earned using the global model is distributed among the banks based on their contributions, the bank launching \ours~can gain extra profit while reducing the shares of others. Such malicious behavior further discourages the long-term participation of banks possessing high-quality training data and undermines the fairness of the federated learning system.
}

\subsection{A Model Poisoning Attack on Contribution Evaluation Methods in FL}\label{sec: attacker's goal}

In what follows, we formally formulate the model poisoning attack on contribution evaluation methods in FL. We denote the set of malicious clients as $\Hat{\Gamma}$.
\textcolor{black}{
For each malicious client $i\in\Hat{\Gamma}$, we use $\mathbf{g}_i^t$ to denote the local model update learned using the local training data without attacks.}
Moreover, we denote the manipulated parameters of the local model update from client $i$ at communication round $t$ as $\mathbf{\hat{g}}_i^t$. We use $\mathcal{E}$ to denote the contribution evaluation method deployed by the server, where $\mathcal{E}(\mathbf{g}_i^t)$ (or $\mathcal{E}(\mathbf{\hat{g}}_i^t)$) denotes the contribution calculated by the server when the client $i$ sends the local model update $\mathbf{g}_i^t$ (or $\mathbf{\hat{g}}_i^t$) to the server. 
The goal of the attacker is to craft $\mathbf{\hat{g}}_i^t$ for each malicious client $i\in\hat{\Gamma}$ such that the accumulated contribution $\sum_{i\in\hat{\Gamma}}\mathcal{E}(\mathbf{\hat{g}}_i^t)$ (or equivalently $\sum_{i\in\hat{\Gamma}}(\mathcal{E}(\mathbf{\hat{g}}_i^t) - \mathcal{E}(\mathbf{g}_i^t))$) is maximized. We call such an attack \emph{model poisoning attack on contribution evaluation methods in FL}. Formally, we formulate the attack as the following optimization problem in the $t$-th communication round:
\begin{align}\label{eq:attack goal}
    \{\mathbf{\hat{g}}_i^t|i\in\hat{\Gamma}\} = \argmax_{\{\mathbf{g}_i|i\in\hat{\Gamma}\}} \text{ } \sum_{i\in\hat{\Gamma}}\mathcal{E}(\mathbf{g}_i).
\end{align}

\subsection{Design Goals}\label{sec:design goals}

We aim to design a model poisoning attack on contribution evaluation methods in FL. 
In particular, we aim to accomplish the following goals in our attack design:

\myparatight{Effective} Our first goal is that the attack should be effective, i.e., it could significantly increase the contributions of malicious clients calculated by the server, compared to the scenario without attacks (i.e., the malicious clients learn their local model updates using the local training datasets).


\myparatight{Efficient} 
Our second goal is that the attack should be efficient, i.e., it should incur small computation and communication costs, compared to the baseline when there is no attack.
The reason for incorporating efficiency into the design goals is that clients in FL are often resource-constrained, e.g., mobile phones and IoT devices \cite{mcmahan2017communication,nguyen2021federated}.

\myparatight{Performance Preserving} 
We note that the attacker's goal is not to disrupt the convergence or performance of the final global model learned from FL. Thus, we aim to design an attack that preserves the performance of the final global model. As a result, the malicious clients, who own low-quality local training data, could obtain a global model with comparable performance to the one learned without attacks.



\myparatight{Aggregation Rule Independent}
We note that many aggregation rules \cite{mcmahan2017communication,blanchard2017machine,yin2018byzantine,xu2021gradient} have been proposed to aggregate local models in FL. We aim to design an attack that is agnostic to the aggregation rules such that our attack is generalizable to a wide range of FL systems.




%% file: method.tex
\section{Description of \ours}\label{sec:attack}
\subsection{Overview of \ours}
\label{Section: overview}

A key challenge in solving the optimization problem in Eq. \eqref{eq:attack goal} is that an attacker does not know $\mathcal{E}(\mathbf{\hat{g}}_i^t)$ for an arbitrary local model update $\mathbf{\hat{g}}_i^t$. 
The reason is that the attacker lacks necessary information such as the local models of other clients or validation dataset of the server that is utilized to calculate contribution (see Section~\ref{Section: Contribution Evaluation in FL} for details).
To address the challenge, our key insight is that a client is more likely to have a high contribution if its local model is more similar to the aggregated global model in each communication round. 
The reason is that the aggregated global model is obtained from local models learned on local training datasets of all clients, including those whose local training data is considered more valuable by the server. 
Based on this insight, we propose \ours, a poisoning attack on contribution evaluation methods in FL.

\ours~consists of two components: \emph{future global model prediction} and \emph{prediction error mitigation}. The future global model prediction component aims to predict the global model in each communication round based on the historic global models (i.e., the global models in previous communication rounds) received by the attacker. 
We note that the attacker could incur certain prediction errors and they would accumulate over communication rounds, resulting in low contributions for the malicious clients. In response, we further propose two strategies to mitigate the impact of errors in our prediction error mitigation component. 
We further discuss how \ours~can improve the attack effectiveness on certain contribution evaluation methods.
Finally, we analyze both the time and space complexities of \ours. Our analysis shows \ours~incurs negligible computation and storage costs. The reason is that it is very efficient to predict the future global model and \ours~only requires saving a few snapshots of historic global models.

\subsection{Detailed Design of \ours}
We first describe the \emph{future global model prediction} component, followed by the \emph{prediction error mitigation} component. 
Then, we discuss how to adapt \ours~to further enhance the effectiveness of our attack.

\subsubsection{Future Global Model Prediction}
\label{Section: Predicting Global Model Updates}

In what follows, we describe how the malicious clients predict the global model update. Note that given the predicted global model update, the predicted global model could be obtained by adding the predicted global model update and the current global model together. 


\myparatight{Predicting the Global Model Updates using the Cauchy Mean Value Theorem}
We use $\hat{\mathbf{g}}^t$ to denote the predicted global model update at communication round $t$.
According to the Cauchy mean value theorem \cite{lang2012real}, the \emph{predicted global model update} $\mathbf{\hat{g}}^t$ at communication round $t$ is calculated as
\begin{align}
    \mathbf{\hat{g}}^t = \mathbf{g}^{t-1} + H^t (\mathbf{w}^t - \mathbf{w}^{t-1}),
\label{equation: estimated global model update}
\end{align}
where $\mathbf{g}^{t-1}$ is the global model update in the communication round $t-1$, $\mathbf{w}^t$ (or $\mathbf{w}^{t-1}$) is the global model at communication round $t$ (or $t-1$), and $H^{t} = \int_{0}^{1}H(\mathbf{w}^{t-1}+z(\mathbf{w}^t-\mathbf{w}^{t-1}))dz$ is an \emph{integrated Hessian matrix}.
Based on Eq. \eqref{equation: estimated global model update}, 
predicting the global model update only requires the current and previous global models, the integrated Hessian matrix, and the previous global model update.
Although each malicious client has background knowledge on 
the previous global model update, previous global model, and current global model, computing the integrated Hessian matrix, however, is computationally expensive. 
In light of this, we utilize the L-BFGS algorithm \cite{byrd1994representations, byrd1995limited}, which is widely used to approximate the integrated Hessian matrix.


\myparatight{Approximating the Integrated Hessian Matrix via the L-BFGS Algorithm}
We approximate the integrated Hessian matrix $H^t$ using the L-BFGS algorithm \cite{byrd1994representations, byrd1995limited}, as outlined in Algorithm \ref{Algorithm: L-BFGS} in Appendix \ref{Appendix:algo}.
The L-BFGS algorithm takes two buffers $\Delta \mathbf{W}^t = [\Delta \mathbf{w}^{t-m}, \Delta \mathbf{w}^{t-m+1}, \cdots, \Delta \mathbf{w}^{t-1}]$ and $\Delta \mathbf{G}^t = [\Delta \mathbf{g}^{t-m}, \Delta \mathbf{g}^{t-m+1}, \cdots, \Delta \mathbf{g}^{t-1}]$ as inputs, and approximates the integrated Hessian matrix, where $\Delta \mathbf{w}^t = \mathbf{w}^{t} - \mathbf{w}^{t-1}$ is the change in the global model and $\Delta \mathbf{g}^t = \mathbf{g}^{t} - \mathbf{g}^{t-1}$ is the change in global model update.
In practical implementation~\cite{byrd1994representations}, the L-BFGS algorithm consumes an additional vector $\mathbf{v}$ of appropriate dimension as an input, and returns a product $H^t\mathbf{v}$, termed \emph{Hessian-vector product}, as the output.
This is sufficient for the malicious clients to predict the global model update since Eq. \eqref{equation: estimated global model update} only requires a Hessian-vector product $H^t  (\mathbf{w}^t - \mathbf{w}^{t-1})$ to predict the global model update $\hat{\mathbf{g}}^t$.
In the remainder of this paper, we represent the L-BFGS algorithm as $\textsc{L-BFGS}(\Delta \mathbf{W}^t, \Delta \mathbf{G}^t, \mathbf{v})$.
\textcolor{black}{We note that the L-BFGS algorithm is employed by the server to estimate clients' local models, as countermeasures against model poisoning attacks in FL in existing studies~\cite{zhang2022fldetector,cao2023fedrecover}. 
This paper, however, considers scenarios where malicious clients use the L-BFGS algorithm to estimate the global model, thereby elevating their contribution evaluated by the server.}

\subsubsection{Prediction Error Mitigation}\label{sec:mitigation}
\myparatight{Insufficiency of Future Global Model Prediction}
We remark that it is inadequate for the malicious clients to only predict the global model update.
The reason is that the L-BFGS algorithm may incur large prediction errors, i.e., the deviation between the predicted global model update and its true value, at some communication rounds.
As we will demonstrate in Section \ref{sec:experiment}, the prediction error results in low contribution evaluated by the server for the malicious clients. 
We identify two major reasons for prediction errors.
The first issue arises because the L-BFGS algorithm requires historical information from past global models and their updates to construct buffers $\Delta \mathbf{W}^t$ and $\Delta \mathbf{G}^t$. 
To compensate for the absence of this historical information, we adopt a strategy used in previous studies \cite{cao2023fedrecover}, \emph{preliminary iteration}, during the initial communication rounds, where malicious clients either learn from their local datasets or use the previous round's global model updates as proxies for their current local updates. 
In the meantime, the malicious clients collect the global models and global model updates to construct the buffers $\Delta \mathbf{W}^t$ and $\Delta \mathbf{G}^t$. 
After that, the malicious clients proceed with the L-BFGS algorithm as discussed in Section \ref{Section: Predicting Global Model Updates}.
The second reason for prediction errors is the potential accumulation of errors over successive communication rounds. 
To address this, we develop a \emph{threshold-based filtering} strategy to mitigate the impact of error accumulation.


\myparatight{Threshold based Filtering} 
As the L-BFGS algorithm cannot predict the exact global model update, it is unavoidable that there would be prediction errors. Moreover, those errors could accumulate over communication rounds, which could lower the contributions of the malicious clients. 
We note that the malicious clients may not be aware when a large prediction error occurs since it does not have access to the true global model update for the future communication rounds, and thus cannot calculate the prediction errors.
To tackle this challenge, we develop a threshold based filtering strategy to estimate whether the predicted global model update incurs a large prediction error. Our intuition is that the prediction error is more likely to be larger if the magnitude of the Hessian-vector product is larger.
Therefore, if the $\ell_2$-norm of the Hessian-vector product is less than a threshold, i.e.,
\begin{align}
    \|\textsc{L-BFGS}(\Delta \mathbf{W}^t, \Delta \mathbf{G}^t, \mathbf{w}^t - \mathbf{w}^{t-1})\| \leq \tau,
\label{Equation threshold-based filtering}
\end{align}
where $\|\cdot\|$ represents $\ell_2$-norm and $\tau$ is a threshold (we defer the detailed discussion on it), 
we consider that the prediction error is tolerable, and thus the malicious clients use the predicted global model update as their local model updates. 
Otherwise, each malicious client calculate its local model update by using its local training dataset or utilizing the global model update from the previous communication round.

A key question in our design is how to set the threshold. Our idea is that the magnitude of the predicted global model update should be on a similar scale as the previous global model updates. Thus, we set the threshold $\tau$ as $l\|\mathbf{w}^{t} - \mathbf{w}^{t-1}\|$, where $l$ is a positive coefficient that can be tuned by each malicious client to control its tolerance on prediction errors. Our experimental results show that $l = 1$ is sufficient to mitigate prediction errors.

\subsubsection{Strategies to Enhance \ours}\label{sec:strategies to enhance}
In this subsection, we discuss how to adapt \ours~to further enhance its effectiveness when certain a contribution evaluation method is employed by the server.

\myparatight{Adaptation to Contribution Evaluation Methods using Cosine Distance}
In the following, we discuss how to adapt \ours~when the server uses the cosine distance between the local model and global model to measure each client's contribution.
This class of methods is widely used in existing studies~\cite{xu2020reputation, xu2021gradient, shi2022fedfaim, jiang2023fair} since cosine distance is efficient to calculate.
In this case, we can rewrite the optimization in Eq. \eqref{eq:attack goal} as the following equivalent formulation:
\begin{equation}\label{eq:distance based}
    \{\hat{\mathbf{g}}_i^t|i\in\hat{\Gamma}\}=\argmax_{\{\mathbf{g}_i|i\in\hat{\Gamma}\}} \text{ }  \sum_{i\in\hat{\Gamma}}(1-\cos(\mathbf{g}_i,\mathbf{g}^t)).
\end{equation}
Here, $\cos(\mathbf{a},\mathbf{b}) = 1 - \frac{\mathbf{a}\cdot \mathbf{b}}{\|\mathbf{a}\|\|\mathbf{b}\|}$ is the cosine distance function, where $\mathbf{a}\cdot\mathbf{b}$ represents the inner product between vectors $\mathbf{a}$ and $\mathbf{b}$. Global model update $\mathbf{g}^t$ is obtained following the aggregation rule (Step III in Section \ref{Section: Federated Learning}).

According to Eq. \eqref{eq:distance based}, we note that the malicious clients can enhance the attack effectiveness by sending an amplified prediction of the global model update $c\hat{g}_i^t$ to the server as their local model updates, where $c>1$ is an amplifying coefficient. 
The insight is that increasing the magnitude of the predicted global model update will make it more dominant compared with other clients' local model updates.
As a result, the angle deviation between the predicted global model update and the aggregated global model update decreases, leading to a smaller cosine distance between them.
Furthermore, such dominance effects will accumulate over communication rounds, thereby elevating the malicious clients' contributions perceived by the server.
We finally remark that introducing the coefficient $c$ does not lower the accuracy of the global model, which will be demonstrated in Section \ref{sec:experiment}.

\myparatight{Adaptation to Contribution Evaluation Methods with Validation Dataset}
Some contribution evaluation methods \cite{wang2020principled, wang2019measure, zhang2021refiner, chen2020dealing, lyu2020collaborative} require the server to have a validation dataset, and use the validation accuracy or loss of local models to measure contributions.
In this case, the malicious clients can execute the L-BFGS algorithm multiple times within one communication round. 
We term this operation as the \emph{local evolution}.
The insight behind local evolution is to mimic the normal training process over multiple epochs at one communication round. 
When the L-BFGS algorithm yields limited prediction error, the local evolution will allow the malicious clients to craft local model updates of higher validation accuracy and hence increase the associated contributions. 

\subsection{Complete Algorithm}

Algorithm \ref{Algorithm: Our Attack} in Appendix \ref{Appendix:algo}  shows our complete algorithm for attacking contribution evaluation methods in FL.
The attacker can initiate attack at any communication round $t$.
If communication round $t$ is a preliminary iteration, then the malicious clients compute their local model updates by learning from the local training data or sending the previous global model update to the server (see Section \ref{sec:strategies to enhance}).
If communication round $t$ is not a preliminary iteration, then the malicious clients leverage the L-BFGS algorithm to predict the next global model update (see Section \ref{Section: Predicting Global Model Updates}).
Given the predicted global model update, the malicious clients estimate whether the prediction error can be tolerated or not using the threshold based filtering as shown in Eq. \eqref{Equation threshold-based filtering}.
If the prediction error is tolerable by the malicious clients (Eq. \eqref{Equation threshold-based filtering} holds true), then they set the prediction of global model update to be their local model updates which will be sent to the server.
Otherwise, the malicious clients compute the local model updates using their local training data or the previous global model update.

\subsection{Complexity Analysis}

\myparatight{Time Complexity}
According to \cite{byrd1994representations}, the time complexity to calculate the Hessian-vector product using Algorithm \ref{Algorithm: L-BFGS} is $\mathcal{O}\left(m^3\right)+6 m p+p$, where $p$ is the model size and $m$ is the buffer length.
If the malicious clients does not launch the attack and learn the local model update using the local training dataset over $e$ epochs, then the time complexity of learning the local model update using the local training dataset $\mathcal{D}_i$ over $e$ epochs is $6|\mathcal{D}_i|eR(p)$, where $R(p)$ is the time complexity for forward propagation \cite{griewank2008evaluating}. 
Although the expression of $R(p)$ is architecture dependent, we note that the time complexity of \ours~is in general significantly less than that of learning the local model update using local training data \cite{freire2022computational}, achieving the design goal of being efficient in Section \ref{sec:design goals}. 
For instance, when the global model is a CNN (see Appendix \ref{appx:model-structure} for the detailed architecture), the computation time of \ours~using an RTX 6000 Ada GPU is 0.004s per communication round.
By contrast, training the same model using the local training dataset of a client takes 1.10s, which is 270$\times$ slower compared to \ours.

\myparatight{Space Complexity}
The space complexity for each malicious client is $\mathcal{O}\left(mp\right)$ when it launches \ours. Note that in practice, we typically choose $m=2$ or $m=3$. Therefore, the proposed attack in this paper imposes a small storage constraint on the malicious clients. For instance, when $m=3$ and the architecture of the global model is CNN, our attack only requires 42.43MB of additional storage space.

%% file: theory.tex
\section{Theoretical Analysis}\label{sec:theoretical}

In this section, we characterize the strategies to enhance the attack effectiveness.
Specifically, we focus on the cases where cosine distance is used by the server to measure contributions. 
All the proofs can be found in Appendix \ref{Appendix: proof}.
The effectiveness of amplifying the predicted global model update with coefficient $c$ is stated as follows.

\begin{proposition}
\label{lemma not decrease the utility}
Let $\mathbf{g'} = \mathcal{A}(\mathbf{g}_1, \cdots, c  \hat{\mathbf{g}}_i, \cdots , \mathbf{g}_N)$ and $\mathbf{g} = \mathcal{A}(\mathbf{g}_1, \cdots, \hat{\mathbf{g}}_i, \cdots , \mathbf{g}_N)$ be the global model updates obtained using the predicted global model update $c \hat{\mathbf{g}}$ and $\hat{\mathbf{g}}$, respectively.
When $c\geq1$, we have the following relationship: 
\begin{equation}
    \cos(\mathbf{g'}, c  \hat{\mathbf{g}}_i) \leq \cos(\mathbf{g}, \hat{\mathbf{g}}_i).
\end{equation}
\end{proposition}
\begin{remark}
    Proposition \ref{lemma not decrease the utility} shows that the cosine distance between the amplified global model update $c \mathbf{g}_i$ and the global model update $\mathbf{g}'$ is no larger than the cosine distance before applying the amplification with coefficient $c\geq 1$. 
    Therefore, our strategy developed for  contribution evaluation using cosine distance will not degrade the attack effectiveness.
\end{remark}

Proposition \ref{lemma not decrease the utility} yields the following corollary.
\begin{corollary}\label{lemma rank non-decreasing}
Let $\mathbf{g'}$ and $\mathbf{g}$ be defined as in Proposition \ref{lemma not decrease the utility}.
If $\cos(\mathbf{g},\hat{\mathbf{g}}_i) \leq \cos(\mathbf{g},\mathbf{g}_j)$, then $\cos(\mathbf{g}',c\hat{\mathbf{g}}_i) \leq \cos(\mathbf{g}',\mathbf{g}_j)$.
\end{corollary}
\begin{remark}
    Corollary \ref{lemma rank non-decreasing} indicates that if a malicious client $i$ surpasses the contribution of another client $j$ by sending the predicted global model update as its local model update, then amplifying the predicted global model update will not make the contribution of malicious client $i$ less than client $j$.
    Therefore, the malicious client $i$ is always perceived as a high-contributing client compared to client $j$ (note that a smaller cosine distance means a higher contribution).
\end{remark}

We finally show that the parameter $c$ can be tuned such that the contributions of malicious clients become larger than any arbitrary client.

\begin{proposition}\label{lemma minimum c}
Let $\mathbf{g'}$ and $\mathbf{g}$ be defined as in Proposition \ref{lemma not decrease the utility}.
Suppose that $\cos(\mathbf{g},\hat{\mathbf{g}}_i)>\cos(\mathbf{g},\mathbf{g}_j)$ holds for some client $j$ and malicious client $i$, and $\mathcal{A}(\mathbf{g}_1, \cdots , \mathbf{g}_N)=\sum_{j\in\Gamma}\alpha_j\mathbf{g}_j$, where $\alpha_j\in[0,1]$ is a weight coefficient and $\sum_{j\in\Gamma}\alpha_j=1$.
If the malicious client $i$ chooses the coefficient $c$ such that $c \geq \frac{\|\hat{\mathbf{g}}_i\| \mathbf{g} \cdot \mathbf{g}_j - \|\mathbf{g}_j\| \mathbf{g} \cdot \hat{\mathbf{g}}_i}{\alpha_i \|\hat{\mathbf{g}}_i\|( \|\mathbf{g}_j\| \|\hat{\mathbf{g}}_i\| - \hat{\mathbf{g}}_i \cdot \mathbf{g}_j)} + 1$,
then $\cos(\mathbf{g}',c\hat{\mathbf{g}}_i) \leq \cos(\mathbf{g}',\mathbf{g}_j)$.
\end{proposition}

\begin{remark}
    Proposition \ref{lemma minimum c} focuses on servers that utilize linear combinations as the aggregation rule, which are widely used in practice \cite{mcmahan2017communication,blanchard2017machine,yin2018byzantine,xu2021gradient}.
    Proposition \ref{lemma minimum c} shows that even if the predicted global model may not be sufficient for a malicious client $i$ to surpass the contribution of another client $j$, choosing a proper coefficient $c$ allows the malicious client $i$ to make a higher contribution (evaluated by the server) compared to client $j$.
\end{remark}

%% file: exp.tex
\section{Empirical Evaluations}\label{sec:experiment}

\label{Section: Experiment}
We perform extensive experiments to evaluate \ours. 
In Section \ref{exp-set}, we show the experimental setup. 
Section \ref{exp-main} presents the results.
Ablation analysis is presented in Section \ref{Section: Ablation}.

\subsection{Experimental Setup}
\label{exp-set}
\myparatight{Datasets and Models} We consider three benchmark datasets MNIST \cite{MNIST}, CIFAR-10 \cite{cifar10}, and Tiny-ImageNet \cite{tinyimagenet}. Specifically, MNIST is a 10-class digit image classification dataset, which contains 60,000 training images and 10,000 testing images of dimension $28\times28$ in grayscale. CIFAR-10 is a 10-class dataset with 50,000 training images and 10,000 testing images uniformly distributed across the classes, where the size of each image is $32\times32\times 3$. Tiny-ImageNet is a color image classification dataset covering 200 classes, with 100,000 training images, 10,000 validation images, and 10,000 testing images. Each image in Tiny-ImageNet is of dimension $64\times 64\times 3$.
We use two Convolution Neural Network (CNN) variants on MNIST and CIFAR-10 datasets respectively, and a pre-trained VGG11 model \cite{vgg} on Tiny-ImageNet. The model structures are shown in Appendix \ref{appx:model-structure}. 

\myparatight{Data Partition}
For each dataset, we consider one homogeneous data partition (denoted as UNI) and two heterogeneous data partitions (denoted as POW and CLA) among clients by following previous studies on contribution evaluation methods in FL \cite{xu2021gradient, lyu2020collaborative,xu2020reputation}.
The heterogeneous data partitions yield non-i.i.d. data distributions.
We detail each data partition as follows.
\begin{itemize}
    \item \textbf{UNI}: This data partition uniformly splits the training images in each dataset among all clients, yielding an i.i.d. data distribution among the clients.
    \item \textbf{POW}: Following \cite{xu2021gradient, lyu2020collaborative,xu2020reputation}, the sizes of local training datasets of all clients are sampled from a parameterized power law distribution. We set the parameter of power law distribution as two. This leads to a data-size heterogeneous setting among the clients. For example, on MNIST, the numbers of training images of 10 clients are 110, 219, 328, 437, 546, 655, 764, 873, 982, and 1,086, respectively. The details for the power law distribution can be found in Appendix~\ref{appx:data-partition}.
    \item \textbf{CLA}: Following \cite{xu2021gradient, lyu2020collaborative,xu2020reputation}, we use CLA to create a class imbalance setting where the local training datasets of different clients cover heterogeneous numbers of classes. For example, on MINST, the local training datasets of 10 clients contain training images from 6, 6, 7, 7, 8, 8, 9, 9, 10, and 10 classes, respectively. We show the details in Appendix \ref{appx:data-partition}.
\end{itemize}


\myparatight{FL Setup} By default, we assume there are $N = 10$ clients. Note that we set $N=10$ since some contribution evaluation methods calculate the Shapley value to measure contributions, whose computation cost grows exponentially as the number of clients increases.
As some contribution evaluation methods require the server to have a validation dataset to compute contributions, we reserve 20\% of the data samples from the training images of MNIST, CIFAR-10, and Tiny ImageNet as the validation dataset.
Each client uses stochastic gradient descent (SGD) to update its local model for $\epsilon$ epochs with a batch size of 128. We set $\epsilon = 3$ for MNIST and CIFAR-10, and $\epsilon = 5$ for Tiny-ImageNet, considering that the classification tasks on MNIST and CIFAR-10 are easier than that on Tiny-ImageNet. 
Following previous studies~\cite{xie2019dba,li2020federated,zhao2018federated}, we set the learning rate $\eta$ as $ 0.03$ for MNIST, $0.05$ for CIFAR-10, and $0.001$ for Tiny-ImageNet, with the learning rate exponentially decaying at rate $\gamma = 0.995$. The total number of communication rounds is $T=60$.


\myparatight{Evaluation Metrics}
We define the following two metrics to demonstrate the effectiveness of \ours.
The first metric is called \textit{contribution score}, which measures the fraction of contribution from each individual client. Formally, the contribution score for the client $i$ is computed as follows:
\begin{equation}
    CS_i = \frac{\sum_{t =1}^T e_i^t}{\sum_{j \in \Gamma} \sum_{t =1 }^T e_j^t},
\end{equation}
where $\Gamma$ is the set of all clients and $e_i^t$ (or $e_j^t$) is the contribution computed by the server using a contribution evaluation method for the client $i$ (or $j$) in the $t$-th communication round. 

We note that elevation in the contribution scores does not necessarily imply that the malicious clients' contributions can surpass those of other clients.
Therefore, we propose the second metric named \emph{rank gain} to measure the change in the ranks of a client's contribution score with and without attacks. 
Formally, the rank gain for the client $i$ is computed as follows:
\begin{equation}
   \Delta R_i = \widehat{R}_i - R_i,
\end{equation}
where $\widehat{R}_i$ and $R_i$ represent the ranking (in ascending order) of the contribution score of the client $i$ among all clients with and without attack, respectively. An attack is more effective when $CS_i$ and $\Delta R_i$ are larger for a malicious client $i$.

We use \emph{ACC} to measure the performance preserving property of \ours. In particular, ACC measures the classification accuracy of the final global model on testing inputs. Additionally, we will compare the computation cost of \ours~with the attack free setting (i.e., a malicious client uses its local training dataset to learn a local model update) to demonstrate the efficiency of \ours. 


\myparatight{Compared Baselines} 
We note that there are no existing studies on attacking contribution evaluation methods in FL. In response, we generalize several existing methods as baselines~\cite{lin2019free,xu2021validation,bagdasaryan2020backdoor}. 
First, we consider the scenario where the malicious clients do not launch any attack and follow the procedure outlined in Section \ref{Section: Federated Learning} to learn their local model updates (denoted as \emph{Attack Free}).
In addition to Attack Free, we compare \ours~with the following three baselines~\cite{lin2019free,xu2021validation,bagdasaryan2020backdoor}:

\begin{itemize}
    \item \textbf{Delta Weight}: In this baseline \cite{lin2019free}, the malicious client $i$ crafts local model updates as $\mathbf{g}_i^t = \mathbf{w}^{t-1} - \mathbf{w}^{t} + \delta$, where $\mathbf{w}^{t-1} - \mathbf{w}^{t}$ is the global model update in the previous communication round and each entry of $\delta$ follows a zero-mean Gaussian distribution with a standard deviation $\sigma=5\times 10^{-5}$. 
    \item \textbf{Data Augment}: This baseline utilizes data augmentation to increase the sizes of local training datasets of the malicious clients. In particular, each malicious client randomly rotates, scales, and crops the data samples from its local training dataset to form a new dataset, and then merges the newly generated dataset with the original one.
    At each communication round, each malicious client learns its local model update using the augmented dataset, aiming to increase its contribution \cite{xu2021validation}.
    \item \textbf{Scaling Attack}: In this baseline, each malicious client first learns a local model update using its local training dataset. Then the magnitudes of these local model updates are amplified using a scaling attack \cite{bagdasaryan2020backdoor}.
    In our experiments, the scaling factor is set to 2.
\end{itemize}

\input{float_element/fig_cs-rank_cla}

\myparatight{Contribution Evaluation Methods} 
We consider five state-of-the-art contribution evaluations that can be employed by the server.
We briefly introduce these methods as follows. The detailed description can be found in Appendix \ref{app:contribution evaluation method}.
\begin{itemize}
    \item \textbf{FedSV} \cite{wang2020principled}: FedSV is a game theory based contribution evaluation. 
    It measures the contribution of each client at each communication round following two steps. It first calculates an empirical loss evaluated on a validation dataset held by the server. Then FedSV computes contributions by splitting the empirical loss among all clients using the Shapley value. 
    FedSV uses FedAvg \cite{mcmahan2017communication} as the aggregation rule.
    \item \textbf{LOO} \cite{wang2020principled, zhang2021refiner}: This is a game theory based contribution evaluation. 
    To quantify the contribution from a client $i$, it computes two global models at each communication round, with and without the client's local model update.
    Then the client's contribution is calculated as the difference of the values of an empirical loss function evaluated on a validation dataset using these two global models.
    LOO uses FedAvg \cite{mcmahan2017communication} as the aggregation rule.
    \item \textbf{CFFL} \cite{lyu2020collaborative}: This is an individual performance based contribution evaluation. The server calculates the contribution of a client $i$ using the accuracy of its local model evaluated on a validation dataset. The aggregation rule of CFFL is a variant of FedAvg \cite{mcmahan2017communication}, which considers both the data size and the number of classes of a client. Specifically, when the data size is imbalanced, the aggregation rule follows FedAvg. When the class numbers are imbalanced, the aggregation rule assigns weights to the local model updates of clients according to the number of classes in their local training datasets.
    \item \textbf{GDR} \cite{xu2021gradient}: This is an individual performance based contribution evaluation. The server utilizes the cosine distance between the aggregated global model updates and local model updates to estimate the Shapley value, and assigns the Shapley value to each client as its contribution. 
    GDR uses a weighted sum of all local model updates to compute the global model update, where the weight associated with each client's local model update is the rolling mean of its contribution.
    \item \textbf{RFFL} \cite{xu2020reputation}: This is an individual performance based contribution evaluation. 
    The contribution of a client is quantified by the cosine similarity between the client's local model update and the aggregated global model update.
    RFFL uses a similar aggregation rule as RFFL.
\end{itemize}

\myparatight{\ours~Setup} By default, we consider a single malicious client. We note that when a client has a high contribution even if there is no attack, it is very challenging to improve the ranking of its contribution score. In response, by default, we select a client whose contribution is the lowest without attacks as the malicious client.
Unless otherwise mentioned, the buffer length is set to $m=3$, the threshold is chosen as $l=1$ to mitigate prediction error, and the tunable amplifying coefficient is set as $c=1$. 
When the server employs contribution evaluation methods using cosine distance (GDR and RFFL) and validation datasets (FedSV, LOO, and CFFL), we set the local evolution rounds to one and two, respectively.
During the preliminary iteration and when the L-BFGS algorithm incurs a large prediction error, the malicious client executes the delta weight attack (using the global model update in the previous communication round with Gaussian noise).
In Section \ref{Section: Ablation}, we perform ablation analysis and evaluate alternative strategies when the  L-BFGS incurs a large prediction error.

\input{float_element/table_main_acc}
\subsection{Experimental Results}\label{exp-main}

We evaluate \ours~using the metrics in Section \ref{exp-set}.
When the context is clear, we drop the subscript $i$ for the malicious client.

\myparatight{\ours\ is Effective and Aggregation Rule Independent} 
Figure~\ref{fig:cs-vs-rank-cla} and~\ref{fig:cs-vs-rank-UNI},~\ref{fig:cs-vs-rank-POW} (in Appendix \ref{app:additional exp}) compare the contribution score and rank again of the malicious client when using \ours~and baselines. 
We have the following key observations.
First, the contribution score and rank gain of \ours~consistently outperform those of all baselines.
For example, when the server employs CFFL \cite{lyu2020collaborative} as the contribution evaluation method and the data partition is CLA, the malicious client is perceived with contribution score $0.108$ and rank gain $8$ by the server for the classification task on Tiny-ImageNet dataset. 
This moves the malicious client from being the lowest-contributing to becoming the second highest-contributing client.
All the baselines in this case, however, give rank gain zero, indicating that the malicious client is still evaluated as the lowest-contributing client by the server under those baseline attacks.
Our second observation is that under non-i.i.d. data distribution, some baselines yield the same rank gain as \ours, e.g., when the server employs RFFL and the data partition is CLA.
However, we note that the malicious client attains the highest contribution score using \ours~in these scenarios.
Therefore, \ours~is more effective compared with all baselines under non-i.i.d. data distributions.

We note that \ours~is effective across all contribution evaluation methods in Figure~\ref{fig:cs-vs-rank-cla}, which utilize different aggregation rules.
This indicates that \ours~is independent of the aggregation rules used in FL. 


\input{float_element/table-art}

\myparatight{\ours\ Preserves the Performance} 
We show the accuracy on testing inputs of the final global model in Table \ref{tab-effect-acc}.
We observe that the accuracy of the final global model under \ours~remains within a negligible $1\%$ deviation from highest ACC in the worst-case.
Therefore, \ours~preserves the performance of the final global model learned in FL.
Note that \ours~can potentially lead to the higher ACC under some contribution evaluation methods than Attack Free, e.g., FedSV, CFFL, and GDR.
The reason is that \ours~replaces the malicious clients' original local model updates (that should have been learned from their low-quality local training data) with the predicted global model update which integrates updates from other clients learned using local training data of higher qualities.


\input{float_element/table_ab-nodenum}

\myparatight{\ours\ is Efficient} 
We compare the computation cost of \ours~with Attack Free (when the malicious client utilizes its local training dataset to learn a local model update). Table \ref{tab:art} shows the ratio between the computation cost of learning a local model update using a local training dataset and \ours, i.e., (computation cost of learning a local model update)/(computation cost of \ours). 
Our key observation is that \ours~is significantly more efficient than learning a local model update using the local training dataset. The reason is that it is very efficient to predict the future global model.
Note that \ours~does not incur extra communication cost since the malicious client simply replaces its local model update with the predicted global model update.



\subsection{Ablation Analysis}
\label{Section: Ablation}

We perform ablation analysis on the CIFAR-10 dataset under  CFFL and RFFL contribution evaluation methods. 
Here CFFL and RFFL are chosen as the representative contribution evaluation methods for servers with and without validation datasets, respectively.

\myparatight{Effect of Client Number $N$} Table \ref{tab:numnode} evaluates the effect of client number $N$. 
We compare the contribution score $CS$ and the rank gain $\Delta R$ of \ours~with Attack Free (the malicious client uses its local training dataset to learn a local model update). We have the following observations.  
First, \ours~is consistently effective. In particular,  \ours~consistently makes the malicious client, who has the lowest contribution score without attack, the high-contributing client regardless of the client number $N$.
In our experiments, the malicious client is evaluated as the highest-contributing client ($\Delta R=9,19,49$) in 14 out of 18 settings.
This observation aligns with our results shown in Figure \ref{fig:cs-vs-rank-cla}, and indicates that the design of \ours \  is insensitive to FL systems with different numbers of clients.
We note that \ours~consistently preserves the ACC, i.e., the ACC under \ours~is similar to that of Attack Free. In other words, \ours~preserves the performance of the final global model for FL with different number of clients.

\input{float_element/table-abnormal-fix}



\myparatight{Effect of the Threshold Based Filtering} 
We evaluate the effect of the threshold based filtering discussed in Section \ref{sec:mitigation}, which is used when \ours~potentially incurs high prediction error. 
We compare the ACC, $CS$, $\Delta R$ with and without the threshold based filtering in Table \ref{tab:ab-abnormal-fix}.
We observe that when the data distribution is non-i.i.d. (e.g., POW data partition), the threshold based filtering can significantly improve the attack effectiveness and the accuracy of the final global model (ACC) under CFFL.
Hence, the threshold based filtering is necessary to the design of \ours.
\input{float_element/table_amp_coeff}

\myparatight{Effect of Amplifying Coefficient $c$} In Table \ref{tab:amp-coeff}, we evaluate the effect of the amplifying coefficient $c$. We observe that as we increase the coefficient $c$, the contribution score of the malicious client increases under all data partitions when the server utilizes cosine distance to measure the contributions (RFFL).
This observation aligns with our theoretical analysis in Section \ref{sec:theoretical}.
Moreover, \ours~preserves the ACC under all data partitions with different choices of amplifying coefficient $c$.


\myparatight{More Experiments} Due to space constraint, we defer the results of the attack effectiveness under UNI and POW data partitions under all datasets (in Figure \ref{fig:cs-vs-rank-UNI} and \ref{fig:cs-vs-rank-POW}), effects of strategies for preliminary iteration and threshold based filtering (in Table~\ref{tab:ab-warmup}), buffer length (in Figure \ref{fig:ab-mem-len}), local evolution (in Figure \ref{fig:ab-lcaol-evolution}), the fraction of clients selected by the server in each communication round (in Table \ref{tab:selected-clients}), and the fraction of malicious clients (in Figure \ref{fig:attacker-number-cs-appx}) to Appendix \ref{app:additional exp}. 
\textcolor{black}{We also show experimental results when the attacker does not know the contribution evaluation methods deployed by the server in Appendix \ref{app:additional exp}.} 
In summary, our results show that \ours~is relatively insensitive to these factors, and is effective in boosting the malicious clients' contributions perceived by the server.


%% file: float_element/fig_cs-rank_cla.tex
\begin{figure*}[h]
    \centering
    \subfloat{
    \centering
    \includegraphics[width=0.7\linewidth]{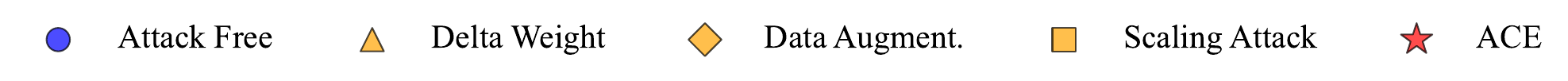}
    }
    
    \setcounter{subfigure}{0}
    \subfloat[FedSV]{
        \label{fig:capparatus}
        \centering
        \includegraphics[width=0.19\linewidth]{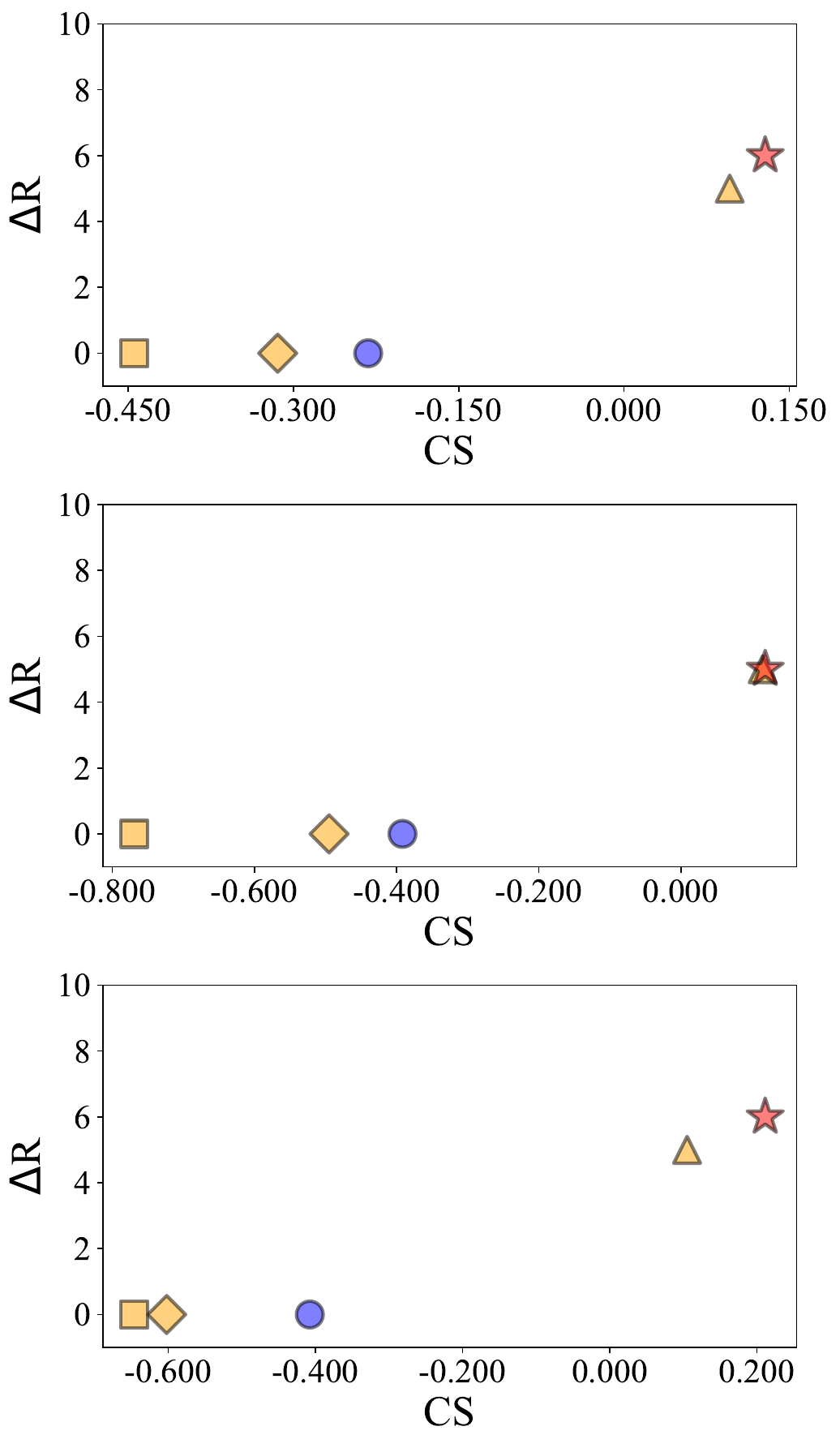}
    }
    \hfill
    \subfloat[LOO]{
        \label{fig:capparatus}
        \centering
        \includegraphics[width=0.19\linewidth]{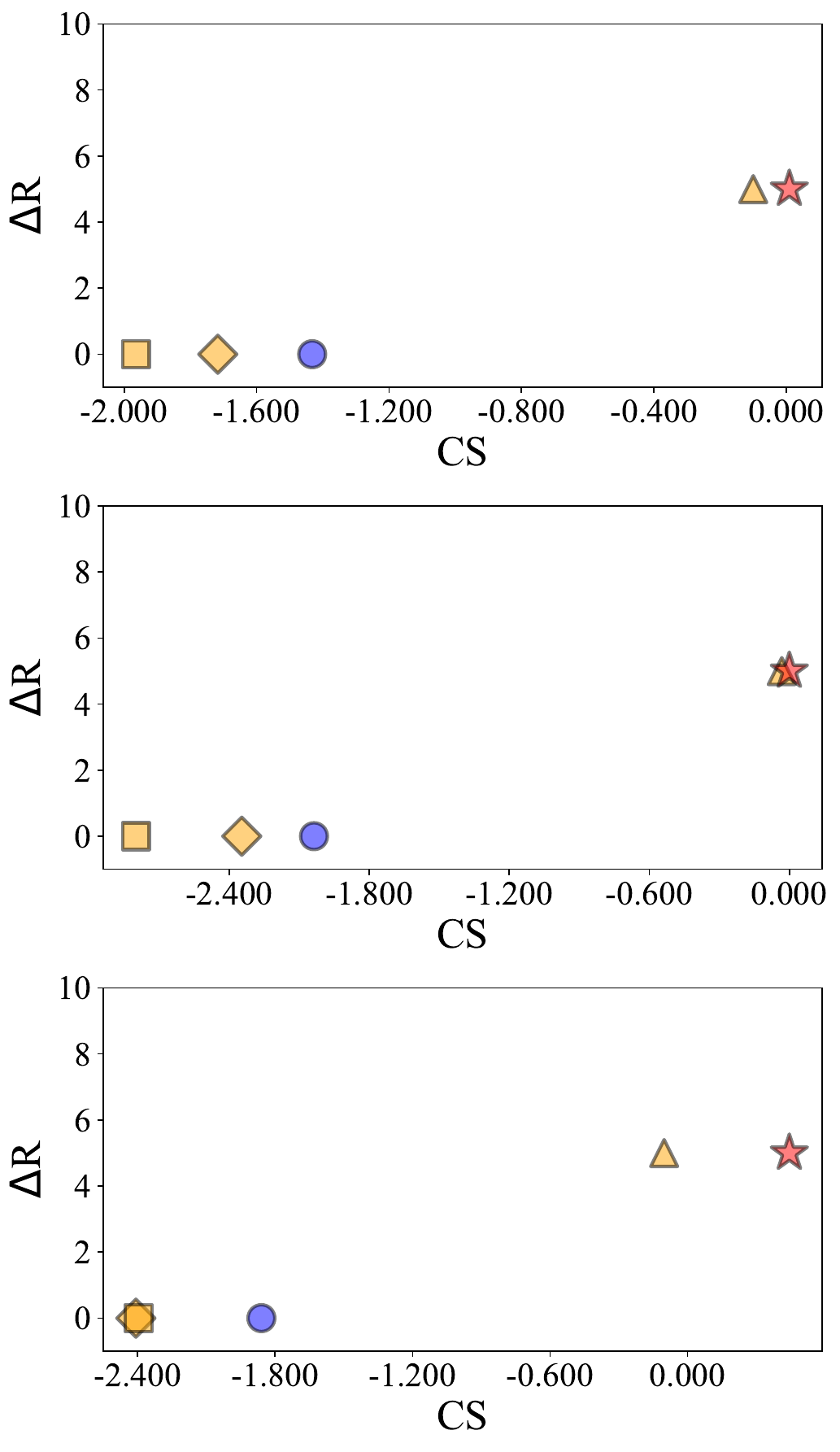}
    }
    \hfill
    \subfloat[CFFL]{
        \label{fig:capparatus}
        \centering
        \includegraphics[width=0.19\linewidth]{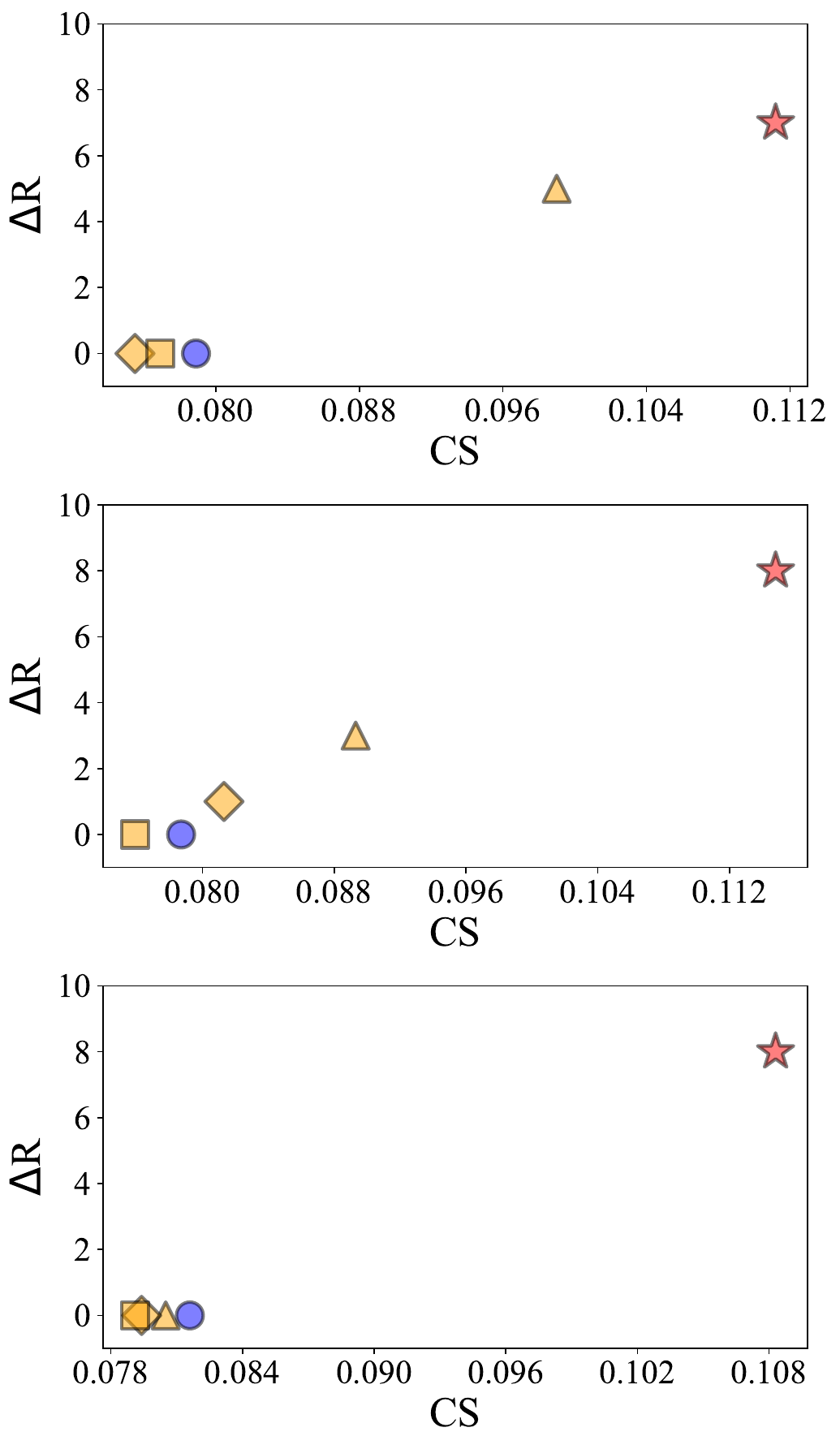}
    }
    \hfill
    \subfloat[GDR]{
        \label{fig:capparatus}
        \centering
        \includegraphics[width=0.19\linewidth]{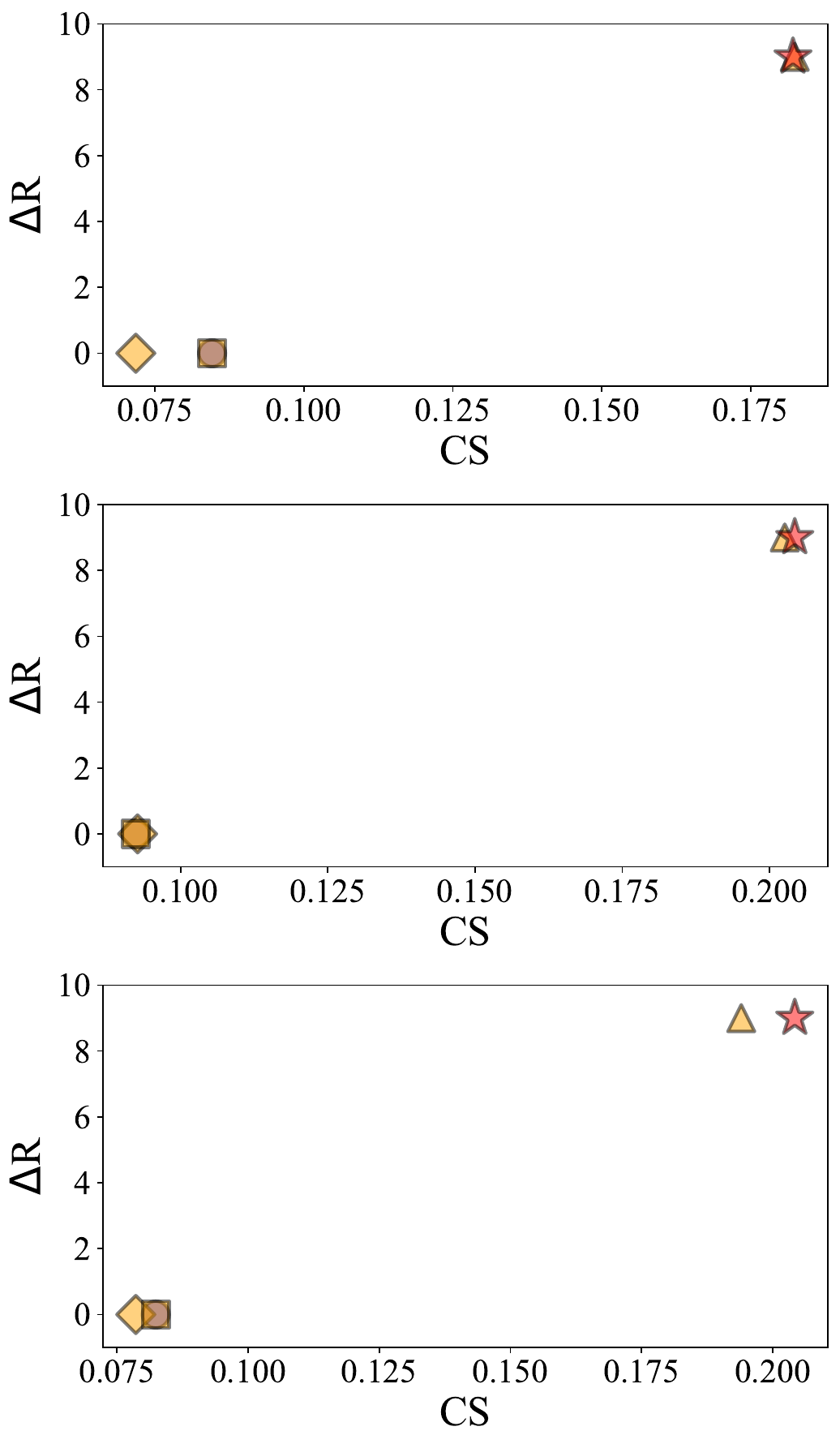}
    }
    \hfill
    \subfloat[RFFL]{
        \label{fig:capparatus}
        \centering
        \includegraphics[width=0.19\linewidth]{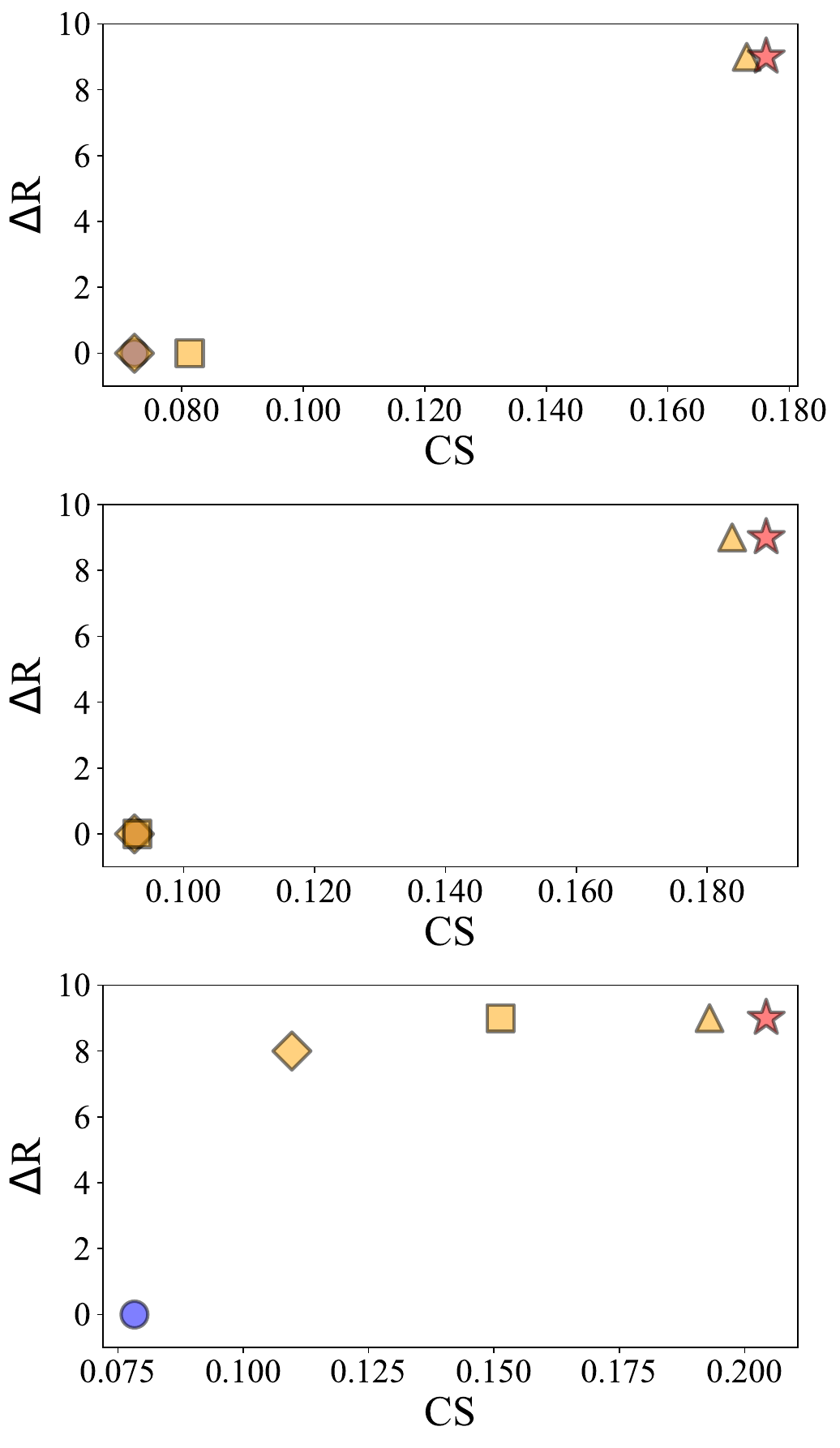}
    }
    \caption{Comparing the contribution score $CS$ and rank gain $\Delta R$ of the attacker when using \ours~and baselines under three datasets, i.e., MNIST (first row), CIFAR-10 (second row), and Tiny-ImageNet (third row), and five contribution evaluation methods, i.e., FedSV, LOO, CFFL, GDR, and RFFL. The data partition method is CLA (a heterogeneous setting). Our results show \ours~is more effective than baselines. The results for data partitions UNI and POW are in Figure \ref{fig:cs-vs-rank-UNI} and \ref{fig:cs-vs-rank-POW} of Appendix \ref{app:additional exp}.
}
    \label{fig:cs-vs-rank-cla}
\end{figure*}

%% file: float_element/table_main_acc.tex
\newcolumntype{g}{>{\columncolor{gray!15}}c}

\begin{table*}[!h]\renewcommand{\arraystretch}{1.0}
\centering
\caption{This table summarizes the ACC of the final global model learned with \ours~and all baselines, evaluated under three data partitions (UNI, POW, and CLA), three datasets (MNIST, CIFAR-10, and Tiny ImageNet), and five contribution evaluation methods (FedSV, LOO, CFFL, GDR, and RFFL).
The accuracy of the final global model under \ours~remains within a negligible $1\%$ deviation from highest ACC in the worst-case.
Thus \ours~preserves the accuracy of the final global model. }
\label{tab-effect-acc}

\resizebox{\linewidth}{!}{%
\fontsize{4}{5}\selectfont
\begin{tabular}{c|c|ggg|ggg|ggg}
\toprule 
Contribute &  \multirow{2}{*}{} &
  \multicolumn{3}{c|}{MNIST} & 
  \multicolumn{3}{c|}{CIFAR-10} &
  \multicolumn{3}{c}{Tiny-ImageNet} 
   \\ \rowcolor{white}
Evaluation& \multirow{-2}{*}{Attack} & 
  UNI & POW & CLA & UNI & POW & CLA & UNI & POW & CLA 
   \\ \midrule \rowcolor{white}
 & Attack Free & 95.86\% & 	95.69\% & 	89.89\% & 	71.16\% & 	70.82\% & 	56.32\% & 	46.37\% & 	47.84\% & 	44.98\%  
   \\ \rowcolor{white}
 & Delta Weight & 95.68\% & 	95.46\% & 	90.39\% & 	70.89\% & 	71.02\% & 	57.16\% & 	46.10\% & 	47.80\% & 	45.27\% 
   \\ \rowcolor{white}
 & Data Augment. & 95.87\% & 	95.67\% & 	89.88\% & 	71.63\% & 	70.27\% & 	56.05\% & 	46.77\% & 	48.37\% & 	45.26\% 
   \\ \rowcolor{white}
 & Scaling Attack & 95.85\% & 	95.81\% & 	89.66\% & 	71.58\% & 	71.01\% & 	55.29\% & 	46.59\% & 	48.07\% & 	45.01\% 
   \\ \multirow{-5}{*}{FedSV} 
 & \textbf{\ours} & 95.81\% & 	95.53\% & 	91.27\% & 	71.30\% & 	71.45\% & 	57.60\% & 	46.35\% & 	48.23\% & 	45.94\% 
   \\ \midrule \rowcolor{white}
 & Attack Free & 95.86\% & 	95.69\% & 	89.89\% & 	71.16\% & 	70.82\% & 	56.32\% & 	46.37\% & 	47.84\% & 	44.98\% 
   \\ \rowcolor{white}
 & Delta Weight & 95.88\% & 	95.69\% & 	90.39\% & 	70.89\% & 	71.02\% & 	57.16\% & 	46.10\% & 	47.80\% & 	45.27\% 
   \\ \rowcolor{white}
 & Data Augment. & 95.94\% & 	95.73\% & 	89.88\% & 	71.63\% & 	70.27\% & 	56.05\% & 	46.77\% & 	48.37\% & 	45.26\% 
   \\ \rowcolor{white}
 & Scaling Attack & 95.78\% & 	95.66\% & 	89.66\% & 	71.58\% & 	71.01\% & 	55.29\% & 	46.59\% & 	48.07\% & 	45.01\% 
   \\ \multirow{-5}{*}{LOO}
 & \textbf{\ours} & 96.06\% & 	95.51\% & 	91.27\% & 	71.30\% & 	71.45\% & 	57.60\% & 	46.35\% & 	48.23\% & 	45.94\%

   \\ \midrule \rowcolor{white}
 & Attack Free &  96.83\% & 	94.71\% & 	79.43\% & 	71.84\% & 	60.65\% & 	49.99\% & 	51.77\% & 	48.23\% & 	39.96\%
   \\ \rowcolor{white}
 & Delta Weight & 96.58\% & 	91.89\% & 	82.26\% & 	70.66\% & 	59.37\% & 	50.62\% & 	51.30\% & 	44.18\% & 	40.54\%
   \\ \rowcolor{white}
 & Data Augment. &  97.44\% & 	94.49\% & 	79.19\% & 	73.08\% & 	60.93\% & 	50.62\% & 	51.92\% & 	47.83\% & 	40.04\%
   \\ \rowcolor{white}
 & Scaling Attack & 97.01\% & 	94.59\% & 	79.28\% & 	71.55\% & 	60.41\% & 	49.91\% & 	52.22\% & 	44.23\% & 	39.87\%
   \\ \multirow{-5}{*}{CFFL}
 & \textbf{\ours} &  96.61\% & 	95.35\% & 	83.18\% & 	70.44\%	&62.03\%&	52.45\% & 	51.53\% & 	49.20\% & 	42.02\%
 
   \\ \midrule \rowcolor{white}
 & Attack Free & 96.26\% & 	96.23\% & 	85.41\% & 	70.97\% & 	71.33\% & 	56.66\% & 	51.80\% & 	51.96\% & 	44.78\%
   \\ \rowcolor{white}
 & Delta Weight & 96.84\% & 	96.43\% & 	89.02\% & 	70.32\% & 	70.76\% & 	59.18\% & 	52.19\% & 	52.57\% & 	46.01\%
   \\ \rowcolor{white}
 & Data Augment. & 96.43\% & 	96.18\% & 	87.42\% & 	72.01\% & 	71.12\% & 	57.38\% & 	51.79\% & 	52.04\% & 	44.84\%
   \\ \rowcolor{white}
 & Scaling Attack & 96.26\% & 	96.23\% & 	85.42\% & 	71.01\% & 	71.36\% & 	56.63\% & 	51.84\% & 	51.89\% & 	44.78\%
   \\ \multirow{-5}{*}{GDR}
 & \textbf{\ours} &  96.78\% & 	96.53\% & 	89.12\% & 	70.27\% & 	70.60\% & 	59.23\% & 	52.64\% & 	52.77\% & 	46.61\%
   \\ \midrule \rowcolor{white}
 & Attack Free & 96.78\% & 	96.85\% & 	92.67\% & 	71.78\% & 	71.03\% & 	57.66\% & 	52.35\% & 	52.43\% & 	46.72\%
   \\ \rowcolor{white}
 & Delta Weight & 96.66\% & 	96.85\% & 	91.83\% & 	70.69\% & 	71.07\% & 	56.95\% & 	51.89\% & 	52.49\% & 	46.84\%
   \\ \rowcolor{white}
 & Data Augment. & 96.25\% & 	96.08\% & 	92.67\% & 	71.84\% & 	71.04\% & 	57.60\% & 	51.83\% & 	52.50\% & 	46.31\%
   \\ \rowcolor{white}
 & Scaling Attack & 95.96\% & 	95.97\% & 	91.73\% & 	71.73\% & 	71.07\% & 	56.60\% & 	50.84\% & 	52.50\% & 	46.17\%
   \\ \multirow{-5}{*}{RFFL}
   & \textbf{\ours} &  96.64\% & 	96.87\% & 	92.30\% & 	70.72\% & 	70.90\% & 	57.36\% & 	51.75\% & 	52.31\% & 	46.54\%
   \\ \bottomrule
\end{tabular}%
}

\end{table*}

%% file: float_element/table-art.tex
\begin{table}[!t]\renewcommand{\arraystretch}{1.0}
\centering
\caption{This table shows the ratio between the computation costs of using a local training dataset to learn a local model update and \ours. The data partition method is UNI. We observe that \ours~is significantly more efficient.}
\label{tab:art}
\resizebox{\columnwidth}{!}{%
    \begin{tabular}{c | c c c c c}
    \toprule 
    Dataset & FedSV & LOO & CFFL &	GDR	& RFFL \\ \midrule
    MNIST & \ \ 30.88$\times$ & \ \ 30.88$\times$ & \ \ 7.48$\times$ & 16.15$\times$ & \ \ 18.26$\times$ \\
    CIFAR-10 & 270.81$\times$ & 270.81$\times$ &  21.25$\times$ & 86.48$\times$ & 101.44$\times$ \\
    Tiny-ImageNet & \ \ 35.35$\times$ &  \ \ 35.35$\times$ & 13.26$\times$ & 29.22$\times$ & \ \ 24.79$\times$ \\
    \bottomrule
    \end{tabular}%
}

\end{table}

%% file: float_element/table_ab-nodenum.tex

\newcommand{\data}[1]{\small{\textit{\textcolor{black}{#1}}}}

\begin{table}[!t]
\centering
\caption{This table evaluates the effect of the number of clients in FL. 
Regardless of the number of clients, \ours~consistently makes the attacker the highest-contributing agent. 
AF is the abbreviation for Attack Free.}
\label{tab:numnode}
\resizebox{\columnwidth}{!}{%
    \begin{tabular}{c   c   c|  c c l  | c}
    \toprule 
    \multirow{2}{*}{\makecell{Contribution \\evaluation}} & \multirow{2}{*}{\makecell{Data \\ partition}} &\multirow{2}{*}{\makecell{\#Clients}} & \multicolumn{2}{c}{$CS$} & \multirow{2}{*}{\makecell{Relative \\ improv.}} & \multirow{2}{*}{$\Delta R$} \\  
    
    & &  & AF & \ours &  & \\ 
    \midrule
    \multirow{9}{*}{CFFL} & \multirow{3}{*}{UNI} & 10  &	0.099	& 0.105 &\data{+6.41\%}	 &	9 \\
    & & 20  &	0.050 &	0.053 &	\data{+7.14\%}	&19 \\
    & & 50  &	0.020 &	0.023 	& \data{+19.69\%}	& 49 \\ \cmidrule{2-7}
    
    & \multirow{3}{*}{POW} & 10 &	0.086	& 0.106 &	\data{+22.73\%}	&8\\
    & & 20 &0.043&	0.056	& \data{+30.02\%}	&19\\
    & & 50 &	0.016	&0.024 &	\data{+54.03\%}	&49\\ \cmidrule{2-7}
    
    & \multirow{3}{*}{CLA}& 10 &0.079	&0.115	&\data{+45.80\%}	&8 \\ 
    & & 20 	&0.040	&0.058	& \data{+47.21\%} &	18 \\
    & & 50 &	0.016&	0.024&	\data{+49.94\%}&	49\\ \midrule 
    
    \multirow{9}{*}{RFFL} & \multirow{3}{*}{UNI} & 10 &	0.096&	0.192 &	\data{+100.26\%} &	9 \\
    & & 20 	&0.048	&0.124	&\data{+160.88\%} &	19 \\
    & & 50 &0.018&	0.060&	\data{+237.81\%} &	49 \\ \cmidrule{2-7}
    & \multirow{3}{*}{POW} & 10 &	0.045&	0.196	&\data{+335.49\%}&	9 \\ 
    & & 20 &	0.021	&0.125&	\data{+502.81\%}	&19 \\
    & & 50 &	0.007&	0.064	& \data{+872.81\%}	&49\\ \cmidrule{2-7}
    & \multirow{3}{*}{CLA}& 10  &	0.096 &	0.189 &	\data{+97.41\%} &	9 \\ 
    & & 20 	&0.045&	0.123&	\data{+174.98\%} &	16\\
    & & 50 & 0.017	 & 0.064 &	\data{+284.43\%}&	49\\ 
    \bottomrule
    \end{tabular}%
}

\end{table}

%% file: float_element/table-abnormal-fix.tex
\begin{table}[ht]
\centering
\caption{This table shows the effect of the threshold based filtering. We compare the ACC, $CS$, and $\Delta R$ of \ours~with and without the threshold based filtering. 
We observe that solely relying on future global model prediction is not sufficient for attack effectiveness. The prediction error mitigation with threshold based filtering is necessary to guarantee the effectiveness of \ours. 
}
\label{tab:ab-abnormal-fix}
\resizebox{\columnwidth}{!}{%
\begin{tabular}{cc|cc |cc | cc}
\toprule 
\multirow{2}{*}{\makecell{Contri.\\Eval.}}& \multirow{2}{*}{Metric} & \multicolumn{2}{c|}{UNI} & \multicolumn{2}{c|}{POW}  & \multicolumn{2}{c}{CLA}\\ 
& & with  & without & with & without & with & without \\ \midrule 
\multirow{3}{*}{CFFL} & ACC & 70.44\% & 70.50\% & 62.03\% & 56.87\% & 52.45\% & 52.51\%  \\
& $CS$ & 0.1051 & 0.1049 & 0.1055 & 0.0526 & 0.1148 & 0.1149 \\
& $\Delta R$ & 9 & 9 & 8 & 0 & 8 & 8 \\ \midrule
\multirow{3}{*}{RFFL} & ACC & 70.72\% & 70.72\% & 70.90\% & 70.90\% & 57.36\% & 57.36\% \\
&$ CS$ & 0.1917 & 0.1917 & 0.1963 & 0.1963 & 0.1890 & 0.1890 \\
& $\Delta R$  & 9 & 9 & 9 & 9 & 9 & 9  \\ \bottomrule
\end{tabular}%
}
\end{table}

%% file: float_element/table_amp_coeff.tex
\begin{table}[!t]\renewcommand{\arraystretch}{1.0}
\centering
\caption{This table presents the effect of amplifying coefficient $c$ when the server employs RFFL as the contribution evaluation method under all data partitions.
The contribution score of the malicious client increases as the amplifying coefficient increases. 
AF is the abbreviation of Attack Free.}
\label{tab:amp-coeff}

\resizebox{\columnwidth}{!}{%
    \begin{tabular}{c c | c c c cc}
    \toprule 
    \makecell{Data \\ Partition} & Metric & AF  & $c$ = 1 & $c$  = 1.5 & $c$  = 2 & $c$  = 2.5 \\ \midrule
    \multirow{2}{*}{UNI} & ACC & 71.78\% & 70.72\% & 70.90\% & 70.83\% & 70.59\% \\
    & $CS$ & 0.096 & 0.192 & 0.199 & 0.205 & 0.207 \\ \midrule
    \multirow{2}{*}{POW}  & ACC & 71.03\% & 70.90\% & 70.78\% & 70.70\% & 70.73\%  \\
    & $CS$ & 0.045 & 0.196 & 0.204 & 0.211 & 0.212 \\ \midrule
    \multirow{2}{*}{CLA}  & ACC & 57.66\% & 57.36\% & 58.33\% & 59.70\% & 60.13\%  \\
    & $CS$ & 0.093 & 0.189 & 0.196 & 0.203 & 0.205\\ \bottomrule
    \end{tabular}%
}

\end{table}

%% file: defense.tex
\section{Countermeasures to \ours~and Evaluations}\label{sec:defense}


\subsection{Countermeasures to \ours}

We focus on defense developed for FL against model poisoning attacks to thwart \ours. 
According to \cite{shi2022challenges}, existing defenses against model poisoning attacks can be divided into three categories: \textit{performance based defense} \cite{li2019abnormal,xie2019zeno,cao2019distributed,cao2020fltrust}, \textit{distance based defense}  \cite{blanchard2017machine,xia2019faba,wan2021shielding,fung2018mitigating,cao2019understanding,zhang2022fldetector,fang2022aflguard,chu2022securing} and \textit{statistics based defense} \cite{yin2018byzantine,munoz2019byzantine,guerraoui2018hidden,xie2018generalized,pillutla2022robust,mozaffari2023every}. As demonstrated in Section \ref{sec:experiment}, \ours~successfully deceives the methods that utilize validation accuracy to measure contribution (e.g., FedSV and LOO), which invalidates the performance based defenses.
We thus mainly focus on distance and statistics based defenses. \textcolor{black}{In particular, we choose Multi-Krum \cite{blanchard2017machine}, Trimmed-Mean \cite{yin2018byzantine}, FABA \cite{xia2019faba}, Sniper \cite{cao2019understanding}, and Foolsgold \cite{fung2018mitigating} as representative countermeasures. These defenses do not require a validation dataset in the server and are commonly employed by the community to defend against poisoning and Sybil attacks in FL.}
\begin{itemize}
    \item \textbf{Multi-Krum \cite{blanchard2017machine}.}
\textcolor{black}{We choose Multi-Krum \cite{blanchard2017machine} from the category of distance based defense as the first countermeasure to \ours, which has been widely used to mitigate model poisoning attacks in FL.}
Multi-Krum focuses on identifying and eliminating local model updates from malicious clients by analyzing the distance between local model updates from different clients. We extend it to defend against \ours. 
In particular, the server computes the Euclidean distance between each pair of local model updates at each communication round to measure how similar or dissimilar the local model updates are to each other.
The server then considers the $k$ most dissimilar local model updates, i.e., $k$ local model updates that has the largest sum of distance to other local model updates, to be sent by the malicious clients. 
\item \textbf{Trimmed-Mean \cite{yin2018byzantine}.}
\textcolor{black}{We extend Trimmed-mean from the class of statistics based defense as the second countermeasure to \ours.}
In each communication round, Trimmed-Mean first sorts each dimension of the local model updates from all clients. Subsequently, it identifies the largest $k$ and smallest $k$ entries for each dimension in the sorted local model updates, considering these entries to be potentially sent by malicious clients.
Finally, the top $k$ clients that are identified by the server most frequently across all dimensions are considered to be malicious by the server.

\color{black}
\item \textbf{FABA \cite{xia2019faba}.} \textcolor{black}{We apply FABA \cite{xia2019faba}, an efficient defense algorithm against Byzantine attacks in FL as the third countermeasure.} In each communication round, the server first computes the mean of local model updates, and then computes the difference between the mean and each local model update. The server subsequently identifies the local model update with the largest difference from the mean as malicious.

\item \textbf{Sniper \cite{cao2019understanding}.} 
\textcolor{black}{We consider a clustering-based approach, Sniper, to defend against \ours.}
Sniper clusters benign local model updates by solving a maximum clique problem in each communication round. Specifically, the server first calculates the Euclidean distances between each pair of local model updates, then constructs a graph such that each local model update is a vertex in the graph. An edge exists if the Euclidean distance between two local model updates is smaller than a pre-determined threshold. In the end, the server finds the maximum clique in the graph and identifies vertices (local model updates) in the clique as benign. The remaining local model updates are identified as malicious.

\item \textbf{Foolsgold \cite{fung2018mitigating}.} \textcolor{black}{We evaluate \ours~against Foolsgold \cite{fung2018mitigating}, a Sybil detection countermeasure.} Foolsgold maintains aggregate historical updates of local model updates from each client to better estimate the similarity of the overall contributions made by clients. Foolsgold detects malicious clients by calculating the pairwise cosine similarity of historical updates as a representation of how strongly two clients are acting similarly. The clients with high pairwise cosine similarity are identified as malicious.  

\color{black}
%

\item \textbf{Random Guess.} 
The last countermeasure considered in this paper is Random Guess.
In particular, the server randomly selects $k$ local model updates as manipulated by the malicious clients. 
\end{itemize}
We note that $k$ is a hyper-parameter for those countermeasures.
We consider a strong defense scenario where the server knows the total number of local model updates (which is used to set $k$) from the malicious clients.

\subsection{Evaluations of Countermeasures to \ours}
We empirically evaluate the six detection countermeasures against \ours~under our default setting.

\myparatight{Evaluation Metrics} 
We use three metrics, \textbf{precision}, \textbf{recall}, and \textbf{F1-Score} to evaluate the detection performance of Multi-Krum, Trimmed-Mean, and Random Guess when the attacker launches \ours. 
\begin{itemize}
    \item \textbf{Precision.} Precision is defined as the fraction of local model updates that are indeed from the malicious clients among all predicted ones.
    \item \textbf{Recall.} Recall is the fraction of local model updates from the malicious clients that are successfully predicted by a countermeasure. 
    \item \textbf{F1-Score.} F1-Score measures the harmonic mean of precision and recall, i.e., $\text{F1-Score} = 2 \cdot \frac{\text{Precision} \times \text{Recall}}{\text{Precision} + \text{Recall}}$.
\end{itemize}
We note that each defense will detect local model updates from the malicious clients in each communication round. Thus, we report the average precision, recall, and F1-Score overall all communication rounds.

\myparatight{Evaluation Results}
We report the precision, recall, and F1-score when the server utilizes Multi-Krum, Trimmed-Mean, \textcolor{black}{FABA, Sniper, Foolsgold}, or Random Guess to mitigate \ours\ in Table \ref{tab:detect_all} on CIFAR-10, where the contribution evaluation methods are CFFL and RFFL. For UNI data distribution, we observe that the malicious client is rarely detected by any of these countermeasures, indicating that none of these are effective and adequate to thwart \ours. 
In particular, the performances of these detection methods are worse than the naive Random Guess.
The reason is that the local model updates sent by the malicious client do not significantly diverge from the global model update, leading the server to perceive the malicious client as a benign client. \textcolor{black}{For POW and CLA, two more realistic data partitions, there is only a marginal increase in detection performance. However, the performance remains worse than Random Guess, highlighting the stealthiness of \ours~in non-i.i.d. settings. We defer the evaluation results when parameter $c\geq 1$ to Appendix \ref{app:additional exp}.}

\textcolor{black}{
We note that the insights of detecting malicious clients behind Multi-Krum, Trimmed Mean, FABA, Sniper, and Foolsgold can be classified into two categories, whereas neither of these was adequate to detect \ours.
The inadequacy of Multi-Krum, Trimmed Mean, FABA, and Sniper can be attributed to their insight that malicious clients likely to send local model updates significantly different from other clients' updates. 
Foolsgold identifies malicious clients whose local model updates are too similar to others as malicious.
However, as discussed in Section \ref{sec: attacker's goal}, our \ours~does not exhibit either excessive similarity or dissimilarity to benign clients, and thus can evade detection by all these countermeasures.
This highlights the urgent need to develop new mitigation strategies to defend against \ours.
}

\input{float_element/table-detection}

%% file: float_element/table-detection.tex
\begin{table}[ht]
\color{black}

\centering
\caption{\textcolor{black}{This table summarizes the precision, recall, and F1-score when Multi-Krum, Trimmed-Mean, FABA, Sniper, Foolsgold, or Random Guess is employed to mitigate \ours~under both i.i.d. (i.e., UNI) and non-i.i.d. (i.e., POW and CLA) data distributions. We observe that none of these countermeasures are effective and adequate to defend \ours. None of these countermeasures are effective and adequate to defend \ours in all data distributions. If Precision and Recall are 0, F1-Score is not defined and denoted as N/A.}}

\resizebox{\columnwidth}{!}{%
\begin{tabular}{c| c | c c c | ccc}
\toprule 
\multirow{2}{*}{\makecell{Data \\ Partition}} & \multirow{2}{*}{Countermeasure} & \multicolumn{3}{c|}{CFFL} & \multicolumn{3}{c}{RFFL} \\
& & Precision & Recall & F1-Score &  Precision & Recall & F1-Score\\ \midrule
\multirow{6}{*}{UNI} & Multi-Krum &  0.017 & 0.017 & 0.017 &  0.017 & 0.017 & 0.017\\ 
& Trimmed-Mean &  0.017 & 0.017 & 0.017 &  0.017 & 0.017 & 0.017 \\
& FABA &  0.017 & 0.017 & 0.017 &  0 & 0 & N/A \\
& Sniper &  0.017 & 0.017 & 0.017 &  0 & 0 & N/A \\
& Foolsgold &  0 & 0 & N/A & 0 & 0 & N/A \\
& Random Guess &  0.100 & 0.100 & 0.100  & 0.100 & 0.100 & 0.100\\
\midrule
\multirow{6}{*}{POW} & Multi-Krum &  0.017 & 0.017 & 0.017 &  0.017 & 0.017 & 0.017\\ 
& Trimmed-Mean &  0.017 & 0.017 & 0.017 &  0.017 & 0.017 & 0.017 \\
& FABA &  0.017 & 0.017 & 0.017 &  0.017 & 0.017 & 0.017 \\
& Sniper &  0.022 & 0.033 & 0.027 &  0 & 0 & N/A \\
& Foolsgold &  0 & 0 & N/A & 0 & 0 & N/A \\
& Random Guess &  0.100 & 0.100 & 0.100  & 0.100 & 0.100 & 0.100\\ 
\midrule
\multirow{6}{*}{CLA} & Multi-Krum &  0.017 & 0.017 & 0.017 &  0.017 & 0.017 & 0.017\\ 
& Trimmed-Mean &  0.017 & 0.017 & 0.017 &  0 & 0 & N/A \\
& FABA &  0.017 & 0.017 & 0.017 &  0.017 & 0.017 & 0.017 \\
& Sniper &  0.022 & 0.033 & 0.026 &  0 & 0 & N/A \\
& Foolsgold &  0 & 0 & N/A & 0 & 0 & N/A \\
& Random Guess &  0.100 & 0.100 & 0.100  & 0.100 & 0.100 & 0.100\\ 
\bottomrule
\end{tabular}%
}
\label{tab:detect_all}
\end{table}

%% file: discussion.tex
\section{Discussion and Limitation}
\label{sec-discussion}

\myparatight{Untargeted Poisoning and Backdoor Attacks in FL} \textcolor{black}{Attacks against FL have been extensively studied, such as untargeted poisoning attacks~\cite{shejwalkar2021manipulating,fang2020local,karimireddy2020byzantine, xie2020fall,cao2022mpaf,zhou2021deep, tolpegin2020data} and backdoor attacks~\cite{xie2019dba,bagdasaryan2020backdoor,sun2019can,wang2020attack, zhang2022neurotoxin, baruch2019little}.
For image classification tasks, the final global model of FL learned under untargeted poisoning attacks produces incorrect predictions indiscriminately on the testing inputs, whereas the global model learned under backdoor attacks outputs attacker-chosen class when the inputs are embedded with the attacker-chosen triggers. We note that the untargeted poisoning attacks have a different goal from our attack.} Scaling Attack~\cite{bagdasaryan2020backdoor} is a state-of-the-art backdoor attack. We extend it to attack contribution evaluation methods in FL. Our experimental results in Section \ref{sec:experiment} demonstrate that \ours~consistently outperforms Scaling Attack. The reason is that backdoor attacks are not designed to improve the contribution of a malicious client. 

\myparatight{Defenses against Untargeted Poisoning and Backdoor Attacks in FL} To mitigate these attacks, various defenses have been proposed \cite{blanchard2017machine, yin2018byzantine, sun2019can, cao2020fltrust,rieger2022deepsight,wu2020mitigating,ozdayi2021defending,cao2023fedrecover,kumari2023baybfed,chu2022securing,nguyen2022flame,mozaffari2023every,fung2018mitigating} to identify and remove the local model updates from malicious clients.
However, most of these defenses are ineffective to mitigate \ours.
The reason is that these existing defenses identify the malicious clients by detecting abnormal local model updates from the clients.
For example, FLTrust \cite{cao2020fltrust} assigns a trust score to each local model update by comparing its direction with the global model update. 
As the deviation in the directions between the local model update and the global model increases, the client's trust score decreases, indicating that it is more likely to be malicious.
Our design of \ours, however, does not lead to abnormal local model updates. 
Instead, the manipulated local model updates sent by the malicious clients using \ours~tend to be similar to the global model update in order to elevate their contributions.
This makes \ours~particularly challenging to be detected as shown in Section~\ref{sec:defense}. 

\myparatight{Limitation of \ours} One limitation of \ours~is how to select the \emph{optimal} parameters and strategies for \ours.
Our evaluation results in Figure \ref{fig:ab-mem-len} and Table \ref{tab:ab-warmup} show that the effectiveness of \ours~is relatively insensitive to these choices under most of the cases with different datasets, data partitions, and contribution evaluation methods.
In order to further enhance the attack effectiveness of \ours, a future direction is to explore efficient techniques to compute the optimal parameters and strategies.



%% file: conclusion.tex
\section{Conclusion and Future Work}\label{sec:conclusion}

In this paper, we proposed a new model poisoning attack called \ours~on contribution evaluation methods in federated learning (FL).
We showed that the malicious clients in FL could utilize \ours~to predict the future global model update with small errors, and elevate their contributions evaluated by the server in FL.
We empirically evaluated \ours~using five state-of-the-art contribution evaluation methods with three datasets and three data partition methods.
Our results showed that \ours~effectively increased the contributions of malicious clients with negligible costs, and it preserved the performance of the final global model learned from FL.
We also evaluated \ours~against six countermeasures, and showed that none of them can effectively mitigate \ours. An interesting future work is to develop new mitigation strategies against our attack. 

%% file: ack.tex
\section*{Acknowledgement}
We are grateful for the time and insightful comments provided by the reviewers and shepherd, which have significantly enhanced the quality of our work.

This work is partially supported by the Air Force Office of Scientific Research (AFOSR) under grant FA9550-23-1-0208, National Science Foundation (NSF) under grants IIS 2229876, Office of Naval Research (ONR) under grant N00014-23-1-2386, DARPA GARD, the National Aeronautics and Space Administration (NASA) under grant No.80NSSC20M0229, Alfred P. Sloan Fellowship, and the Amazon research award.

This work is supported in part by funds provided by the National Science Foundation, Department of Homeland Security, and IBM. 
Any opinions, findings, and conclusions or recommendations expressed in this material are those of the author(s) and do not necessarily reflect the views of the NSF or its federal agency and industry partners.

%% file: appx.tex

\begin{appendices}


\section{Additional Experimental Results}\label{app:additional exp}

\input{float_element/fig_cs-rank_uni-pow}

\input{float_element/fig_mem_length}

\input{float_element/fig_local_envolution}

\input{float_element/table_abnormal-warmup}

\input{float_element/table_selected_clients}

\input{float_element/fig_attacker_number_cs2}

\input{float_element/table-c-detection}

\input{float_element/fig_unknown_evaluation}

In this appendix, we present more experimental results.

\myparatight{\ours~is Effective under UNI and POW data partitions} 
We show the contribution score versus the rank gain in Figure \ref{fig:cs-vs-rank-UNI} and \ref{fig:cs-vs-rank-POW} under UNI and POW data partitions, respectively.
We observe that \ours~consistently outperforms all baselines when the data distribution is non-i.i.d. (e.g., POW).
We note that, in a few cases, \ours~exhibits slightly lower attack performance compared with the baseline using data augmentation when the data is uniformly distributed among all clients.
The reason is that UNI data partition yields an i.i.d. data distribution among the clients.
By executing data augmentation, the malicious clients expand their local training datasets and disrupt the i.i.d. pattern, thereby gaining extra advantage compared with other clients.


\myparatight{Effect of the Strategies for Preliminary Iteration \& Threshold based filtering} 
We evaluate the effect of the strategies that can be taken for preliminary iteration and threshold based filtering in Table \ref{tab:ab-warmup}. In particular, we consider two strategies for the malicious clients: denoted as \emph{strategy one} ($s_1$) and \emph{strategy two} ($s_2$).
By using strategy one, the malicious clients execute the delta weight attack.
When taking strategy two, the malicious clients learn local model updates using their local training datasets.
These two strategies yield four possible combinations of strategies that can be taken for preliminary iteration and threshold based filtering as follows:
\begin{itemize}
    \item $s_1\times s_1$: Strategy one for both preliminary iteration and  threshold based filtering. 
    \item $s_1\times s_2$: Strategy one for preliminary iteration and strategy two for threshold based filtering.
    \item $s_2\times s_1$: Strategy two for preliminary iteration and strategy one for threshold based filtering.
    \item $s_2\times s_2$: Strategy two for both preliminary iteration and  threshold based filtering.
\end{itemize}
We observe that overall the combination $s_2\times s_2$ provides the malicious clients the best performance in terms of accuracy, contribution score, and rank gain.

\myparatight{Effect of Buffer Length $m$} 
We evaluate the effect of buffer length $m$ by varying it from two to five.
We compare ACC and the rank gain with Attack Free (buffer length $m=0$) in 
Figure \ref{fig:ab-mem-len}.
We observe that \ours~with non-zero buffer lengths significantly improve the rank gain, and hence the contributions of the malicious clients, without degrading ACC.
Furthermore, as we vary the buffer length from two to five, \ours~exhibits a consistent rank gain.

\myparatight{Effect of Local Evolution} 
Figure \ref{fig:ab-lcaol-evolution} evaluates the effect of local evolution (see Section \ref{sec:strategies to enhance}) when validation accuracy is used to measure contribution.
We have the following observations. 
First, \ours~is effective to increase the rank gains for the malicious clients by predicting the future global model.
However, running the L-BFGS algorithm with different number of rounds can lead to distinct contribution score. 
For CFFL, the local evolution decreases the contribution score under all three data partitions. 
We observe that predicting the global model update for future two communication rounds yields the best performance under FedSV.
While the contribution scores generally increases when the malicious clients run more rounds of the L-BFGS algorithm under UNI data partition for LOO, the pattern on POW and CLA data partitions is similar to CFFL and FedSV.

\myparatight{Effect of Client Selection}
In Table \ref{tab:selected-clients}, we evaluate the effect of the fraction of clients selected at each communication round.
We vary the fraction of clients that is selected by the server from $50\%$ to $100\%$.
We make the following observations.
First, \ours~is consistently effective under all fractions of client selections in terms of ACC, CS, and $\Delta R$.
Therefore, \ours~is insensitive to the client selection used by the server.
Furthermore, as the fraction of clients being selected at each communication round increases, the effectiveness of \ours~increases.
The reason is that the malicious clients can make more precise predictions of the global model updates, and therefore make the local model updates of malicious clients better aligned with the global model update.

\myparatight{Effect of the Fraction of Malicious Clients}
In the following, we evaluate the effect of the fraction of malicious clients.
We vary the fraction of malicious clients from $10\%$ to $30\%$, and present the accumulated CS (the summation of contribution scores of all malicious clients) in Figure \ref{fig:attacker-number-cs-appx}.
We observe that the accumulated CS using \ours~is always higher than the Attack Free case under all data partitions when RFFL is used for contribution evaluation. 
This indicates that \ours~is effective under all settings as we vary the fraction of malicious clients.

\color{black}


\myparatight{Effect of Different $c$ on Defense Performances} We present the precision, recall and F1-score when the server utilizes Multi-Krum, Trimmed-Mean, \textcolor{black}{FABA, Sniper, Foolsgold} to mitigate \ours~ in Table \ref{tab:c-detection}, where we apply \ours~under UNI data partition with RFFL contribution evaluation method employed. We observe that the defense methods show marginal performance gain as $c$ increases. 
This indicates the stealthiness of \ours~towards the choice of $c$.

\myparatight{Experimental Results When the Attacker is Unware of the Contribution Evaluation Methods} In Figure \ref{fig:transfer-exp}, we evaluate the scenario where the attacker does not know the specific contribution evaluation method used by the server. 
Specifically, the attacker makes a guess on the contribution evaluation method used by the server, which may not necessarily be identical to the method employed.
Our results indicate that the client launching \ours~can still successfully elevate its contribution evaluated by the server even when it is unaware of the contribution evaluation method.

\color{black}

\section{More Experiment Details}
\label{appx:more-exp-detail}
\subsection{Model Structures}
\label{appx:model-structure}
We give the details of the CNN models for experiments on MNIST and CIFAR-10 in Table \ref{tab:cnn-model}. For Tiny-ImageNet, we use the pretrained \textit{VGG11} implementation from PyTorch\footnote{\url{https://pytorch.org/vision/0.16/models/generated/torchvision.models.vgg11.html}}.

\input{float_element/table-cnn}

\subsection{Data Partitions}
\label{appx:data-partition}

\begin{figure}[htbp]
  \centering
  \includegraphics[width=0.8\linewidth]{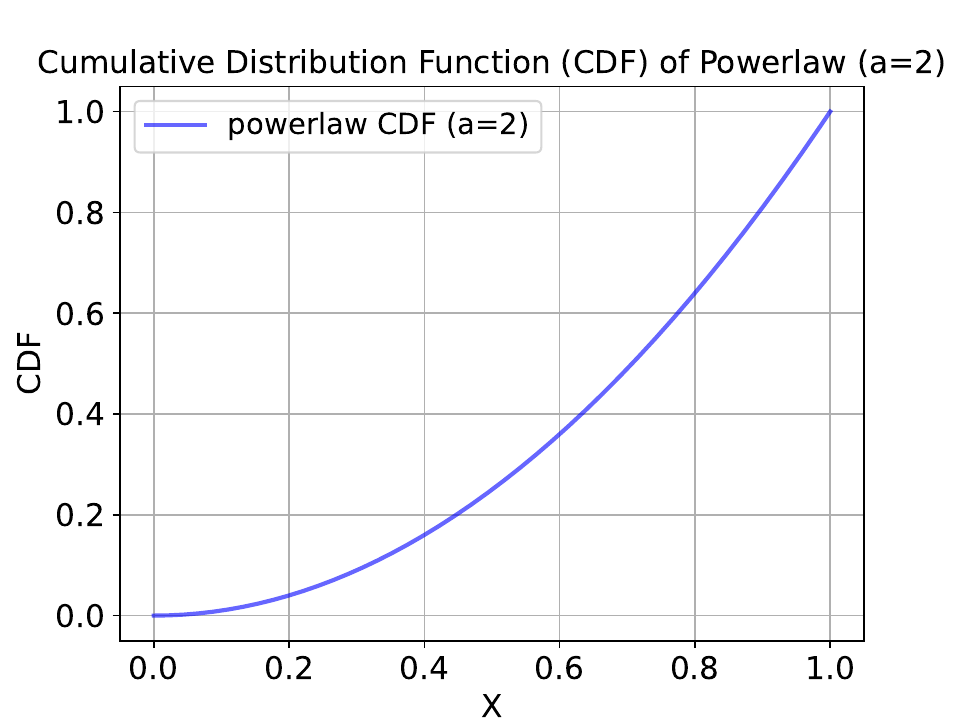}
  \caption{This figure shows the cumulative distribution function of a parameterized power law distribution with parameter $a = 2$.}
  \label{fig:cdf of pow}
\end{figure}

\myparatight{POW} The data partition method POW distributes the training images of each dataset to the clients using a parameterized power law distribution.
The probability density function for the parameterized power law distribution is given as follows
\begin{equation}
    f(x;a) = ax^{a-1},
\end{equation}
where $0 \leq x \leq 1$ is a random variable, and $a>1$ is a shape parameter.
Let $F(x;a)$ be the corresponding cumulative density function, which is illustrated in Figure \ref{fig:cdf of pow}.
Then the data partition method POW requires that the number of clients that owns no less than $x$ fraction of the training images follows $F(x;a)$.
In our experiments, we set $a = 2$. For the CIFAR-10 and Tiny-ImageNet datasets, the numbers of training samples of the ten clients are $731, 1458, 2184, 2911, 3637, 4364, 5090, 5817, 6543$ and  $7265$, respectively. 

\myparatight{CLA}
The data partition method CLA splits the data samples into the clients' local training datasets as follows. 
We first categorize the dataset based on labels. Then CLA specifies the numbers of classes that should be covered by the local training dataset of each client. We then sample a subset of data samples these specified classes, and allocate these data samples to the local training dataset of the client. 
Using CLA, the numbers of classes of 10 clients are $6,6,7,7,8,8,9,9,10$, and $10$ for the CIFAR-10 dataset, and $100,111,122,133,144,155,166,177,188$, and $200$ for Tiny-ImageNet dataset, respectively.

\subsection{Details on Contribution Evaluation Methods}\label{app:contribution evaluation method}

In our experiments, we consider five state-of-the-art contribution evaluations that can be employed by the server.
We detail each of these methods in the following.
\begin{itemize}
    \item \textbf{FedSV} \cite{wang2020principled}: FedSV is a game theory based contribution evaluation method. The contribution of client $i$ is then calculated as 
    \begin{equation*}
        \mathcal{E}(\mathbf{g}_i^t) = \frac{1}{\left|\Gamma\right|} \sum_{S \subseteq \Gamma \backslash\{i\}} \frac{U^t(S \cup\{i\})-U^t(S)}{\left(\begin{array}{c}
\left|\Gamma\right|-1 \\
|S|
\end{array}\right)},
    \end{equation*}
    where the utility function $U^t(S)$ is defined as $U^t(S) = L(\mathcal{D}_s;\mathbf{w}^t) - L(\mathcal{D}_s;\frac{1}{|S|} \sum_{k \in S} \mathbf{w}_k^{t+1})$, $\mathcal{D}_s$ is a validation dataset kept by the server, and $S\subseteq\Gamma\setminus\{i\}$ is a subset of clients not including client $i$.
    FedSV uses FedAvg \cite{mcmahan2017communication} as the aggregation rule.
    \item \textbf{LOO} \cite{wang2020principled, zhang2021refiner}: This is a game theory based contribution evaluation. It measures the contribution of each client $i$ as $\mathcal{E}(\mathbf{g}_i^t)  = L(\mathcal{D}_s;\mathbf{w}_{-i}^t) - L(\mathcal{D}_s;\mathbf{w}^t)$, where $\mathbf{w}_{-i}^t$ denotes the global model without aggregating the local model update from client $i$ at round $t$, and $\mathcal{D}_s$ is the server's validation dataset.
    LOO uses FedAvg \cite{mcmahan2017communication} as the aggregation rule.
    \item \textbf{CFFL} \cite{lyu2020collaborative}: This is an individual performance based contribution evaluation. The contribution of client $i$ is measured by the accuracy of local model from client $i$ using the validation dataset $\mathcal{D}_s$, i.e., $\mathcal{E}(\mathbf{g}_i^t) = \frac{\operatorname{vacc}_i}{\sum_{j \in \Gamma} \operatorname{vacc}_j}$, where $\operatorname{vacc}_i$ is the validation accuracy of client $i$'s local model $\textbf{w}_i^{t+1}$. The aggregation rule of CFFL is a variant of FedAvg \cite{mcmahan2017communication}, which considers both the data size and the number of classes of a client. Specifically, when the data size is imbalanced, the aggregation rule follows FedAvg. When the class number is imbalanced, the aggregation rule is $\mathcal{A}(\mathbf{g}_1^t, \mathbf{g}_2^t, \cdots, \mathbf{g}_{N}^t) = \sum_{i\in\Gamma} \frac{\left|\textit{class}_i\right|}{\sum_j\left|\textit{class}_j\right|}\mathbf{g}_i^t$, where $|\textit{class}|_i$ denotes the number of classes in $\mathcal{D}_i$.
    \item \textbf{GDR} \cite{xu2021gradient}: This is an individual performance based contribution evaluation. It leverages the cosine distance between the aggregated global model updates and local model updates to estimate SV. It then uses the so-called cosine gradient Shapley value as the client contribution, i.e., $\mathcal{E}(\mathbf{g}_i^t) = S_c(\mathbf{u}^t,\mathbf{u}_i^t)$, where $\mathbf{u}_i^t := \epsilon \mathbf{g}_i^t / \| \mathbf{g}_i^t\|$ represents the local model update of client $i$ normalized using a coefficient $\epsilon$, and $\mathbf{u}^t$ is the global model update by aggregating the normalized local model updates. The aggregation rule is $\mathcal{A}(\mathbf{g}_1^t, \mathbf{g}_2^t, \cdots, \mathbf{g}_{N}^t) = \sum_{i\in\Gamma} r_i^t \mathbf{g}_i^t$, where $r_i^t$ is a normalized weight coefficient calculated as the rolling mean of $\mathcal{E}(\mathbf{g}_i^t)$, i.e., $r_i^t = \alpha r_i^{t-1} + (1-\alpha) \mathcal{E}(\mathbf{g}_i^t)$, where the relative weight $\alpha \in (0,1)$.
    \item \textbf{RFFL} \cite{xu2020reputation}: This is an individual performance based method. The contribution is quantified by the cosine similarity between the local model update $\mathbf{g}_i^t$ and the aggregated global model update $\mathbf{g}^t$, i.e., $\mathcal{E}(\mathbf{g}_i^t) = S_c(\mathbf{g}^t,\mathbf{g}_i^t)$, where $S_c(\mathbf{a},\mathbf{b})$ denotes the cosine similarity between vectors $\mathbf{a}$ and $\mathbf{b}$. The aggregation rule of RFFL is similar to GDR, which uses $r_i^t$ as the weight of $\mathbf{g}_i^t$.
\end{itemize}


\section{Algorithm Details}
\label{Appendix:algo}

This appendix presents the L-BFGS algorithm and the complete algorithm of \ours.
We summarize the L-BFGS algorithm in Algorithm \ref{Algorithm: L-BFGS}, which takes the buffers $\Delta \mathbf{W}$ and $\Delta \mathbf{G}$ as well as $\mathbf{v} = \mathbf{w}^t-\mathbf{w}^{t-1}$ as inputs, and outputs the Hessian-vector product $H^t(\mathbf{w}^t-\mathbf{w}^{t-1})$ for global model update prediction.
The complete algorithm of \ours~is given in Algorithm \ref{Algorithm: Our Attack}.

\input{float_element/algo_LBFGS}
\input{float_element/algo_attack}

\section{Proofs of Section \ref{sec:theoretical}}
\label{Appendix: proof}

\myparatight{Proof of Proposition \ref{lemma not decrease the utility}}
Since $\cos(\mathbf{g'}, c \cdot \mathbf{\hat{g}}_i) = \cos(\mathbf{g'}, \mathbf{\hat{g}}_i)$, we have
\begin{align*}
& \cos(\mathbf{g}, \mathbf{\hat{g}}_i) - \cos(\mathbf{g'}, c \cdot \mathbf{\hat{g}}_i) = \cos(\mathbf{g}, \mathbf{\hat{g}}_i) - \cos(\mathbf{g'}, \mathbf{\hat{g}}_i) \\
&= (1 - S_c(\mathbf{g}, \mathbf{\hat{g}}_i)) - (1- S_c(\mathbf{g'}, \mathbf{\hat{g}}_i)) = S_c(\mathbf{g'}, \mathbf{\hat{g}}_i) - S_c(\mathbf{g}, \mathbf{\hat{g}}_i)\\
&=\frac{\mathbf{g'} \cdot \mathbf{\hat{g}}_i}{\|\mathbf{g'}\| \|\mathbf{\hat{g}}_i\|} - \frac{\mathbf{g} \cdot \mathbf{\hat{g}}_i}{\|\mathbf{g}\| \|\mathbf{\hat{g}}_i\|} = \frac{\|\mathbf{g}\| \mathbf{g'} \cdot \mathbf{\hat{g}}_i - \|\mathbf{g'}\| \mathbf{g} \cdot \mathbf{\hat{g}}_i}{\|\mathbf{g'}\| \|\mathbf{g}\| \|\mathbf{\hat{g}}_i\|}.
\end{align*}
We denote by $\mathbf{g'}=\mathbf{g}+(c-1)\alpha_i \mathbf{\hat{g}}_i$. For $\mathbf{g} \cdot \mathbf{\hat{g}}_i \geq 0$, we have:
\begin{align*}
&\cos(\mathbf{g}, \mathbf{\hat{g}}_i) - \cos(\mathbf{g'}, c \cdot \mathbf{\hat{g}}_i) \\
& \geq \frac{\|\mathbf{g}\| (\mathbf{g}+(c-1)\alpha_i \mathbf{\hat{g}}_i) \cdot \mathbf{\hat{g}}_i - (\|\mathbf{g}\|+\|(c-1)\alpha_i \mathbf{\hat{g}}_i\|) \mathbf{g} \cdot \mathbf{\hat{g}}_i}{\|\mathbf{g'}\| \|\mathbf{g}\| \|\mathbf{\hat{g}}_i\|}\\
&= \frac{\|\mathbf{g}\| \mathbf{g} \cdot \mathbf{\hat{g}}_i+(c-1)\alpha_i\|\mathbf{g}\| \mathbf{\hat{g}}_i \cdot \mathbf{\hat{g}}_i-\|\mathbf{g}\| \mathbf{g} \cdot \mathbf{\hat{g}}_i-\|(c-1)\alpha_i \mathbf{\hat{g}}_i\| \mathbf{g} \cdot \mathbf{\hat{g}}_i}{\|\mathbf{g'}\| \|\mathbf{g}\| \|\mathbf{\hat{g}}_i\|}\\
&= \frac{(c-1)\alpha_i\|\mathbf{g}\| \|\mathbf{\hat{g}}_i\|^2 -(c-1)\alpha_i \|\mathbf{\hat{g}}_i\| \mathbf{g} \cdot \mathbf{\hat{g}}_i}{\|\mathbf{g'}\| \|\mathbf{g}\| \|\mathbf{\hat{g}}_i\|}\\
&= \frac{(c-1)\alpha_i\|\mathbf{\hat{g}}_i\| (\|\mathbf{g}\|\|\mathbf{\hat{g}}_i\|- \mathbf{g} \cdot \mathbf{\hat{g}}_i)}{\|\mathbf{g'}\| \|\mathbf{g}\| \|\mathbf{\hat{g}}_i\|} \geq 0.
\end{align*}    

The first inequality uses Triangle Inequality, i.e., $\|\mathbf{a}+\mathbf{b}\| \leq \|\mathbf{a}\|+\|\mathbf{b}\|$. The second inequality holds because $c>1$ and $\|\mathbf{g}\|\|\mathbf{\hat{g}}_i\| \geq \| \mathbf{g} \cdot \mathbf{\hat{g}}_i \| \geq \mathbf{g} \cdot \mathbf{\hat{g}}_i$.

Similarly, for $\mathbf{g} \cdot \mathbf{\hat{g}}_i \leq 0$, we have:
\begin{align*}
&\cos(\mathbf{g}, \mathbf{\hat{g}}_i) - \cos(\mathbf{g'}, c \cdot \mathbf{\hat{g}}_i) = \frac{\|\mathbf{g}\| \mathbf{g'} \cdot \mathbf{\hat{g}}_i - \|\mathbf{g'}\| \mathbf{g} \cdot \mathbf{\hat{g}}_i}{\|\mathbf{g'}\| \|\mathbf{g}\| \|\mathbf{\hat{g}}_i\|}\\
& = \frac{\|\mathbf{g}\| (\mathbf{g}+(c-1)\alpha_i \mathbf{\hat{g}}_i) \cdot \mathbf{\hat{g}}_i - \|\mathbf{g'}\| \mathbf{g} \cdot \mathbf{\hat{g}}_i}{\|\mathbf{g'}\| \|\mathbf{g}\| \|\mathbf{\hat{g}}_i\|}\\
& = \frac{\|\mathbf{g}\| \mathbf{g} \cdot \mathbf{\hat{g}}_i + (c-1)\alpha_i \|\mathbf{g}\| \|\mathbf{\hat{g}}_i\|^2 - \|\mathbf{g'}\| \mathbf{g} \cdot \mathbf{\hat{g}}_i}{\|\mathbf{g'}\| \|\mathbf{g}\| \|\mathbf{\hat{g}}_i\|}\\
& = \frac{\|\mathbf{g}'-(c-1)\alpha_i \mathbf{\hat{g}}_i\| \mathbf{g} \cdot \mathbf{\hat{g}}_i + (c-1)\alpha_i \|\mathbf{g}\| \|\mathbf{\hat{g}}_i\|^2 - \|\mathbf{g'}\| \mathbf{g} \cdot \mathbf{\hat{g}}_i}{\|\mathbf{g'}\| \|\mathbf{g}\| \|\mathbf{\hat{g}}_i\|}\\
& \geq \frac{(\|\mathbf{g}'\|+\|(c-1)\alpha_i \mathbf{\hat{g}}_i\|) \mathbf{g} \cdot \mathbf{\hat{g}}_i + (c-1)\alpha_i \|\mathbf{g}\| \|\mathbf{\hat{g}}_i\|^2 - \|\mathbf{g'}\| \mathbf{g} \cdot \mathbf{\hat{g}}_i}{\|\mathbf{g'}\| \|\mathbf{g}\| \|\mathbf{\hat{g}}_i\|}\\
&= \frac{(c-1)\alpha_i \|\mathbf{\hat{g}}_i\| \mathbf{g} \cdot \mathbf{\hat{g}}_i + (c-1)\alpha_i\|\mathbf{g}\| \|\mathbf{\hat{g}}_i\|^2}{\|\mathbf{g'}\| \|\mathbf{g}\| \|\mathbf{\hat{g}}_i\|}\\
&= \frac{(c-1)\alpha_i\|\mathbf{\hat{g}}_i\| (\|\mathbf{g}\|\|\mathbf{\hat{g}}_i\| +  \mathbf{g} \cdot \mathbf{\hat{g}}_i)}{\|\mathbf{g'}\| \|\mathbf{g}\| \|\mathbf{\hat{g}}_i\|} \geq 0.
\end{align*}  
The first inequality holds uses Triangle Inequality, i.e., $\|\mathbf{a}-\mathbf{b}\| \leq \|\mathbf{a}\|+\|\mathbf{b}\|$.  The second inequality holds because $c>1$ and $\|\mathbf{g}\|\|\mathbf{\hat{g}}_i\| \geq \| \mathbf{g} \cdot \mathbf{\hat{g}}_i \| \geq - \mathbf{g} \cdot \mathbf{\hat{g}}_i$.

Combine the above two cases, we have:
\begin{align*}
& \cos(\mathbf{g}, \mathbf{\hat{g}}_i) - \cos(\mathbf{g'}, c \cdot \mathbf{\hat{g}}_i) \\ 
& \geq \frac{(c-1)\alpha_i\|\mathbf{\hat{g}}_i\| (\|\mathbf{g}\|\|\mathbf{\hat{g}}_i\| -  |\mathbf{g} \cdot \mathbf{\hat{g}}_i|)}{\|\mathbf{g}+(c-1)\alpha_i \mathbf{\hat{g}}_i\| \|\mathbf{g}\| \|\mathbf{\hat{g}}_i\|} \geq 0.
\end{align*}
Therefore, $\cos(\mathbf{g'}, c \cdot \mathbf{\hat{g}}_i) \leq \cos(\mathbf{g}, \mathbf{\hat{g}}_i)$.

\myparatight{Proof of Corollary \ref{lemma rank non-decreasing}}
Since $\cos(\mathbf{g},\mathbf{\hat{g}}_i) \leq \cos(\mathbf{g},\mathbf{g}_j)$, we have:
\begin{align*}
& \cos(\mathbf{g},\mathbf{g}_j) - \cos(\mathbf{g},\mathbf{\hat{g}}_i)
= S_c(\mathbf{g},\mathbf{\hat{g}}_i) - S_c(\mathbf{g},\mathbf{g}_j) \\
&= \frac{\mathbf{g} \cdot \mathbf{\hat{g}}_i}{\|\mathbf{g}\| \|\mathbf{\hat{g}}_i\|} - \frac{\mathbf{g} \cdot \mathbf{g}_j}{\|\mathbf{g}\| \|\mathbf{g}_j\|}\\
& = \frac{\|\mathbf{g}_j\| \mathbf{g} \cdot \mathbf{\hat{g}}_i - \|\mathbf{\hat{g}}_i\| \mathbf{g} \cdot \mathbf{g}_j}{\|\mathbf{g}\| \|\mathbf{\hat{g}}_i\| \|\mathbf{g}_j\|} \geq 0.
\end{align*}
Since the denumerator $\|\mathbf{g}\| \|\mathbf{\hat{g}}_i\| \|\mathbf{g}_j\| \geq 0$, the numerator $\|\mathbf{g}_j\| \mathbf{g} \cdot \mathbf{\hat{g}}_i - \|\mathbf{\hat{g}}_i\| \mathbf{g} \cdot \mathbf{g}_j \geq 0$. Denote $\mathbf{g'}=\mathbf{g}+(c-1)\alpha_i \mathbf{\hat{g}}_i$, we have:
\begin{align*}
& \cos(\mathbf{g}',\mathbf{g}_j) - \cos(\mathbf{g}',c\mathbf{\hat{g}}_i) = \cos(\mathbf{g}',\mathbf{g}_j) - \cos(\mathbf{g}',\mathbf{\hat{g}}_i)\\
& = S_c(\mathbf{g}',\mathbf{\hat{g}}_i) - S_c(\mathbf{g}',\mathbf{g}_j) \\
& = \frac{\mathbf{g'} \cdot \mathbf{\hat{g}}_i}{\|\mathbf{g'}\| \|\mathbf{\hat{g}}_i\|} - \frac{\mathbf{g'} \cdot \mathbf{g}_j}{\|\mathbf{g'}\| \|\mathbf{g}_j\|} 
= \frac{\|\mathbf{g}_j\| \mathbf{g'} \cdot \mathbf{\hat{g}}_i - \|\mathbf{\hat{g}}_i\| \mathbf{g}' \cdot \mathbf{g}_j}{\|\mathbf{g'}\| \|\mathbf{\hat{g}}_i\| \|\mathbf{g}_j\|}\\
& = \frac{\|\mathbf{g}_j\| (\mathbf{g}+(c-1)\alpha_i \mathbf{\hat{g}}_i) \cdot \mathbf{\hat{g}}_i - \|\mathbf{\hat{g}}_i\| (\mathbf{g}+(c-1)\alpha_i \mathbf{\hat{g}}_i) \cdot \mathbf{g}_j}{\|\mathbf{g'}\| \|\mathbf{\hat{g}}_i\| \|\mathbf{g}_j\|}\\
& = \frac{\|\mathbf{g}_j\| \mathbf{g} \cdot \mathbf{\hat{g}}_i - \|\mathbf{\hat{g}}_i\| \mathbf{g} \cdot \mathbf{g}_j + (c-1)\alpha_i( \|\mathbf{g}_j\| \mathbf{\hat{g}}_i \cdot \mathbf{\hat{g}}_i - \|\mathbf{\hat{g}}_i\| \mathbf{\hat{g}}_i \cdot \mathbf{g}_j)}{\|\mathbf{g'}\| \|\mathbf{\hat{g}}_i\| \|\mathbf{g}_j\|}\\
& \geq \frac{(c-1)\alpha_i( \|\mathbf{g}_j\|  \mathbf{\hat{g}}_i \cdot \mathbf{\hat{g}}_i - \|\mathbf{\hat{g}}_i\| \mathbf{\hat{g}}_i \cdot \mathbf{g}_j)}{\|\mathbf{g'}\| \|\mathbf{\hat{g}}_i\| \|\mathbf{g}_j\|}\\
& = \frac{(c-1)\alpha_i ( \|\mathbf{g}_j\| \|\mathbf{\hat{g}}_i\|- \mathbf{\hat{g}}_i \cdot \mathbf{g}_j)}{\|\mathbf{g'}\| \|\mathbf{g}_j\|} \geq 0.
\end{align*}
The inequality holds because  $\|\mathbf{g}_j\| \mathbf{g} \cdot \mathbf{\hat{g}}_i - \|\mathbf{\hat{g}}_i\| \mathbf{g} \cdot \mathbf{g}_j \geq 0$.
Therefore, we have $\cos(\mathbf{g}',\mathbf{g}_j) \geq \cos(\mathbf{g}',c\mathbf{\hat{g}}_i)$.

\myparatight{Proof of Proposition \ref{lemma minimum c}}
By the definition of cosine distance, we have
\begin{align*}
& \cos(\mathbf{g}',\mathbf{g}_j) - \cos(\mathbf{g}',c\mathbf{\hat{g}}_i)
= \cos(\mathbf{g}',\mathbf{g}_j) - \cos(\mathbf{g}',\mathbf{\hat{g}}_i) \\
& = \frac{\mathbf{g'} \cdot \mathbf{\hat{g}}_i}{\|\mathbf{g'}\| \|\mathbf{\hat{g}}_i\|} - \frac{\mathbf{g'} \cdot \mathbf{g}_j}{\|\mathbf{g'}\| \|\mathbf{g}_j\|} 
= \frac{\|\mathbf{g}_j\| \mathbf{g'} \cdot \mathbf{\hat{g}}_i - \|\mathbf{\hat{g}}_i\| \mathbf{g}' \cdot \mathbf{g}_j}{\|\mathbf{g'}\| \|\mathbf{\hat{g}}_i\| \|\mathbf{g}_j\|}\\
& = \frac{\|\mathbf{g}_j\| (\mathbf{g}+(c-1)\alpha_i \mathbf{\hat{g}}_i) \cdot \mathbf{\hat{g}}_i - \|\mathbf{\hat{g}}_i\| (\mathbf{g}+(c-1)\alpha_i \mathbf{\hat{g}}_i) \cdot \mathbf{g}_j}{\|\mathbf{g'}\| \|\mathbf{\hat{g}}_i\| \|\mathbf{g}_j\|}\\
& = \frac{\|\mathbf{g}_j\| \mathbf{g} \cdot \mathbf{\hat{g}}_i - \|\mathbf{\hat{g}}_i\| \mathbf{g} \cdot \mathbf{g}_j + (c-1)\alpha_i \|\mathbf{\hat{g}}_i\|( \|\mathbf{g}_j\| \|\mathbf{\hat{g}}_i\| - \mathbf{\hat{g}}_i \cdot \mathbf{g}_j)}{\|\mathbf{g'}\| \|\mathbf{\hat{g}}_i\| \|\mathbf{g}_j\|}\\
& \geq 0.
\end{align*}
Note that since $\cos(\mathbf{g},\mathbf{\hat{g}}_i) > \cos(\mathbf{g},\mathbf{g}_j)$, we have  $\|\mathbf{g}_j\| \|\mathbf{\hat{g}}_i\| - \mathbf{\hat{g}}_i \cdot \mathbf{g}_j > 0$. Thus we have: 
\begin{align*}
c \geq \frac{\|\mathbf{\hat{g}}_i\| \mathbf{g} \cdot \mathbf{g}_j - \|\mathbf{g}_j\| \mathbf{g} \cdot \mathbf{\hat{g}}_i}{\alpha_i \|\mathbf{\hat{g}}_i\|( \|\mathbf{g}_j\| \|\mathbf{\hat{g}}_i\| - \mathbf{\hat{g}}_i \cdot \mathbf{g}_j)} + 1.
\end{align*}

\end{appendices}

%% file: float_element/fig_cs-rank_uni-pow.tex
\begin{figure*}[htbp]
    \centering
    \subfloat{
    \centering
    \includegraphics[width=0.7\linewidth]{figs/main-csvsdr/cs-vs-rank-legend.pdf}
    }
    
    \setcounter{subfigure}{0}
    \subfloat[FedSV]{
        \label{fig:capparatus}
        \centering
        \includegraphics[width=0.19\linewidth]{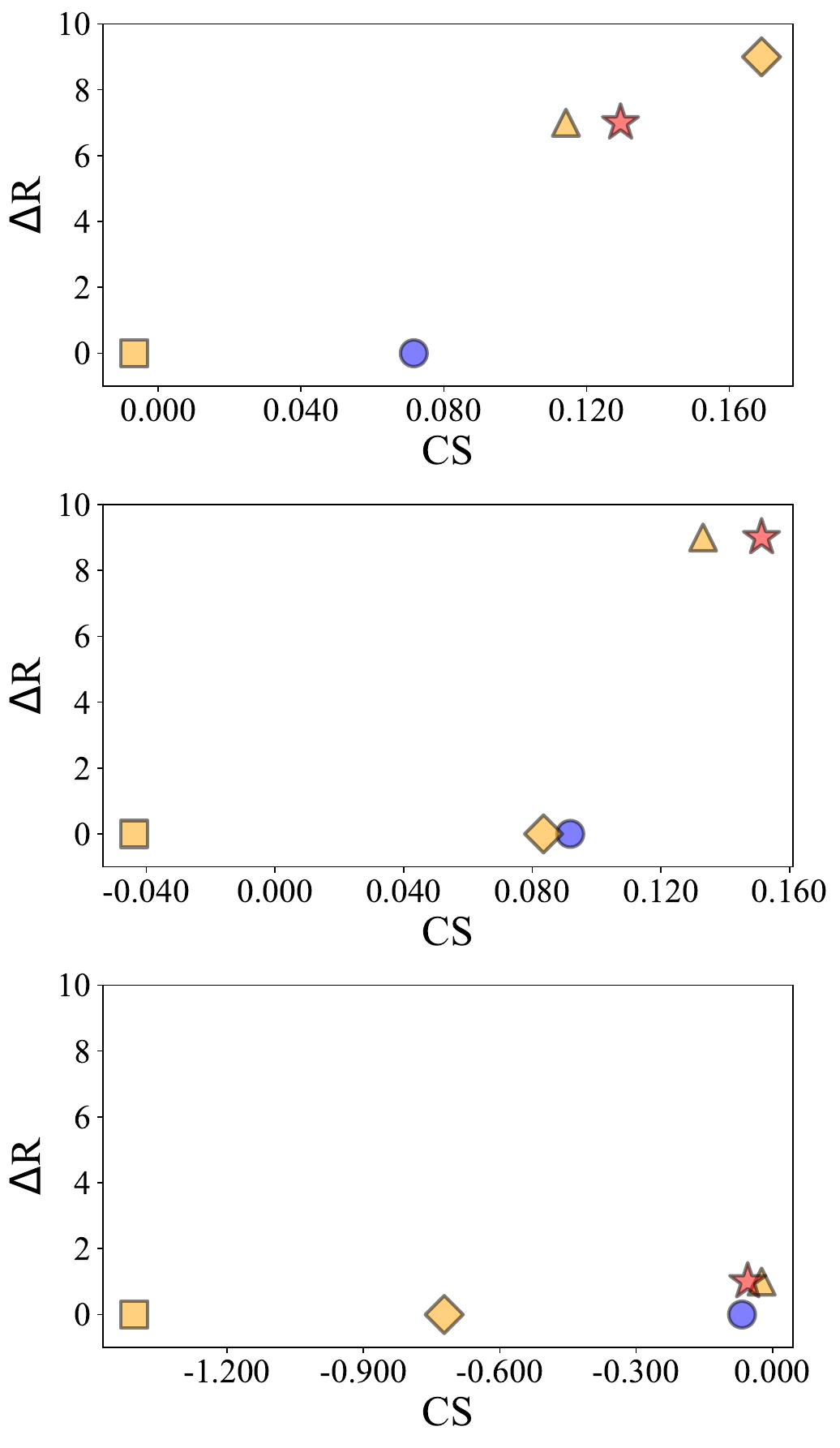}
    }
    \hfill
    \subfloat[LOO]{
        \label{fig:capparatus}
        \centering
        \includegraphics[width=0.19\linewidth]{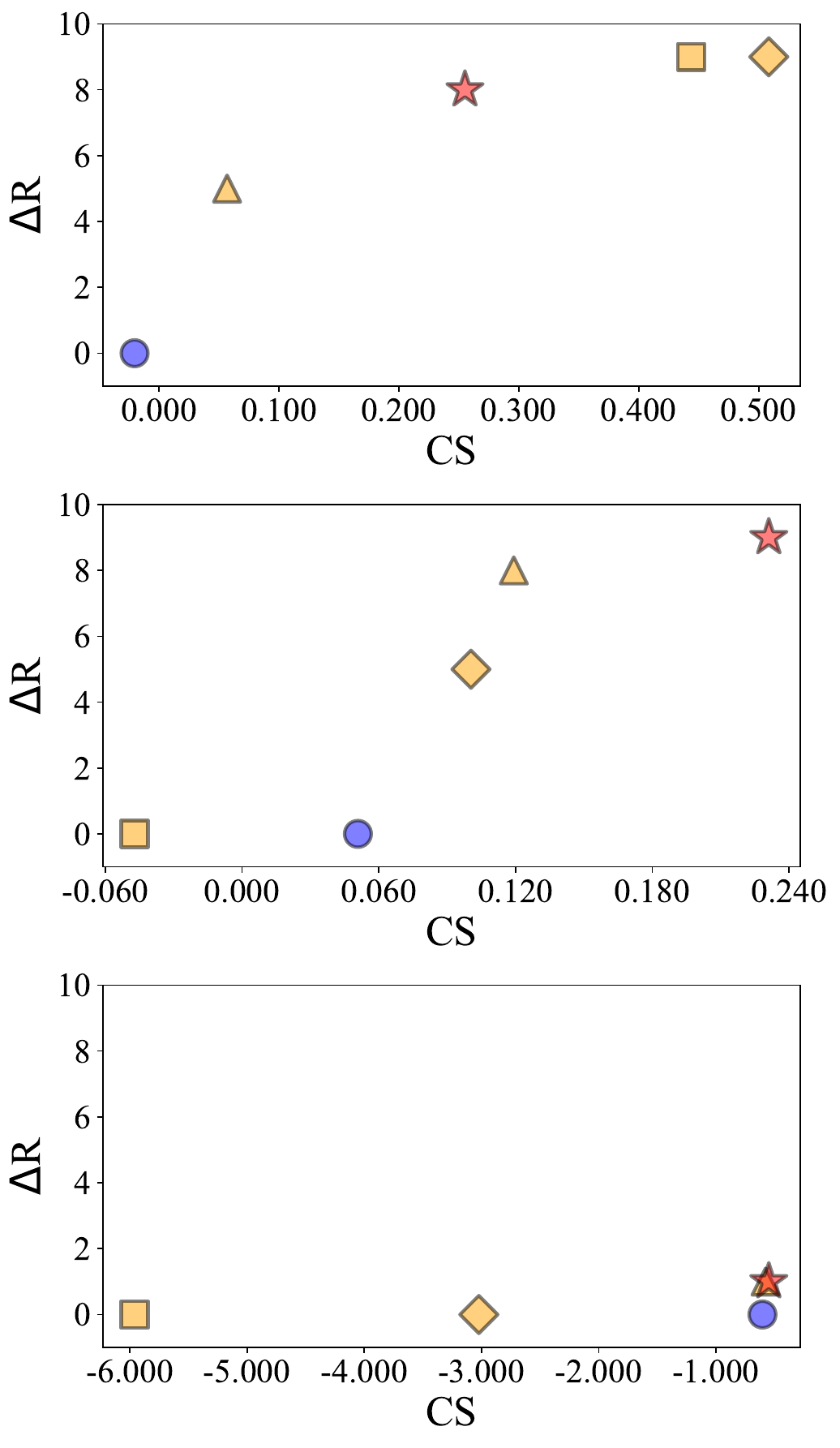}
    }
    \hfill
    \subfloat[CFFL]{
        \label{fig:capparatus}
        \centering
        \includegraphics[width=0.19\linewidth]{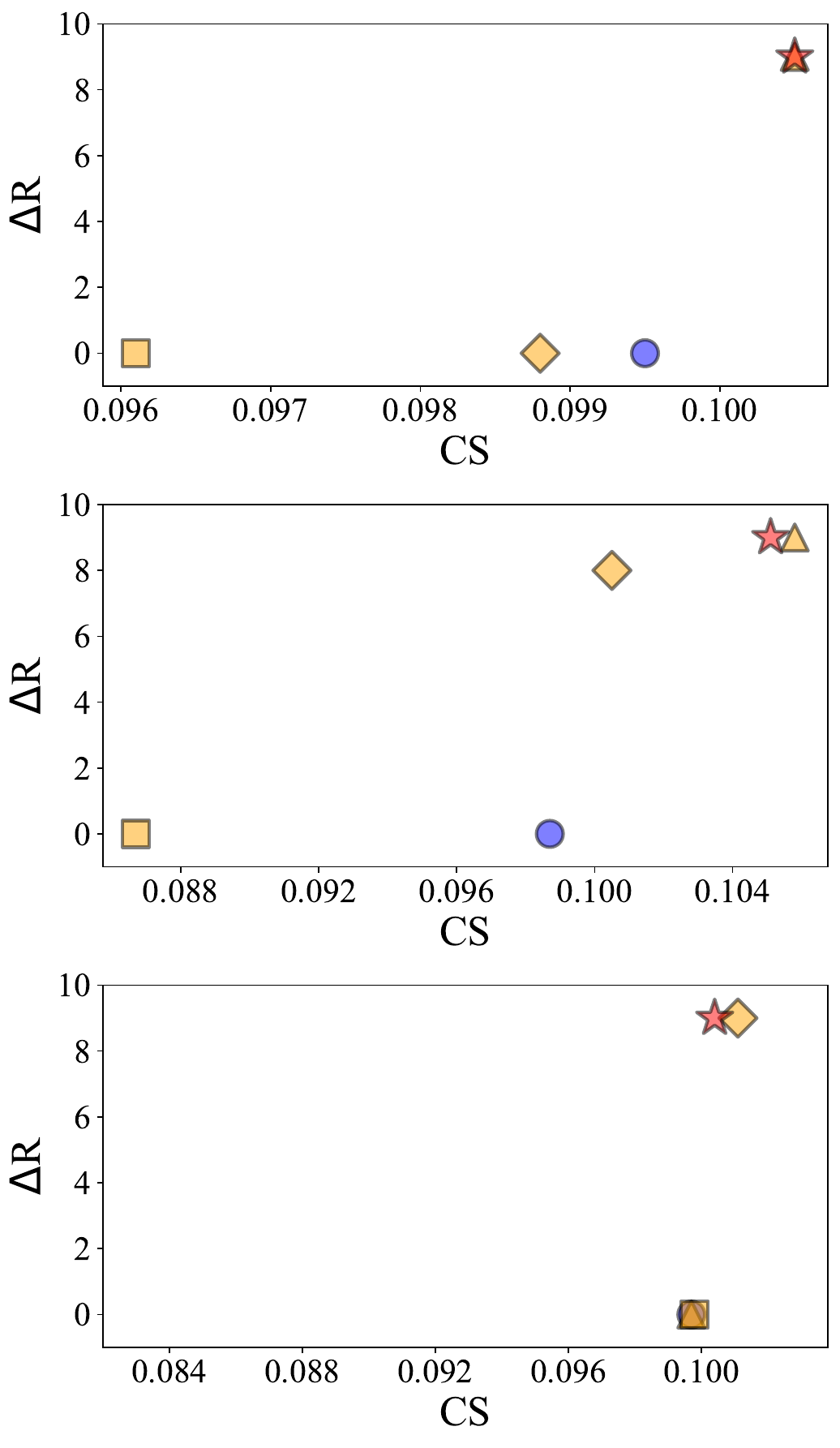}
    }
    \hfill
    \subfloat[GDR]{
        \label{fig:capparatus}
        \centering
        \includegraphics[width=0.19\linewidth]{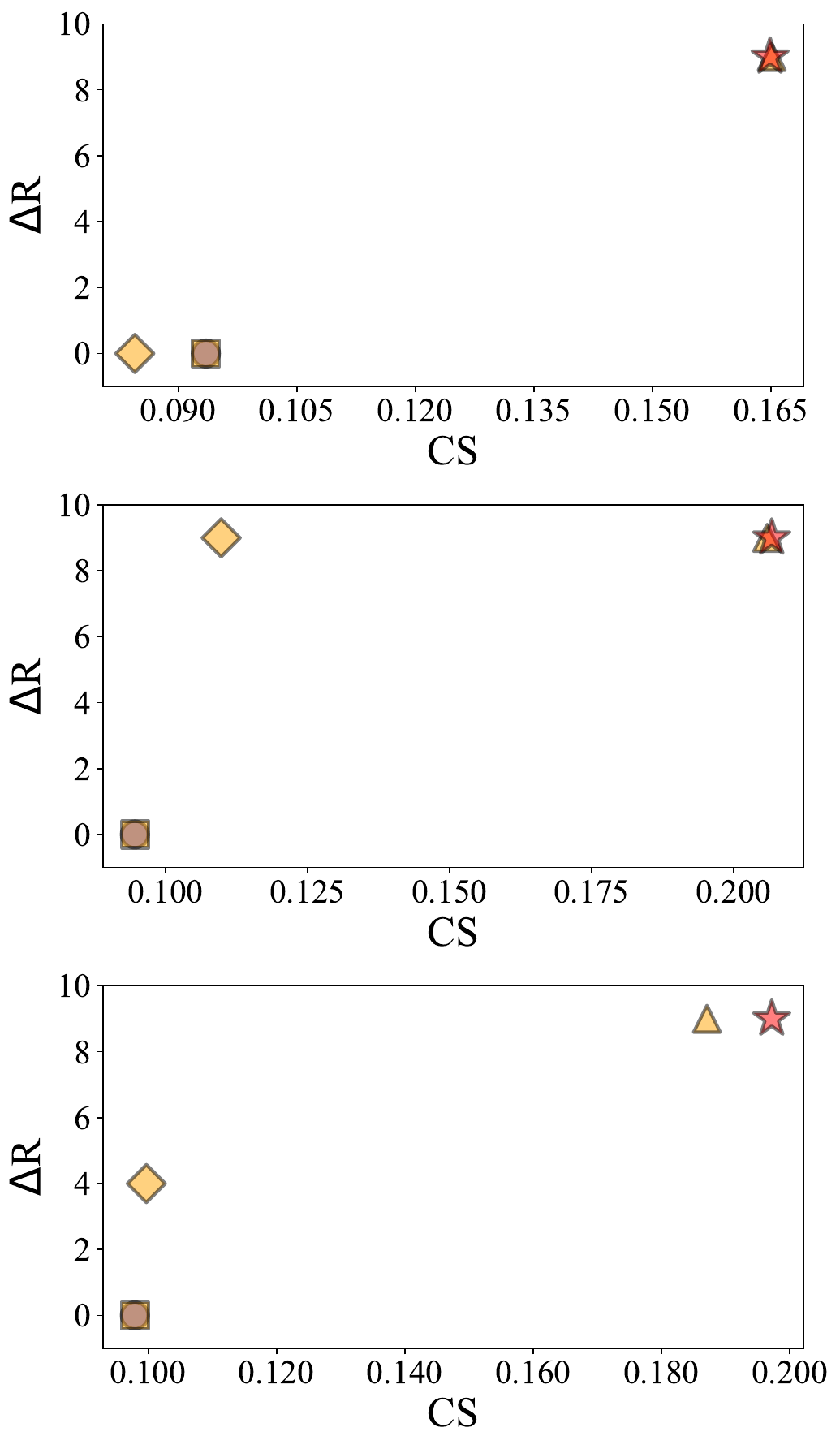}
    }
    \hfill
    \subfloat[RFFL]{
        \label{fig:capparatus}
        \centering
        \includegraphics[width=0.19\linewidth]{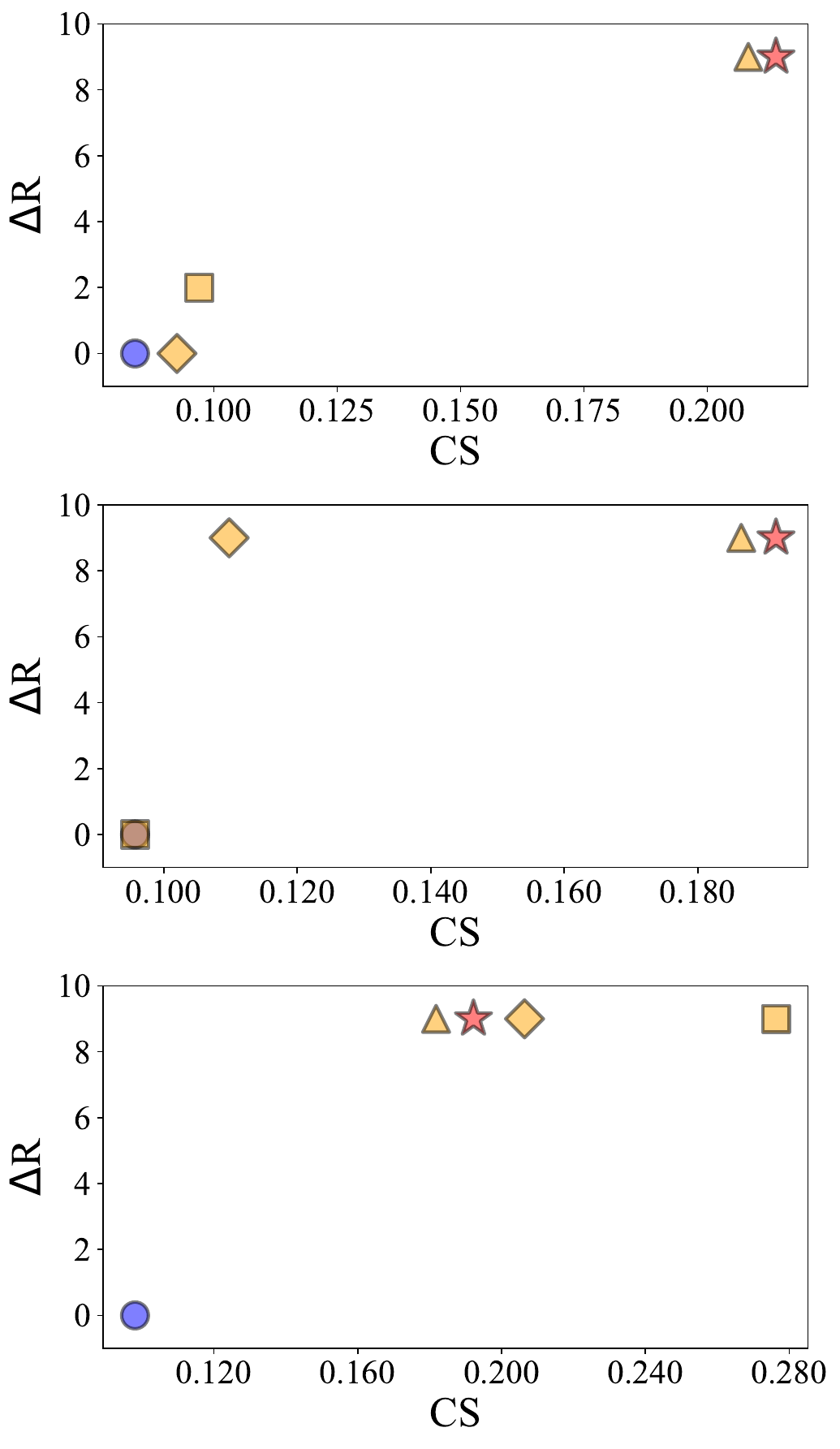}
    }
    \caption{Comparing the contribution score $CS$ and rank gain $\Delta R$ of the attacker when using \ours~and baselines under three datasets, i.e., MNIST (first row), CIFAR-10 (second row), and Tiny-ImageNet (third row), and five contribution evaluation methods, i.e., FedSV, LOO, CFFL, GDR, and RFFL. The data partition method is UNI (i.i.d. data distribution). Our results show \ours~is more effective than baselines under most of the contribution evaluation methods.}
    \label{fig:cs-vs-rank-UNI}
\end{figure*}

\begin{figure*}[htbp]
    \centering
    \subfloat{
    \centering
    \includegraphics[width=0.7\linewidth]{figs/main-csvsdr/cs-vs-rank-legend.pdf}
    }
    
    \setcounter{subfigure}{0}
    \subfloat[FedSV]{
        \label{fig:capparatus}
        \centering
        \includegraphics[width=0.19\linewidth]{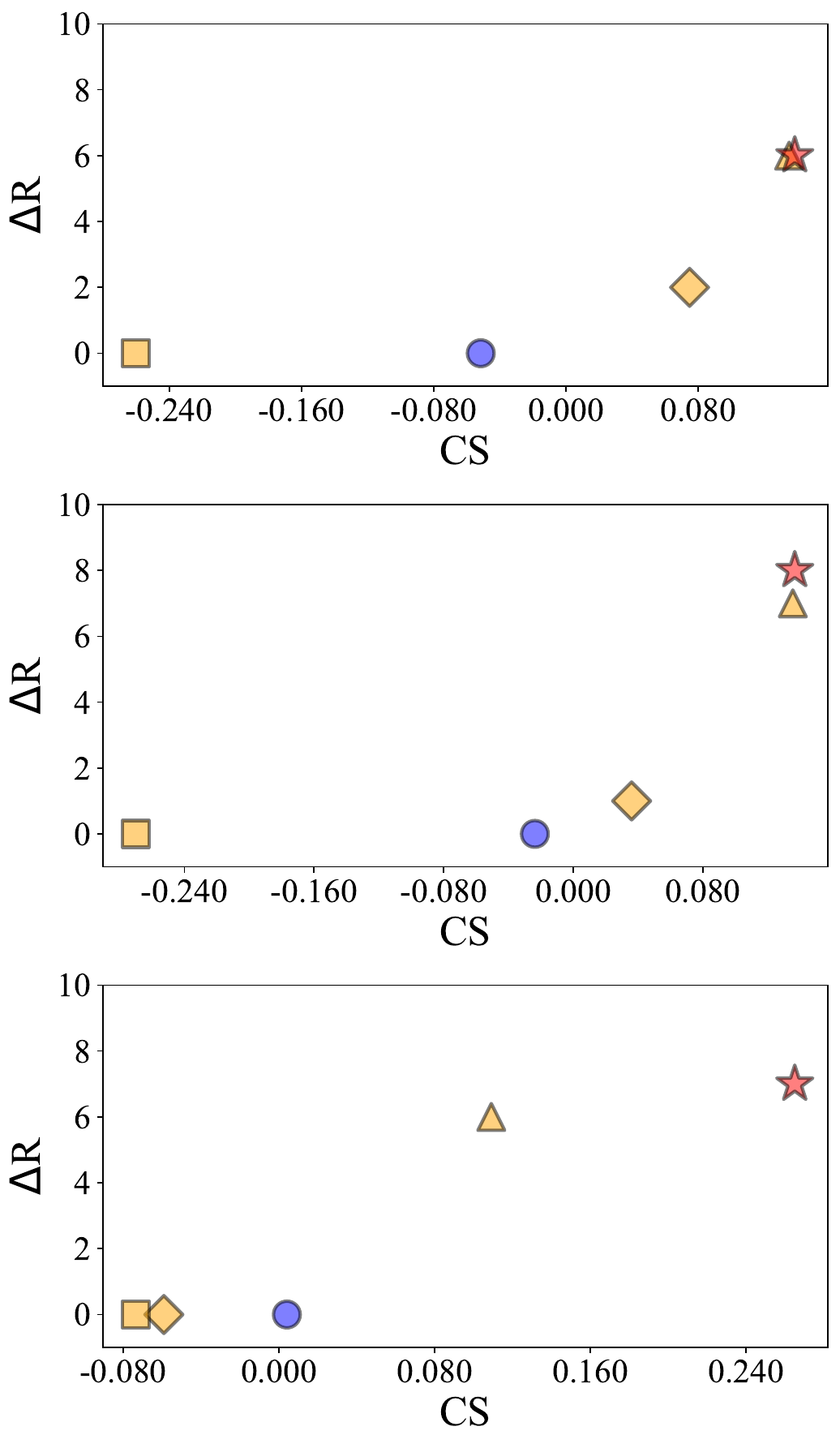}
    }
    \hfill
    \subfloat[LOO]{
        \label{fig:capparatus}
        \centering
        \includegraphics[width=0.19\linewidth]{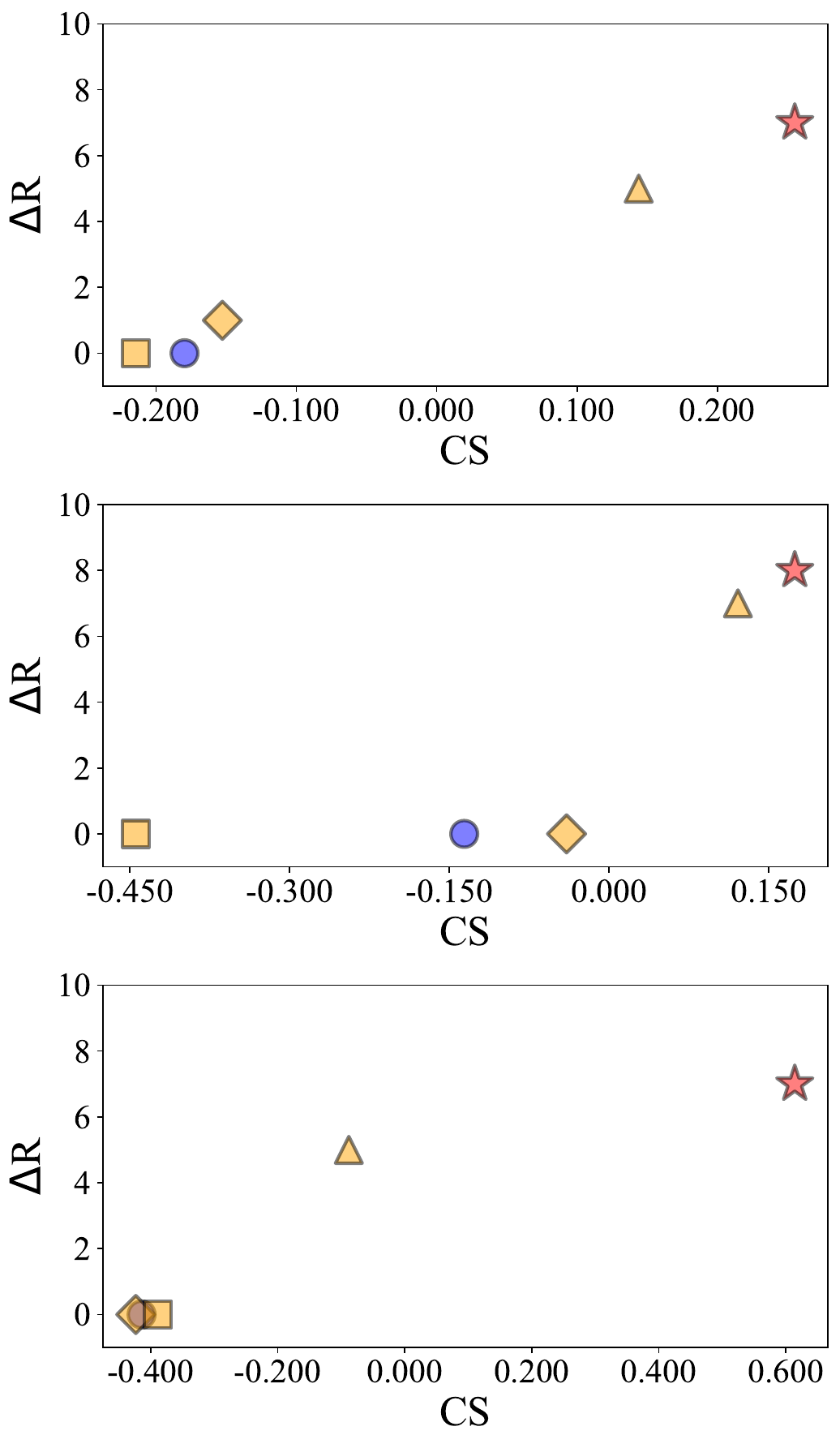}
    }
    \hfill
    \subfloat[CFFL]{
        \label{fig:capparatus}
        \centering
        \includegraphics[width=0.19\linewidth]{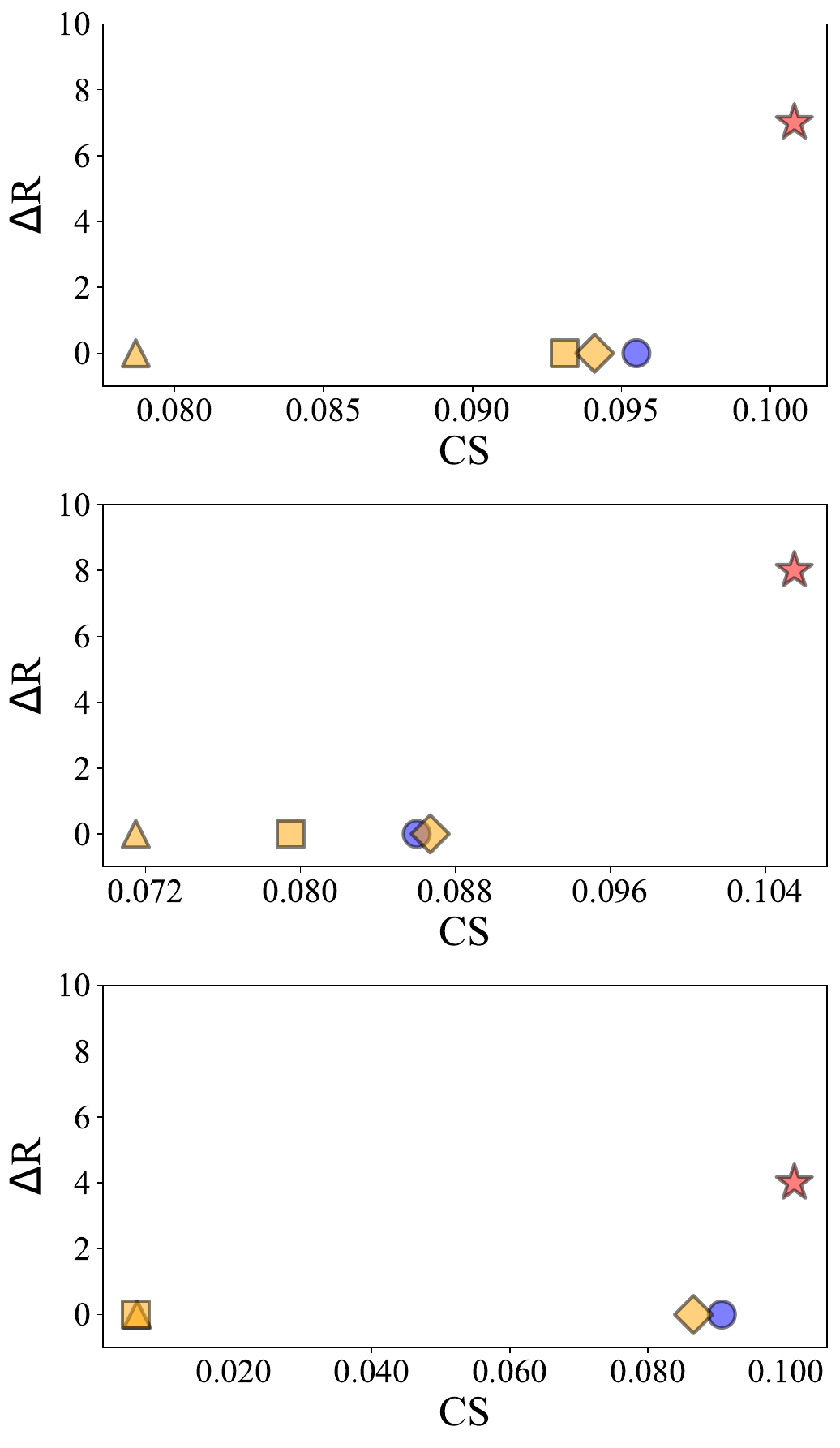}
    }
    \hfill
    \subfloat[GDR]{
        \label{fig:capparatus}
        \centering
        \includegraphics[width=0.19\linewidth]{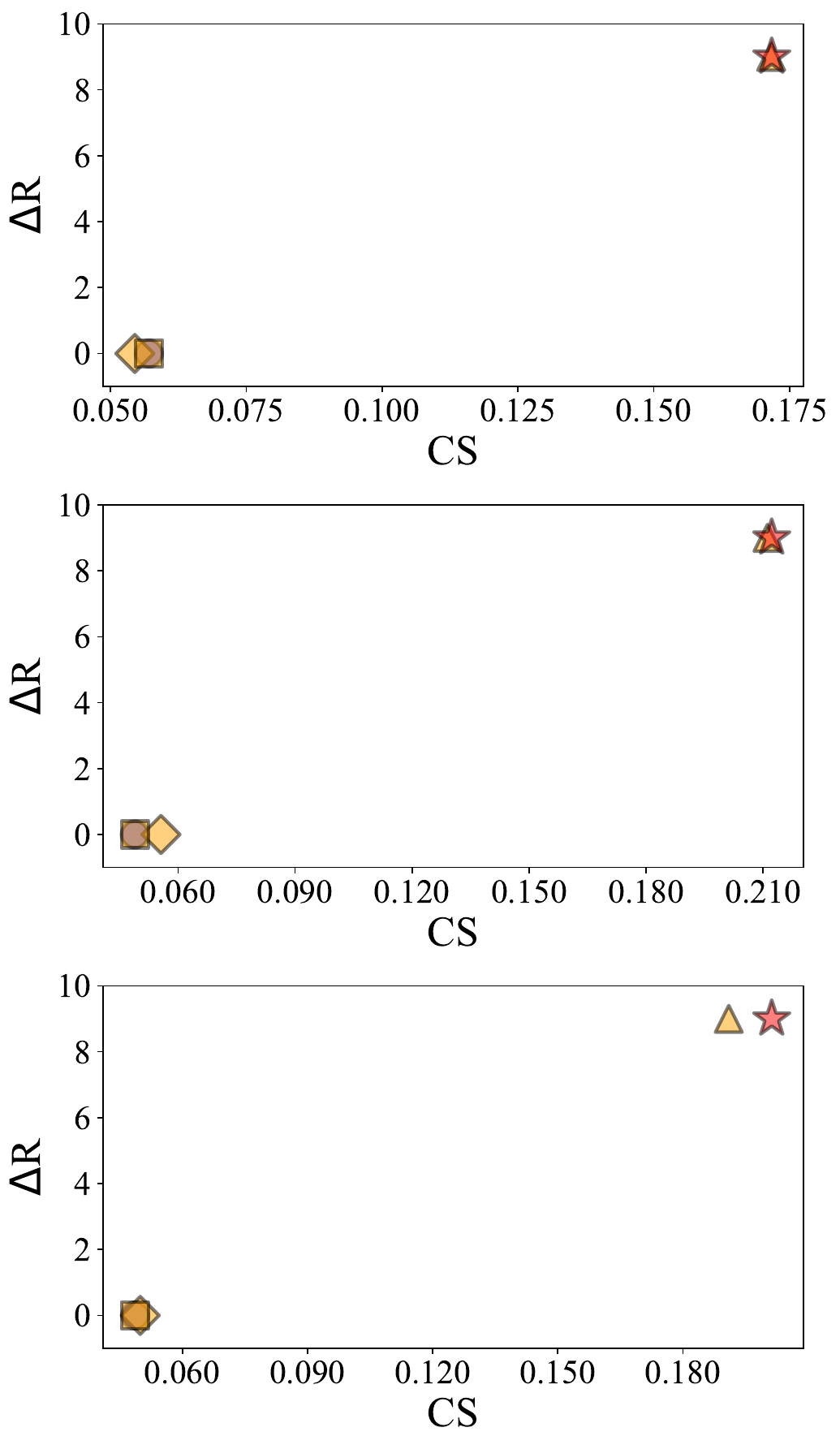}
    }
    \hfill
    \subfloat[RFFL]{
        \label{fig:capparatus}
        \centering
        \includegraphics[width=0.19\linewidth]{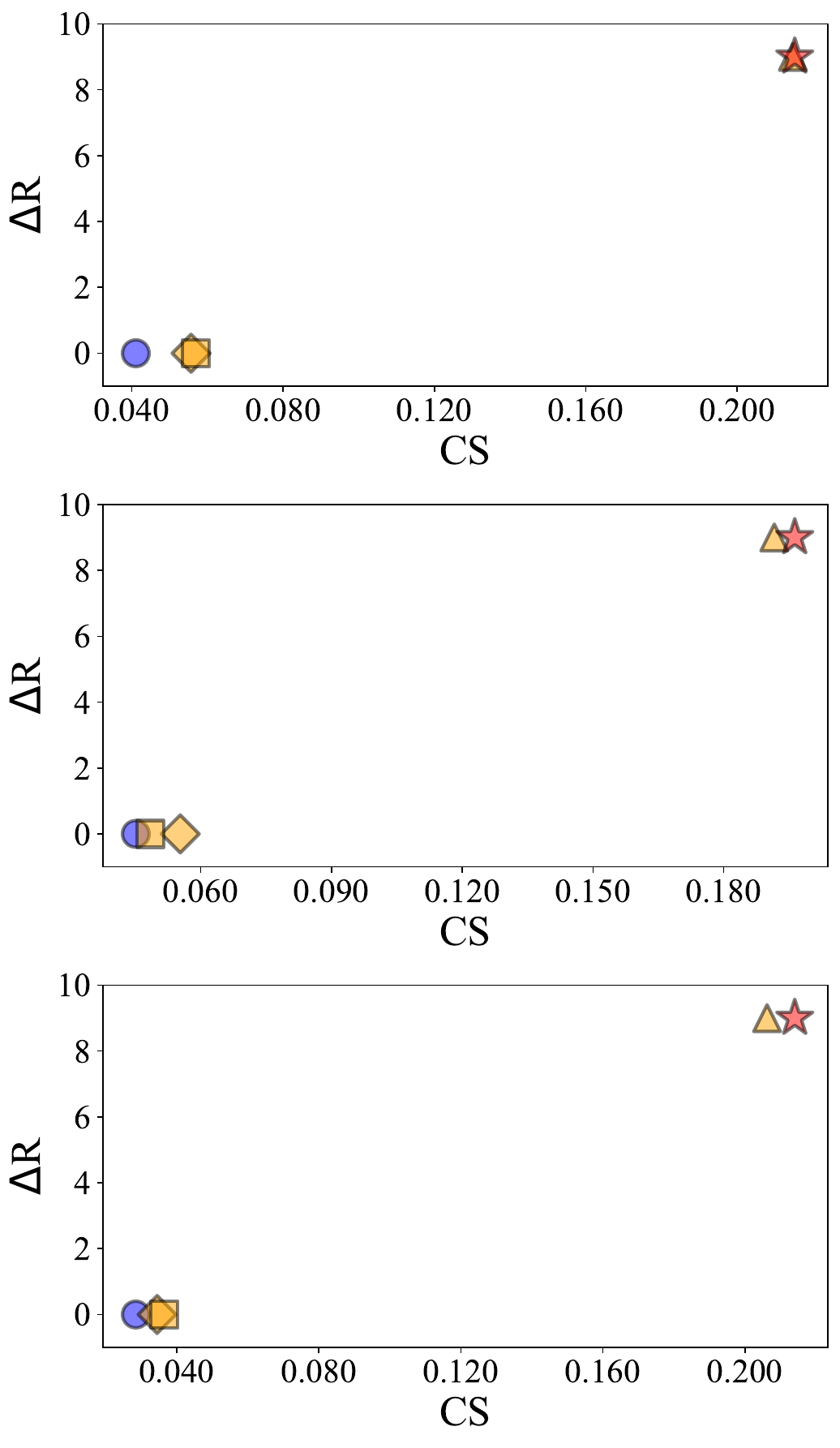}
    }
    \caption{Comparing the contribution score $CS$ and rank gain $\Delta R$ of the attacker when using \ours~and baselines under three datasets, i.e., MNIST (first row), CIFAR-10 (second row), and Tiny-ImageNet (third row), and five contribution evaluation methods, i.e., FedSV, LOO, CFFL, GDR, and RFFL. The data partition method is POW (non-i.i.d. data distribution). Our results show \ours~is consistently more effective than baselines.}
    \label{fig:cs-vs-rank-POW}
\end{figure*}

%% file: float_element/fig_mem_length.tex
\begin{figure*}[htbp]
    \captionsetup[subfloat]{farskip=0pt,captionskip=0pt}
    \centering
    \subfloat[CFFL-UNI]{
        \label{fig:capparatus}
        \centering
        \includegraphics[width=0.32\linewidth]{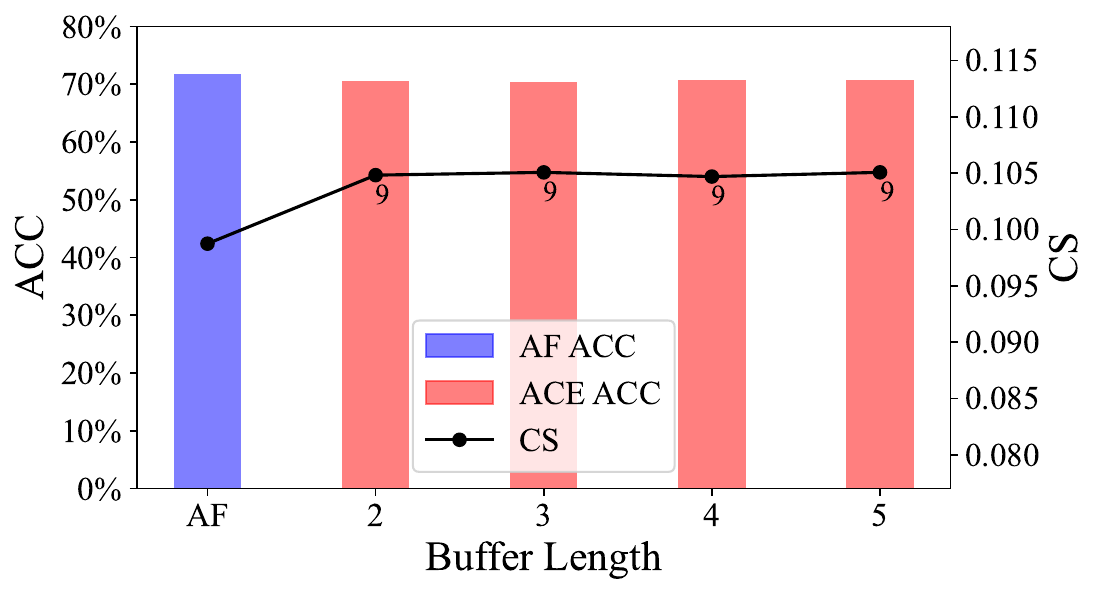}
    }
    \hfill
    \subfloat[CFFL-POW]{
        \label{fig:capparatus}
        \centering
        \includegraphics[width=0.32\linewidth]{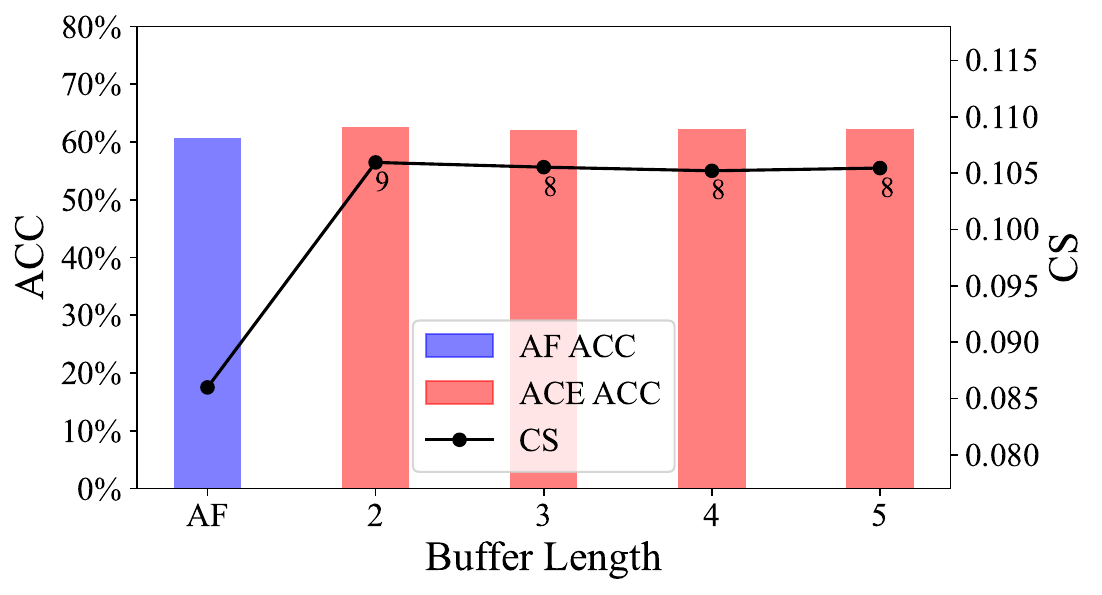}
    }
    \hfill
    \subfloat[CFFL-CLA]{
        \label{fig:capparatus}
        \centering
        \includegraphics[width=0.32\linewidth]{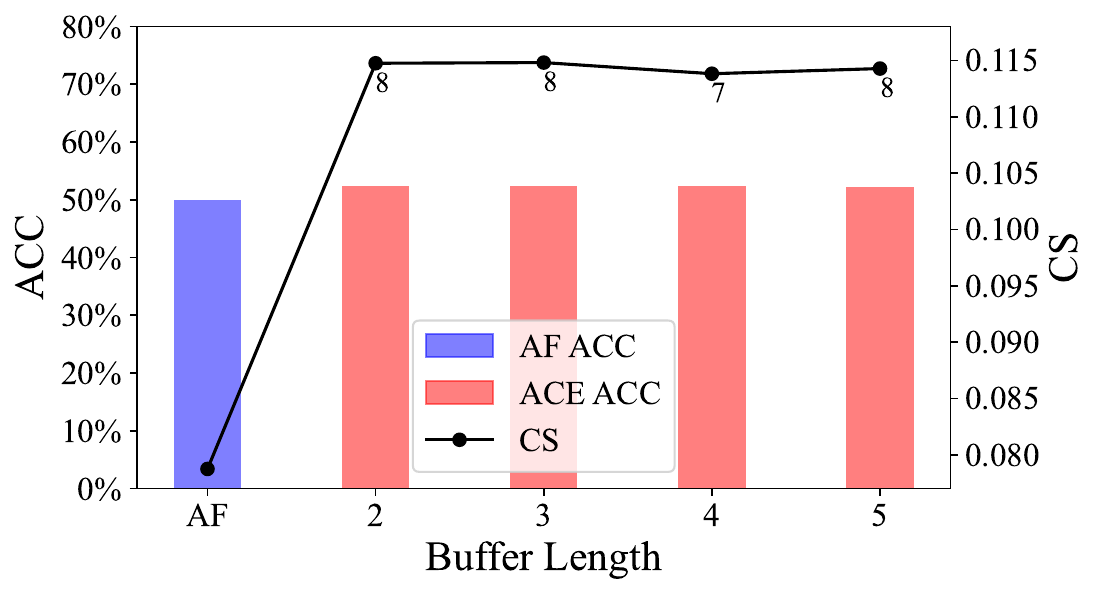}
    }
    \\
    \subfloat[RFFL-UNI]{
        \label{fig:capparatus}
        \centering
        \includegraphics[width=0.32\linewidth]{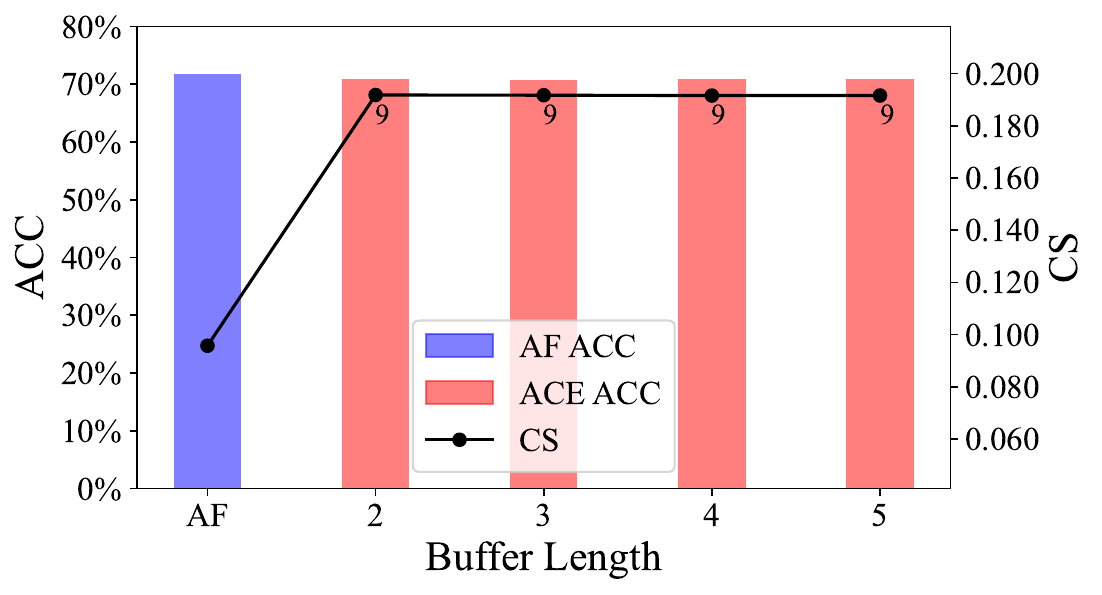}
    }
    \hfill
    \subfloat[RFFL-POW]{
        \label{fig:capparatus}
        \centering
        \includegraphics[width=0.32\linewidth]{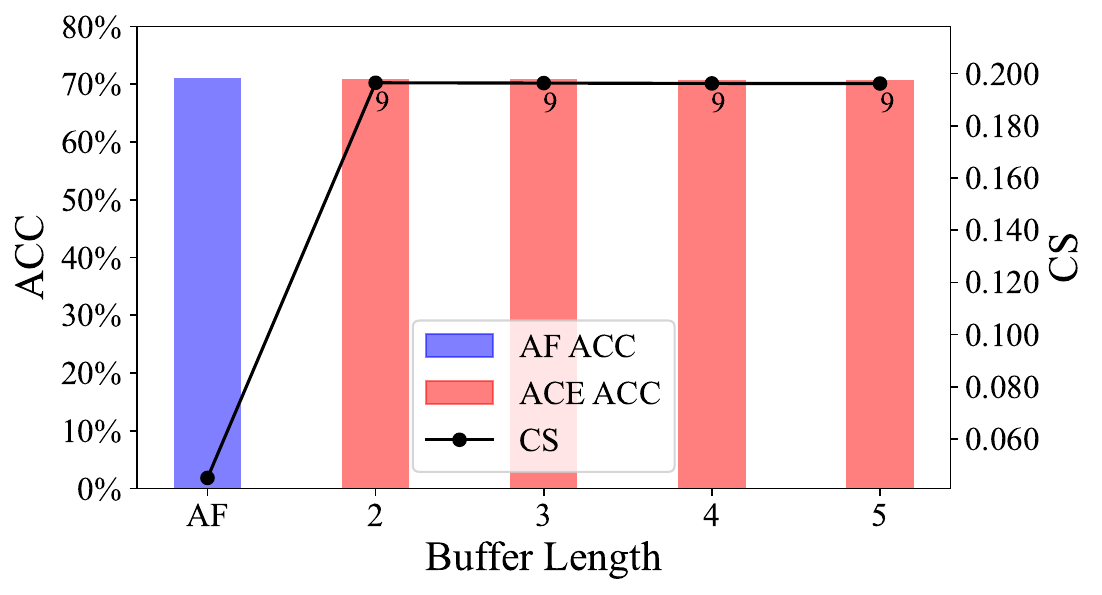}
    }
    \hfill
    \subfloat[RFFL-CLA]{
        \label{fig:capparatus}
        \centering
        \includegraphics[width=0.32\linewidth]{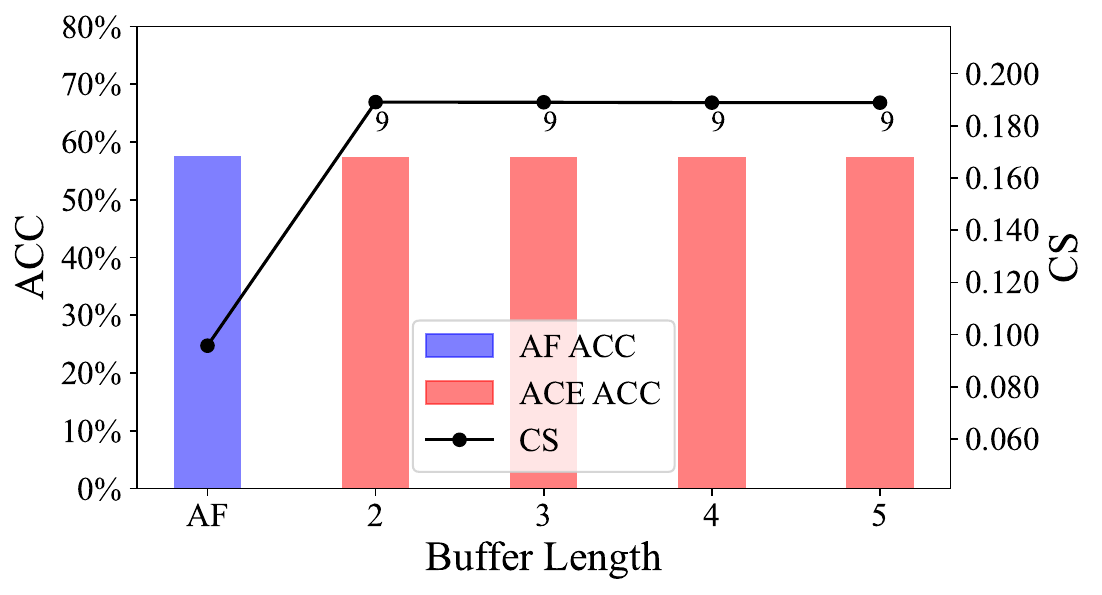}
    }
    
    \caption{Ablation study with different buffer lengths $m=2,3,\ldots,5$. The numbers annotated in the figure are the rank gains. \ours~with non-zero buffer lengths significantly improves the rank gain, without degrading the accuracy. AF is the abbreviation for Attack Free.}
    \label{fig:ab-mem-len}
\end{figure*}

%% file: float_element/fig_local_envolution.tex
\begin{figure*}[htbp]
    \captionsetup[subfloat]{farskip=0pt,captionskip=0pt}
    \centering

    \subfloat[CFFL-UNI]{
        \label{fig:capparatus}
        \centering
        \includegraphics[width=0.32\linewidth]{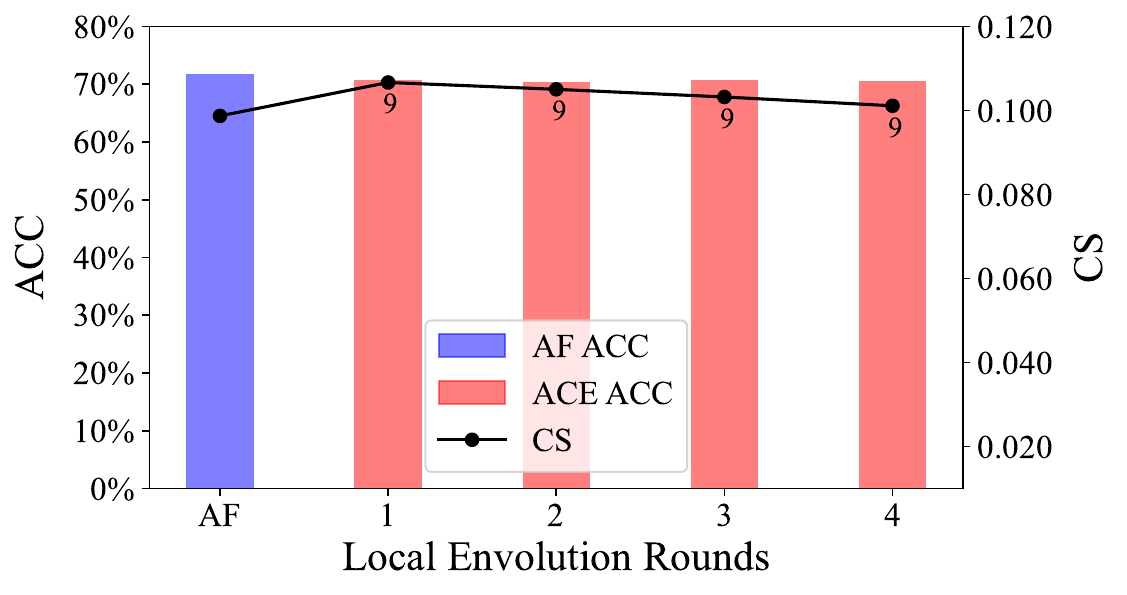}
    }
    \hfill
    \subfloat[CFFL-POW]{
        \label{fig:capparatus}
        \centering
        \includegraphics[width=0.32\linewidth]{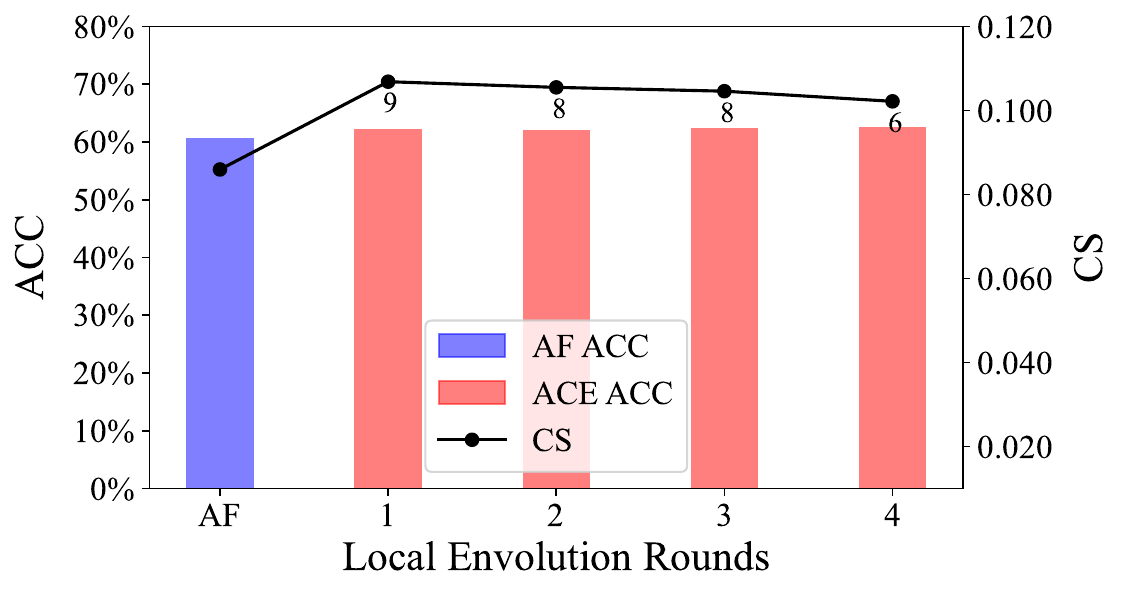}
    }
    \hfill
    \subfloat[CFFL-CLA]{
        \label{fig:capparatus}
        \centering
        \includegraphics[width=0.32\linewidth]{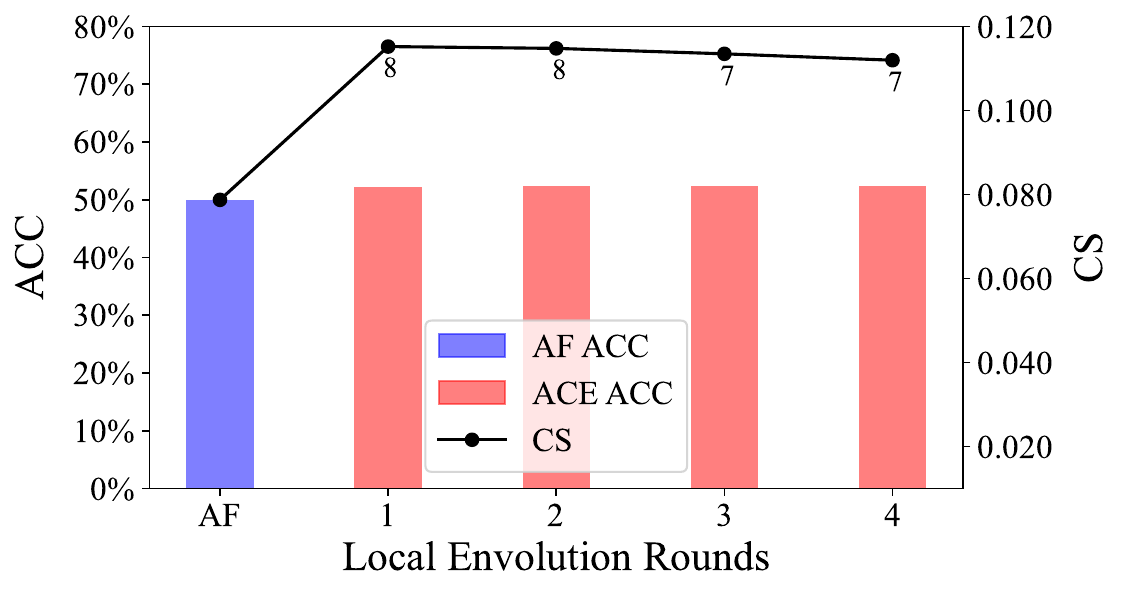}
    }
    \\
    \subfloat[FedSV-UNI]{
        \label{fig:capparatus}
        \centering
        \includegraphics[width=0.32\linewidth]{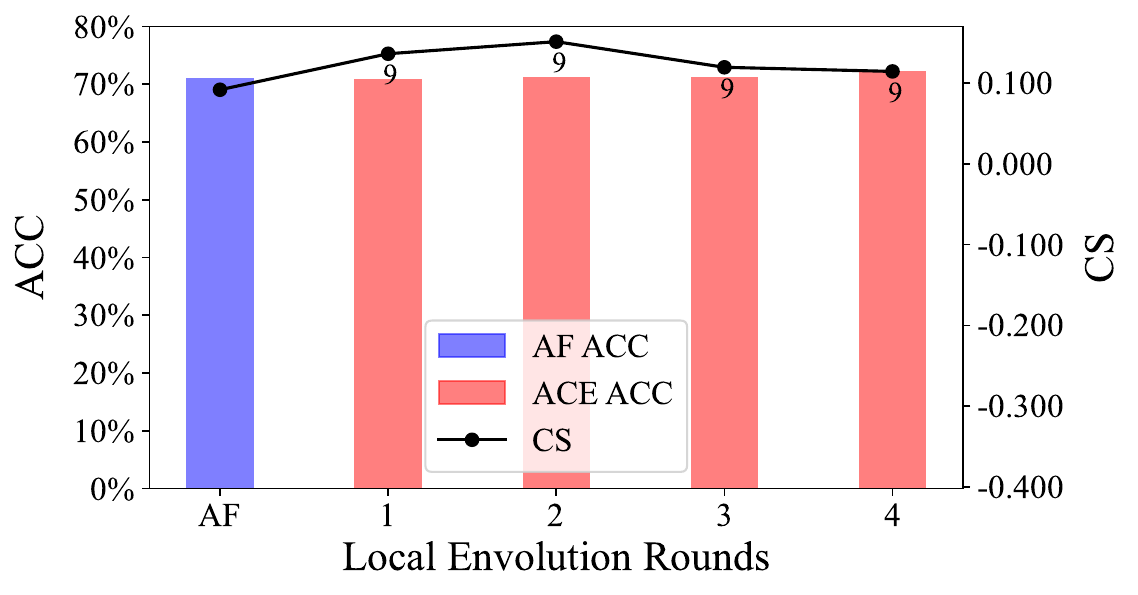}
    }
    \hfill
    \subfloat[FedSV-POW]{
        \label{fig:capparatus}
        \centering
        \includegraphics[width=0.32\linewidth]{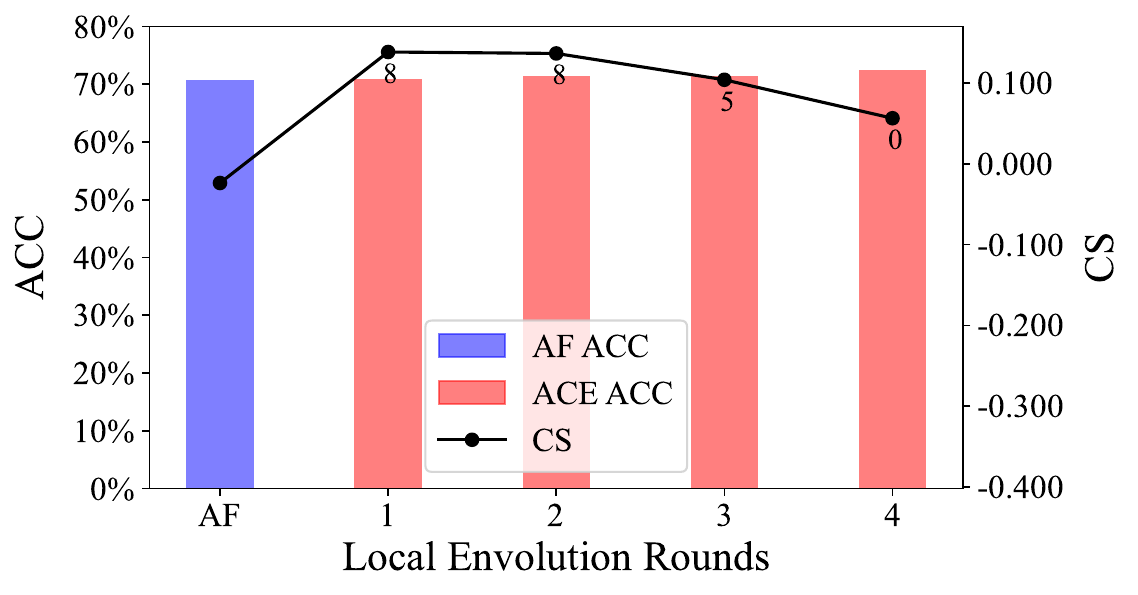}
    }
    \hfill
    \subfloat[FedSV-CLA]{
        \label{fig:capparatus}
        \centering
        \includegraphics[width=0.32\linewidth]{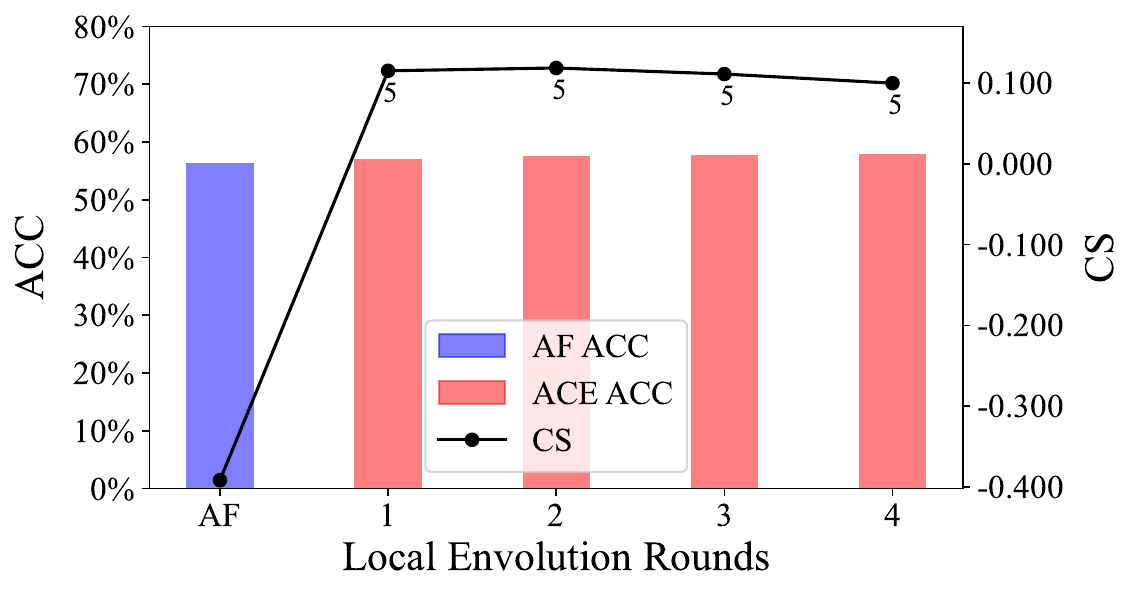}
    }
    \\
    \subfloat[LOO-UNI]{
        \label{fig:capparatus}
        \centering
        \includegraphics[width=0.32\linewidth]{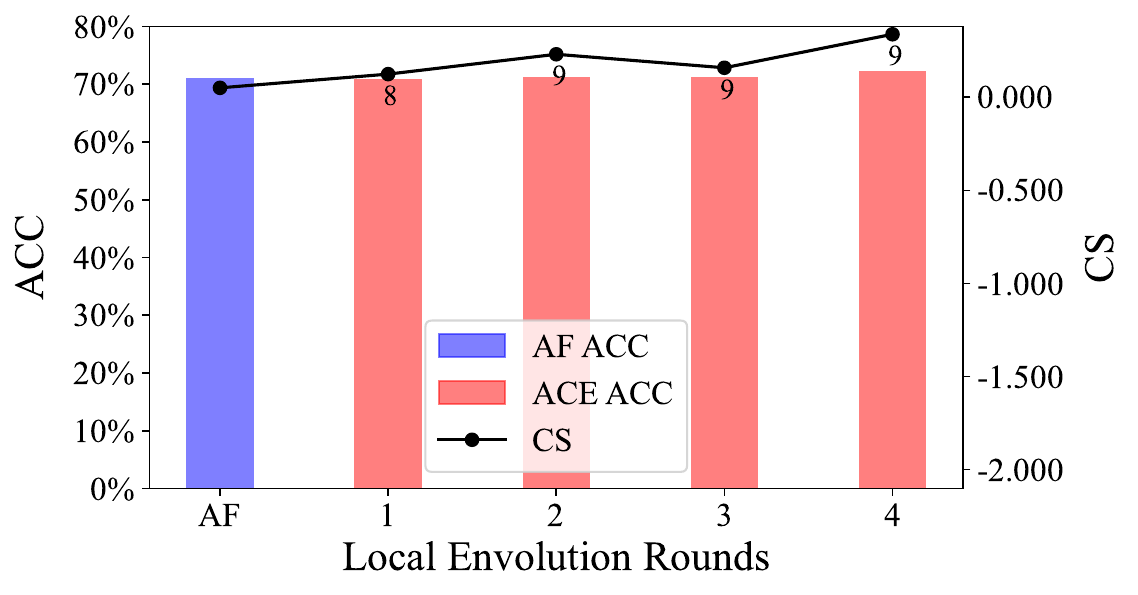}
    }
    \hfill
    \subfloat[LOO-POW]{
        \label{fig:capparatus}
        \centering
        \includegraphics[width=0.32\linewidth]{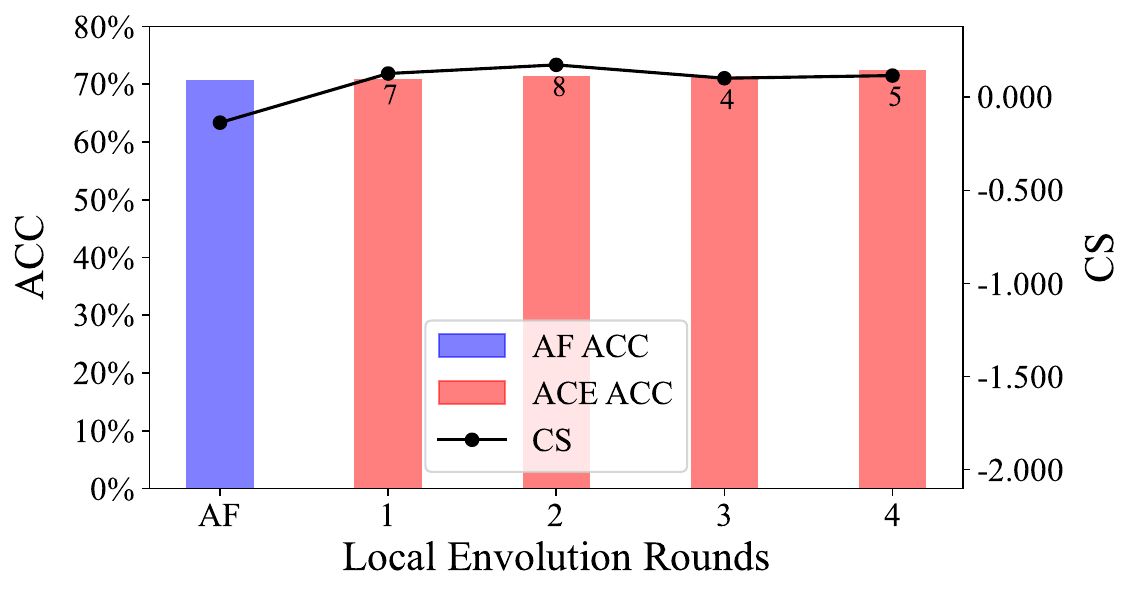}
    }
    \hfill
    \subfloat[LOO-CLA]{
        \label{fig:capparatus}
        \centering
        \includegraphics[width=0.32\linewidth]{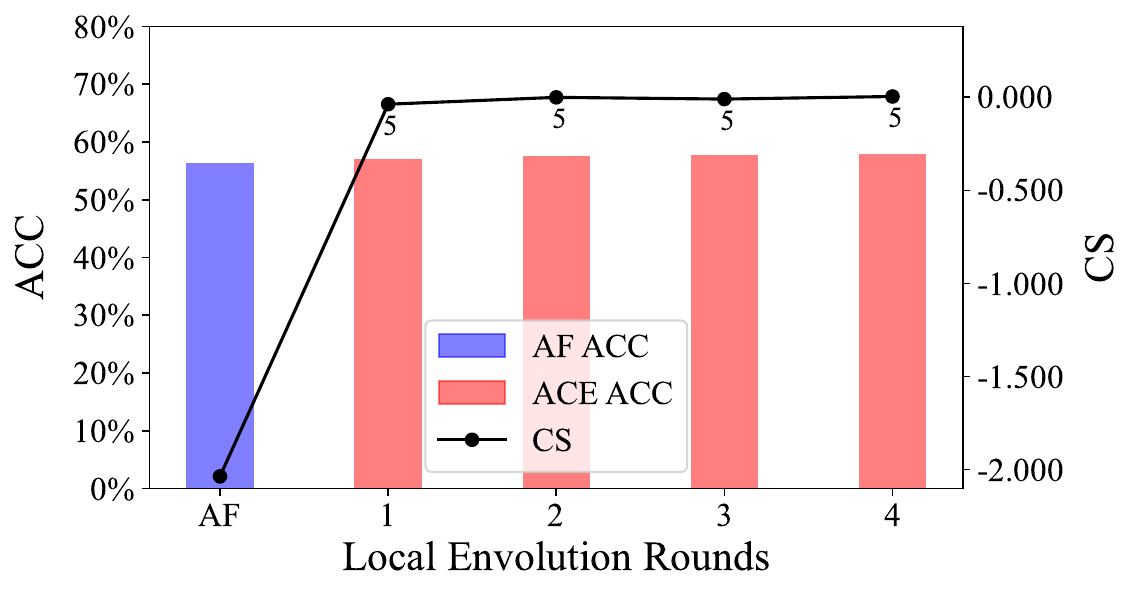}
    }

    \caption{Ablation analysis on the effect of local evolution rounds when CFFL, FedSV, and LOO are used as contribution evaluation methods. The numbers in the figure annotates the rank gain $\Delta R$. AF is the abbreviation for Attack Free.}
    \label{fig:ab-lcaol-evolution}
\end{figure*}

%% file: float_element/table_abnormal-warmup.tex
\begin{table*}[t!]
\centering
\caption{Effect of the strategies that each malicious client take for preliminary iteration and threshold based filtering. In \emph{strategy one} ($s_1$), the attacker could learn the local model update from the local training dataset of a client. In \emph{strategy two} ($s_2$), the attacker could use Delta Weight. We have four combinations as the attacker could either use strategy one or two for preliminary iteration and threshold based filtering. We denote these combinations by $\text{\{strategy for preliminary iteration\}} \times \text{\{strategy for threshold based filtering\}}$.
This leads to combinations $s_1\times s_1$, $s_1\times s_2$, $s_2\times s_1$, and $s_2\times s_2$ .
Overall $s_2\times s_2$ provides the malicious client the best performance in terms of accuracy, contribution score, and rank gain.
}
\label{tab:ab-warmup}
\resizebox{\linewidth}{!}{%
\begin{tabular}{cc || cccc |cccc|cccc}
\toprule 
\multirow{2}{*}{\makecell{Contribution \\
evaluation}} & \multirow{2}{*}{Metric} & \multicolumn{4}{c|}{UNI} & \multicolumn{4}{c|}{POW} & \multicolumn{4}{c}{CLA} \\ 
& & $s_1\times s_1$ & $s_1\times s_2$ & $s_2\times s_1$ & $s_2\times s_2$ &  $s_1\times s_1$ & $s_1\times s_2$ & $s_2\times s_1$ & $s_2\times s_2$ &  $s_1\times s_1$ & $s_1\times s_2$ & $s_2\times s_1$ & $s_2\times s_2$  \\ \midrule
\multirow{3}{*}{CFFL} & ACC & 71.77\% & 71.46\% & 71.50\% & 70.44\% & 62.33\% & 62.35\% & 62.18\% & 62.03\% & 52.12\% & 52.51\% & 52.61\% & 52.45\%  \\
& CS & 0.102 & 0.103 & 0.103 & 0.105 & 0.104 & 0.104 & 0.104 & 0.106 & 0.095 & 0.115 & 0.111 & 0.115\\
& $\Delta$R & 9 & 9 & 9 & 9 & 7 & 7 & 7 & 8 & 3 & 8 & 7 & 8 \\ \midrule
\multirow{3}{*}{RFFL} & ACC & 70.81\% & 70.81\% & 70.72\% & 70.72\% & 70.88\% & 70.88\% & 70.90\% & 70.90\% & 57.34\% & 57.34\% & 57.36\% & 57.36\%  \\
& CS & 0.192 & 0.192 & 0.192 & 0.192 & 0.194 & 0.194 & 0.196 & 0.196 & 0.188 & 0.188 & 0.189 & 0.189 \\
& $\Delta$R & 9&9&9&9&9&9&9&9&9&9&9&9\\ \bottomrule
 
\end{tabular}%
}
\end{table*}

%% file: float_element/table_selected_clients.tex
\newcommand{\moe}[1]{}

\begin{table*}[!t]
\small
\centering
\caption{Effect of the fraction of selected clients by the server in each communication round on \ours~under contribution evaluation methods SV and LOO. Note that other three contribution evaluation methods require the server to select all clients in each communication round. The results show \ours~is consistently effective. 
}
\label{tab:selected-clients}

    \begin{tabular}{c c c || c c  | c c  |  c c}
    \toprule 
    \multirow{2}{*}{\makecell{Contribution\\evaluation}}& \multirow{2}{*}{\makecell{Metrics}} & \multirow{2}{*}{\makecell{Fraction of \\selected clients}} & \multicolumn{2}{c|}{UNI} & \multicolumn{2}{c|}{POW} & \multicolumn{2}{c}{CLA} \\
    & & & Attack Free & \ours & Attack Free & \ours & Attack Free & \ours \\ \midrule
    \multirow{9}{*}{FedSV} & \multirow{3}{*}{ACC} & 50\% & 70.87\% &  	71.73\% 	 & 	69.93\%  	& 	71.69\%  & 	57.30\%  	 & 	59.90\% 	 \\
    
    & & 70\%& 70.86\%	& 71.30\%	& 70.50\%	& 71.39\%	& 55.64\%	& 58.17\%	 \\
    
    & & 100\% & 71.16\% & 71.30\% & 70.82\% & 71.45\% &56.32\% &57.60\%  \\ \cmidrule{2-9}

 & \multirow{3}{*}{CS} & 50\%  & 0.0825	 &0.1157	&-0.0162	 &0.1142	 &-0.1825	 &0.0812	 \\
 
    & & 70\% & 0.0882	& 0.1358	 & -0.0187	 & 0.1316	 & -0.3108	 & 0.0882	 \\
    
    & & 100\% & 0.0918 & 0.1513 & -0.0237 & 0.1367 & -0.3916 & 0.1187 \\ \cmidrule{2-9}

     & \multirow{3}{*}{$\Delta$R} & 50\% & 0	&8	&0	& 6	&0	&5 \\
 
    & & 70\% & 0	 & 9	 & 0	& 8	& 0	 & 5	 \\
    
    & & 100\% & 0 & 9& 0 &8 & 0 &5  \\ \midrule
    
    \multirow{9}{*}{LOO} & \multirow{3}{*}{ACC} & 50\% & 70.87\% &  	71.73\% 	 & 	69.93\%  	& 	71.69\% 	 & 	57.30\%  	 & 	59.90\%   \\
    
     & & 70\%& 70.86\%	& 71.30\%	& 70.50\%	& 71.39\%	& 55.64\%	& 58.17\%	 \\
    
    & & 100\% & 71.16\% & 71.30\% & 70.82\% & 71.45\% &56.32\% &57.60\%  \\ \cmidrule{2-9}
    
    & \multirow{3}{*}{CS} & 50\% & 0.0120	 & 0.2641	 & -0.1088	 & 0.1606	 & -2.0835	 & -0.0693	   \\
    
    & & 70\% & 0.0346	 & 0.2111	 & -0.1362	 & 0.1528	 & -1.9926	& -0.0598	 \\
    
    & & 100\% & 0.0508 & 0.2311 & -0.1361 & 0.1743 & -0.3916 & 0.1187 \\ \cmidrule{2-9}
    & \multirow{3}{*}{$\Delta$R} & 50\% &  0	 & 9	 & 0	 & 7	 & 0	 & 5	   \\
    
    & & 70\% & 0 & 9	 & 0	 & 7	 & 0	 & 5	  \\
    
    & & 100\% & 0 & 9 & 0 & 8 & 0 & 5 \\
    \bottomrule
    \end{tabular}%

\end{table*}

%% file: float_element/fig_attacker_number_cs2.tex


\begin{figure*}[htbp]
    \captionsetup[subfloat]{farskip=0pt,captionskip=0pt}
    \centering
    \hfill
    \subfloat[UNI]{
        \label{fig:capparatus}
        \centering
        \includegraphics[width=0.32\linewidth]{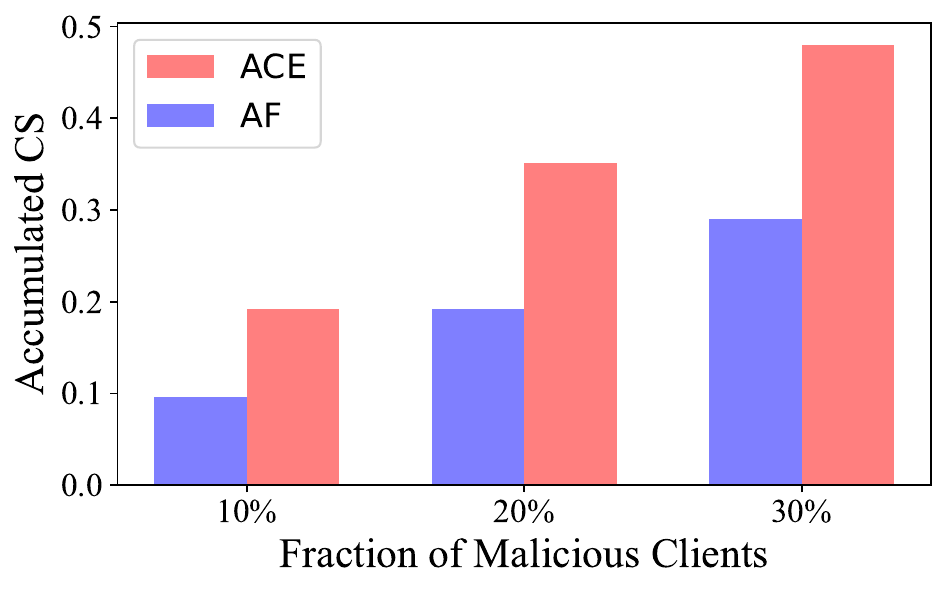}
    }
    \hfill
    \subfloat[POW]{
        \label{fig:capparatus}
        \centering
        \includegraphics[width=0.32\linewidth]{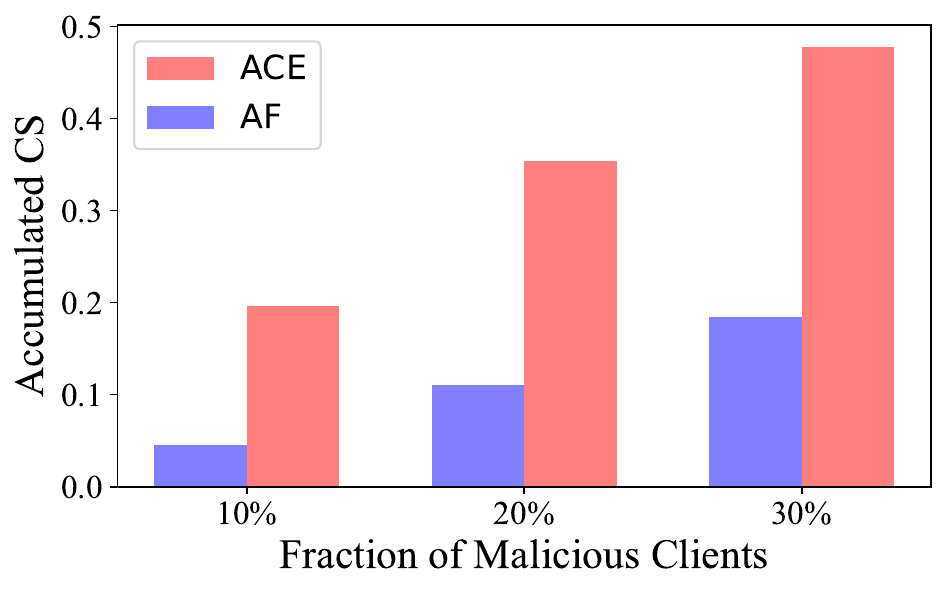}
    }
    \hfill
    \subfloat[CLA]{
        \label{fig:capparatus}
        \centering
        \includegraphics[width=0.32\linewidth]{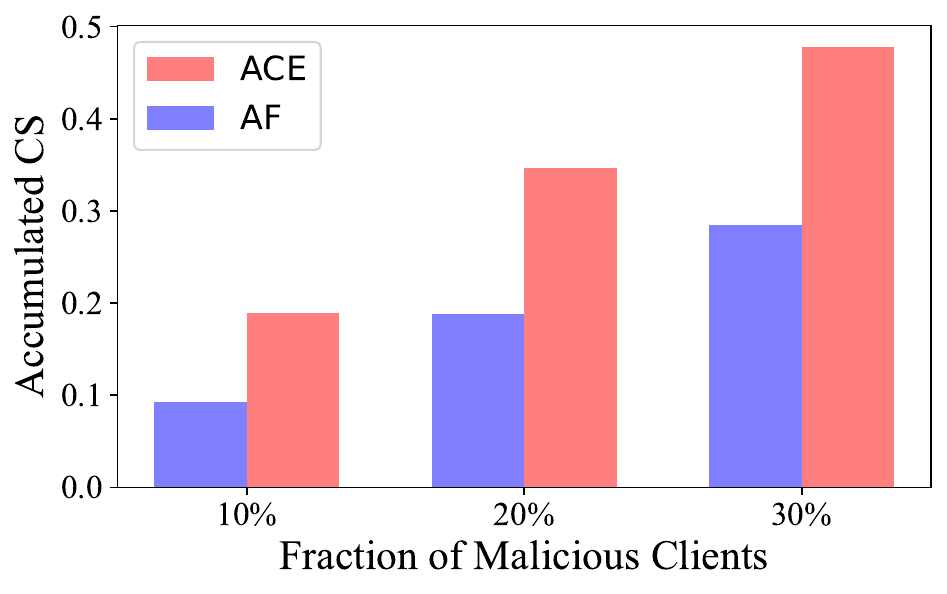}
    }

    \caption{This figure presents the accumulated CS (summation of the contribution scores of all malicious clients) under UNI, POW, and CLA data partitions when the fraction of malicious clients varies from $10\%$ to $30\%$. The server employs RFFL as the contribution evaluation method.
    The results indicate that \ours~is effective under different fractions of malicious clients. }
    \label{fig:attacker-number-cs-appx}
\end{figure*}

%% file: float_element/table-c-detection.tex
\begin{table}[]
\color{black}
    \caption{\textcolor{black}{This table shows the effect of values of $c$ on defense performance under UNI data partition. The contribution evaluation method is RFFL. We observe that the defense methods show marginal performance gain as $c$ increases. This indicates the stealthiness of \ours~towards the choice of $c$. If Precision and Recall are 0, F1-Score is not defined and denoted as N/A.}}
    \centering
    \resizebox{\columnwidth}{!}{
    \begin{tabular}{c c| c c c c} \toprule
        Detection & Metric & $c=1$ & $c=1.5$ & $c=2$ & $c=2.5$  \\ \midrule
         \multirow{3}{*}{Multi-Krum}& Precision &0.017 & 0.017 & 0.017 & 0.017\\
         & Recall & 0.017 & 0.017 & 0.017 & 0.017 \\
         & F1-Score & 0.017 & 0.017 & 0.017 & 0.017 \\ \midrule
         \multirow{3}{*}{Trimmed-Mean}& Precision & 0.017 & 0.017 & 0.017 & 0.017 \\
         & Recall & 0.017 & 0.017 & 0.017 & 0.017\\
         & F1-Score & 0.017 & 0.017 & 0.017 & 0.017\\ \midrule
         \multirow{3}{*}{FABA}& Precision & 0 & 0.017 & 0.017 & 0.017\\
         & Recall & 0 & 0.017 & 0.017 & 0.017 \\
         & F1-Score & N/A & 0.017 & 0.017 & 0.017\\ \midrule
         \multirow{3}{*}{Sniper}& Precision & 0 & 0 & 0 & 0\\
         & Recall & 0 & 0 & 0 & 0 \\
         & F1-Score & N/A & N/A & N/A & N/A \\ \midrule
         \multirow{3}{*}{Foolsgold}& Precision & 0 & 0 & 0 & 0\\
         & Recall & 0 & 0 & 0 & 0 \\
         & F1-Score & N/A & N/A & N/A & N/A\\ \midrule
    \end{tabular}
    }
    \label{tab:c-detection}
\end{table}
\color{black}

%% file: float_element/fig_unknown_evaluation.tex
\begin{figure}
    \centering
    \includegraphics[width=0.8\linewidth]{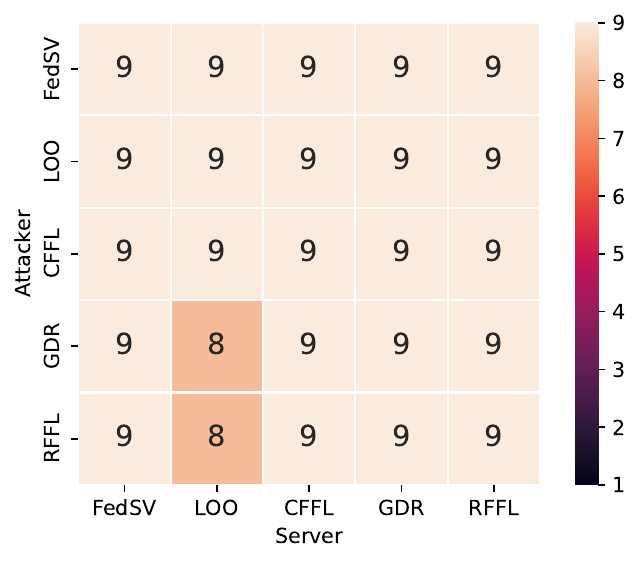}
    \caption{\textcolor{black}{This figure shows the rank gain $\Delta R$ when the attacker is unaware of the contribution evaluation method applied by the server. In this figure, the $y$-axis represents the attacker's guess on the contribution evaluation method used by the server, while the $x$-axis represents the actual contribution evaluation method employed by the server. We observe that the client launching \ours~can still successfully elevate its contribution evaluated by the server even when it is unaware of the contribution evaluation method.}}
    \label{fig:transfer-exp}
\end{figure}

%% file: float_element/table-cnn.tex
\begin{table}[h]
    \centering
    
    \caption{CNN model architectures for MNIST and CIFAR-10 datasets.}
    \begin{tabular}{|c | c|} \hline
        MNIST &  CIFAR-10  \\ \hline 
       Conv3-64 + ReLU  & Conv5-64 + ReLU \\ \hline
       Max Pool, 2x2 & Max Pool, 2x2 \\ \hline
       Conv7-16 + ReLU & Conv5-128 + ReLU \\ \hline
       Max Pool, 2x2 & Max Pool, 2x2 \\ \hline
       FC-64 & FC-64 \\ \hline
       FC-10 & FC-10 \\ \hline
       Softmax & Softmax \\ \hline
    \end{tabular}
    \label{tab:cnn-model}
    
\end{table}

%% file: float_element/algo_LBFGS.tex
\begin{algorithm}[!h]
\caption{L-BFGS Algorithm}
\begin{algorithmic}[1] 
\renewcommand{\algorithmicrequire}{\textbf{Input:}}
\renewcommand{\algorithmicensure}{\textbf{Output:}}
\REQUIRE $\Delta \mathbf{W} = [\Delta \mathbf{w}_{0}, \Delta \mathbf{w}_{1}, \cdots, \Delta \mathbf{w}_{m-1}]$, $\Delta \mathbf{G} = [\Delta \mathbf{g}_{0}, \Delta \mathbf{g}_{1}, \cdots, \Delta \mathbf{g}_{m-1}]$, and a vector $\textbf{v}$.
\ENSURE Approximation of Hessian-vector product $\hat{H} \mathbf{v}$.
\STATE $\mathbf{A}=\Delta \mathbf{W}^T \Delta \mathbf{G}$
\STATE $\mathbf{D}=\operatorname{diag}(\mathbf{A})$ \COMMENT{Diagonal matrix of $\mathbf{A}$}
\STATE $\boldsymbol{L}=\operatorname{tril}(\boldsymbol{A})$ \COMMENT{Lower triangular matrix of $\mathbf{A}$}
\STATE $\sigma=\left(\Delta \mathbf{g}_{m-1}^T \Delta \mathbf{w}_{m-1}\right) /\left(\Delta \mathbf{w}_{m-1}^T \Delta \mathbf{w}_{m-1}\right)$
\STATE $\mathbf{p}=\left[\begin{array}{cc}
-\mathbf{D} & \mathbf{L}^T \\
\mathbf{L} & \sigma \Delta \mathbf{W}^T \Delta \mathbf{W}
\end{array}\right]^{-1}\left[\begin{array}{c}
\Delta \mathbf{G}^T \mathbf{v} \\
\sigma \Delta \mathbf{W}^T \mathbf{v}
\end{array}\right]$
\RETURN $\sigma \mathbf{v}-[\Delta \mathbf{G} \quad \sigma \Delta \mathbf{W}] \mathbf{p}$
\end{algorithmic}
\label{Algorithm: L-BFGS}
\end{algorithm}

%% file: float_element/algo_attack.tex
\begin{algorithm}[!h]
\caption{Complete Algorithm of \ours}
\begin{algorithmic}[1]
\FOR{each round $t$}
    \STATE Receive the current global model $\mathbf{w}^t$ from the server.
    \STATE Update $\Delta \mathbf{W}^{t}$ and $\Delta \mathbf{G}^{t}$.
    \IF{in $\textit{attack rounds}$}
    \label{Algorithm Line: warm up}
        \IF{$t \leq m$}
        \STATE Perform strategy for preliminary iteration to get $\mathbf{\hat{g}}^t_i$.
        \ELSE
        \FOR{$t'$ in local evolution rounds}
            \STATE $\mathbf{v} = \mathbf{w}^{t+t'} - \mathbf{w}^{t+t'-1}$
            \STATE $H^{t+t'} \mathbf{v} = \textsc{L-BFGS}(\Delta \mathbf{W}^{t+t'}, \Delta \mathbf{G}^{t+t'}, \mathbf{v})$
            \IF{$\|H^{t+t'} \mathbf{v}\| \leq l\|\mathbf{v}\|$}
                \IF{$t' = 0$}
                    \STATE Perform strategy for threshold based filtering to get $\mathbf{\hat{w}}^{t+1}$.
                \ELSIF{$t' > 0$}
                    \STATE $t' = t' -1$
                \ENDIF
                \STATE Break
            \ENDIF
            \STATE $\mathbf{g}^{t+t'} = \mathbf{g}^{t+t'-1} + H^{t+t'} \mathbf{v}$
            \STATE $\mathbf{\hat{w}}^{t+t'+1} = \mathbf{\hat{w}}^{t+t'} - \mathbf{g}^{t+t'}$
            \STATE Update $\Delta \mathbf{W}^{t+t'}$ and $\Delta \mathbf{G}^{t+t'}$.
        \ENDFOR
        \STATE $\mathbf{\hat{g}}^t_i = \mathbf{w}^t - \mathbf{\hat{w}}^{t+t'+1}$
        \STATE $\mathbf{\hat{g}}^t_i = c \cdot \mathbf{\hat{g}}^t_i$
        \ENDIF
    \ELSE 
        \STATE $\mathbf{g}^t_i \leftarrow \eta_{i} \nabla L(\mathcal{D}_i,\mathbf{w}^t)$ \COMMENT{Normal training}
    \ENDIF

\ENDFOR
\end{algorithmic}
\label{Algorithm: Our Attack}
\end{algorithm}